\documentclass[journal=jacsat,manuscript=article]{achemso}

\setkeys{acs}{
abbreviations=true,
articletitle=true,
biblabel=plain,
chaptertitle=true,
doi=true,
email=true,
etalmode=truncate,
keywords=true,
maxauthors=100,
super=true
}

\setcitestyle{super,open={},close={}}

\usepackage[utf8]{inputenc}
\usepackage[T1]{fontenc}
\usepackage[english]{babel}
\usepackage[version=4]{mhchem}
\usepackage{siunitx}
\usepackage{graphicx}
\usepackage{longtable}
\usepackage{geometry}
\usepackage{caption}
\usepackage{subcaption}
\usepackage{float}
\usepackage{setspace}
\usepackage{xkeyval}
\usepackage{array}
\usepackage{booktabs}
\usepackage{listings}
\usepackage{lmodern}
\usepackage{mathpazo}
\usepackage{microtype}
\usepackage{hyperref}
\hypersetup{breaklinks,colorlinks=true,linkcolor=black,filecolor=black,urlcolor=black,citecolor=black}

\usepackage{amsmath}
\usepackage{amssymb}
\usepackage{amsfonts}
\usepackage{latexsym}
\usepackage[compact]{titlesec}
\usepackage[usenames,dvipsnames]{xcolor}
\usepackage{mathptmx}
\usepackage{calc}
\usepackage{titletoc}
\usepackage{multirow}
\usepackage{amstext}
\usepackage{tabularx}
\usepackage{adjustbox}
\usepackage[section]{placeins}
\usepackage{natmove}
\usepackage{tablefootnote}
\usepackage{threeparttable}
\usepackage[d]{esvect}
\usepackage{mathtools}
\usepackage{physics}
\usepackage{textcomp}
\usepackage[none]{hyphenat}
\usepackage{xr-hyper}

\vfuzz \hfuzz
\SectionsOn
\SectionNumbersOn
\AbstractOn

\title{Engineering Excitons through Polymorphism and Dimensional Confinement in Low-Dimensional Tellurium}

\author{Gabriel Elyas Gama Araújo}
\affiliation{Federal University of Goi\'as, Institute of Physics, Campus Samambaia, 74960600 Goi\^ania, Brazil}

\author{Alexandre Cavalheiro Dias}
\affiliation{Institute of Physics, Computational Materials Laboratory, University of Bras{\'{i}}lia, Bras{\'{i}}lia, 70910-900, Brazil}
\alsoaffiliation{International Center of Physics, University of Bras{\'{i}}lia, Bras{\'{i}}lia, 70910900, Brazil}

\author{Andreia Luisa da Rosa}
\email{andreialuisa@ufg.br}
\affiliation{Federal University of Goi\'as, Institute of Physics, Campus Samambaia, 74960600 Goi\^ania, Brazil}


\date{\today}

\begin{document}

\begin{abstract}

The interplay between dimensionality, band-edge electronic structure,
and electron--hole interactions governs the optical response of
low-dimensional tellurium, yet the microscopic origin of its excitonic
behavior remains largely unexplored. Here, we investigate the
quasiparticle, excitonic, and optical properties of two-dimensional tellurium polymorphs and one-dimensional helical nanowires using
many-body $GW$ and Bethe--Salpeter equation (BSE) calculations.

Our results reveal a strong dependence of the excitonic response on
band-edge dispersion, crystal symmetry, and dimensional confinement.
$\alpha$-tellurene exhibits comparatively weak and spatially extended
electron--hole correlations, whereas the SOC-induced quasi-flat
band-edge states of $\beta$-tellurene give rise to a strongly bound and
anisotropic near-infrared exciton, with a direct binding energy of
0.40~eV at the $G_{4}W_{0}$@HSE06 level. Momentum-resolved BSE
eigenvectors and real-space exciton wave functions directly reveal the
contrasting localization and anisotropy of these excitonic states.

Remarkably, hydrogen-passivated hexagonal tellurene, previously
identified as a quantum spin Hall phase with $\mathbb{Z}_2=1$, supports
an even larger direct exciton binding energy of 0.51~eV together with a
compact and nearly isotropic in-plane excitonic distribution. This
demonstrates that strong electron--hole correlations are fully
compatible with nontrivial band topology, while the binding strength
and spatial character of the exciton remain strongly dependent on the
underlying band-edge electronic structure and crystal symmetry.
The one-dimensional helical nanowire represents the strong-confinement
limit, exhibiting a direct high exciton binding energy and a
pronounced shift of the optical response toward the ultraviolet.

These results establish tellurium polymorphism as an effective platform
for tuning exciton binding, spatial localization, optical anisotropy,
and spectral response from the near-infrared to the ultraviolet through
structural and dimensional control.

\end{abstract}

\section{Introduction}

Low-dimensional materials exhibit electronic and optical properties that differ substantially from their bulk counterparts owing to reduced dielectric screening and enhanced quantum confinement~\cite{Novoselov2004,Geim2007}. In two-dimensional semiconductors, these effects strengthen electron--hole Coulomb interactions, frequently leading to strongly bound excitons with large binding energies and pronounced optical signatures~\cite{Onida2002RMP,Chernikov2014PRL}. Consequently, excitonic effects play a fundamental role in the determination of the optical response of atomically thin materials.

Tellurium and its two-dimensional derivatives, collectively referred to as tellurene, have recently emerged as attractive platforms for exploring low-dimensional electronic and optical phenomena. Unlike conventional layered van der Waals materials, tellurene can stabilize multiple polymorphs with distinct lattice geometries and electronic structures, including the $\alpha$-, $\beta$-, pentagonal, and hexagonal phases~\cite{Zhu2017PRB,Qiao2018SciBull,Akinwande2019ACS}. The experimental realization of ultrathin tellurene layers and honeycomb tellurene structures has further demonstrated the remarkable structural flexibility and unusual electronic behavior of these systems~\cite{Wang2018NatElectron,liu-2024,hong-2023}.

The electronic properties of tellurene are highly sensitive to dimensionality, crystal symmetry, and spin--orbit coupling (SOC). Previous theoretical and experimental studies have revealed pronounced electronic anisotropy, SOC-driven band reconstruction, and nontrivial topological phases in selected tellurene polymorphs~\cite{liu-2024,araujo-2026}. In our previous work, we demonstrated that the different tellurium polymorphs exhibit markedly distinct band dispersions, SOC-induced electronic reconstructions, and topological characteristics~\cite{araujo-2026}. In particular, SOC drives the formation of quasi-flat band-edge states in $\beta$-tellurene, whereas hydrogen-passivated hexagonal tellurene realizes a quantum spin Hall phase. The reduced dispersion of the $\beta$-tellurene band edges suggests suppressed carrier kinetic energy and therefore raises the possibility of enhanced electron--hole correlations and unconventional excitonic behavior.

Although the electronic structures of tellurene polymorphs are now relatively well established, the microscopic origin of their excitonic and optical response remains largely unexplored. Reduced band dispersion near the fundamental gap enhances the electronic density of states and concentrates the excitonic amplitudes within a restricted region of reciprocal space, thereby modifying optical absorption and dielectric screening~\cite{Grillo2024TellurenePA,Andharia2018}. Because the different tellurium polymorphs span distinct dimensionalities, electronic dispersions, and topological regimes, they provide an ideal platform for identifying which microscopic electronic features govern exciton formation and optical response. Beyond band dispersion, the orbital composition of the band-edge states also influences the optical transition matrix elements, and therefore plays an important role in determining excitonic properties.

A quantitative description of these phenomena requires methods beyond conventional density functional theory (DFT). While DFT provides a reliable description of the electronic ground state, quasiparticle energies and optical excitations require an explicit treatment of many-body interactions. To address this problem, quasiparticle corrections were calculated within the $GW$ approximation~\cite{Hedin1965PR,Hybertsen1986PRB}, while optical excitations and electron--hole interactions were described by solving the Bethe--Salpeter equation (BSE)~\cite{Rohlfing2000PRB,Onida2002RMP}.

Here, we establish the microscopic connection between the band-edge electronic structure and the excitonic optical response across the tellurium family. We combine $GW$ quasiparticle calculations with the Bethe--Salpeter equation to investigate the optical properties of two-dimensional tellurene polymorphs and one-dimensional helical tellurium nanowires. By combining quasiparticle electronic structures, momentum-resolved excitonic eigenvectors, conditional electron--hole wavefunctions, and macroscopic optical spectra, we demonstrate how band-edge dispersion, orbital character, and dimensional confinement jointly determine exciton binding, optical anisotropy, and spectral selectivity across the infrared, visible, and ultraviolet spectral regions.

\section{Computational Details}

First-principles calculations were performed within density functional theory (DFT) using the Vienna \textit{Ab Initio} Simulation Package (VASP).\cite{vasp1,vasp2} The interaction between valence electrons and ionic cores was described using the projector augmented-wave (PAW) method,\cite{paw} while exchange--correlation effects were treated using both the Perdew--Burke--Ernzerhof (PBE) generalized gradient approximation\cite{pbe} and the Heyd--Scuseria--Ernzerhof (HSE06) hybrid functional.\cite{hse06} Long-range dispersion interactions were included through the DFT-D3 method with Becke--Johnson damping.\cite{dftd3} Spin--orbit coupling (SOC) was incorporated self-consistently in all calculations owing to the strong relativistic character of tellurium.

The Kohn--Sham wave functions were expanded in a plane-wave basis set with a kinetic-energy cutoff of 500~eV. All atomic structures were fully relaxed until the residual forces on each atom were smaller than $1\times10^{-6}$~eV/\AA. To avoid spurious interactions arising from periodic boundary conditions, a vacuum spacing of 12~\AA\ was introduced along the $z$ direction for the monolayers and along the transverse directions for the nanowire. The Brillouin zone was sampled using Monkhorst--Pack $k$-point meshes of $18\times18\times1$ for $\alpha$-tellurene, $16\times20\times1$ for $\beta$-tellurene, $14\times14\times1$ for the hydrogen-passivated hexagonal phase, $8\times8\times1$ for the pentagonal phase, and $1\times1\times18$ for the tellurium nanowire. These $k$-point grids were consistently maintained across all stages of calculation, including the self-consistent ground-state DFT evaluations (both PBE and HSE06), $GW$ quasiparticle corrections, and Bethe--Salpeter equation solves.

Quasiparticle energies were calculated using the eigenvalue self-consistent $EVGW_{0}$ scheme (partially self-consistent $GW_{0}$), in which the Green's function $G$ was iteratively updated through four self-consistency cycles ($G_{4}W_{0}$) while keeping the screened Coulomb interaction $W_{0}$ fixed at the initial DFT level. The Kohn--Sham wave functions were expanded in a set of plane waves with a kinetic-energy cutoff of 350~eV. A total of 200 bands per atom were included in the $GW$ calculations to ensure the convergence of the quasiparticle energies. This corresponds to 600 bands for $\alpha$-tellurene (3 atoms), 600 bands for $\beta$-tellurene (3 atoms), 800 bands for the hydrogen-passivated hexagonal phase (4 atoms), 1200 bands for the pentagonal phase (6 atoms), and 600 bands for the tellurium nanowire (3 atoms).

Optical excitations were obtained by solving the Bethe--Salpeter equation\cite{bse_og} (BSE) on top of the converged $G_{4}W_{0}$ quasiparticle energies within the Tamm--Dancoff approximation (TDA)\cite{Tamm1945,Hirata1999CPL}, considering resonant-only transitions. The BSE Hamiltonian kernel was constructed considering a structure-specific active space of occupied valence bands and unoccupied conduction bands: $9$ valence and $8$ conduction bands for $\alpha$-tellurene; $9$ valence and $6$ conduction bands for $\beta$-tellurene; $8$ valence and $8$ conduction bands for the hydrogen-passivated hexagonal phase; $10$ valence and $10$ conduction bands for pentagonal tellurene; and $12$ valence and $8$ conduction bands for the 1D helical nanowire. From the BSE calculations, we computed the complex dielectric function, optical absorption, reflectivity, refractive index, extinction coefficient, and electron energy-loss spectra (EELS). The excitonic states were analyzed through the momentum-resolved BSE eigenvectors and, for representative systems, by means of real-space exciton wave functions.

\section{Results and Discussions}

\subsection{Structural overview}

\begin{figure}[!htb]
    \centering
    \begin{subfigure}[b]{0.32\textwidth}
        \subcaption{}
        \includegraphics[width=\textwidth]{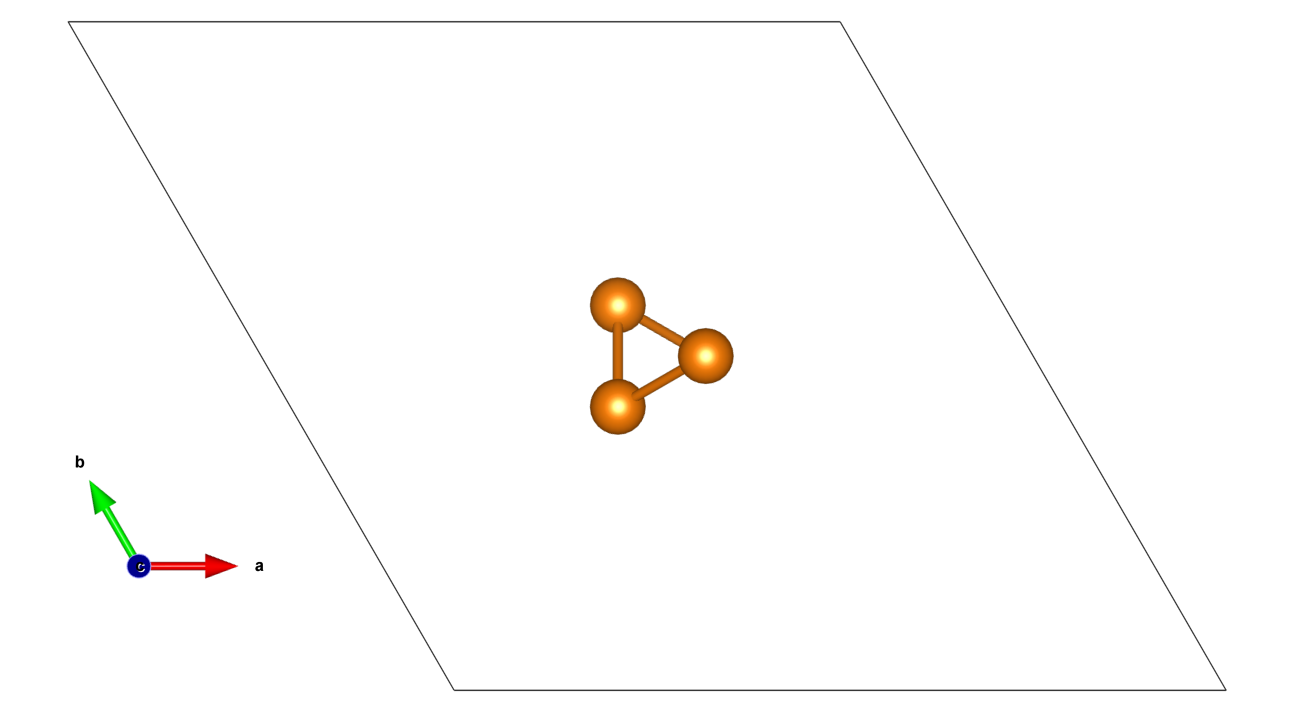}
    \end{subfigure}
    \begin{subfigure}[b]{0.32\textwidth}
        \subcaption{}
        \includegraphics[width=\textwidth]{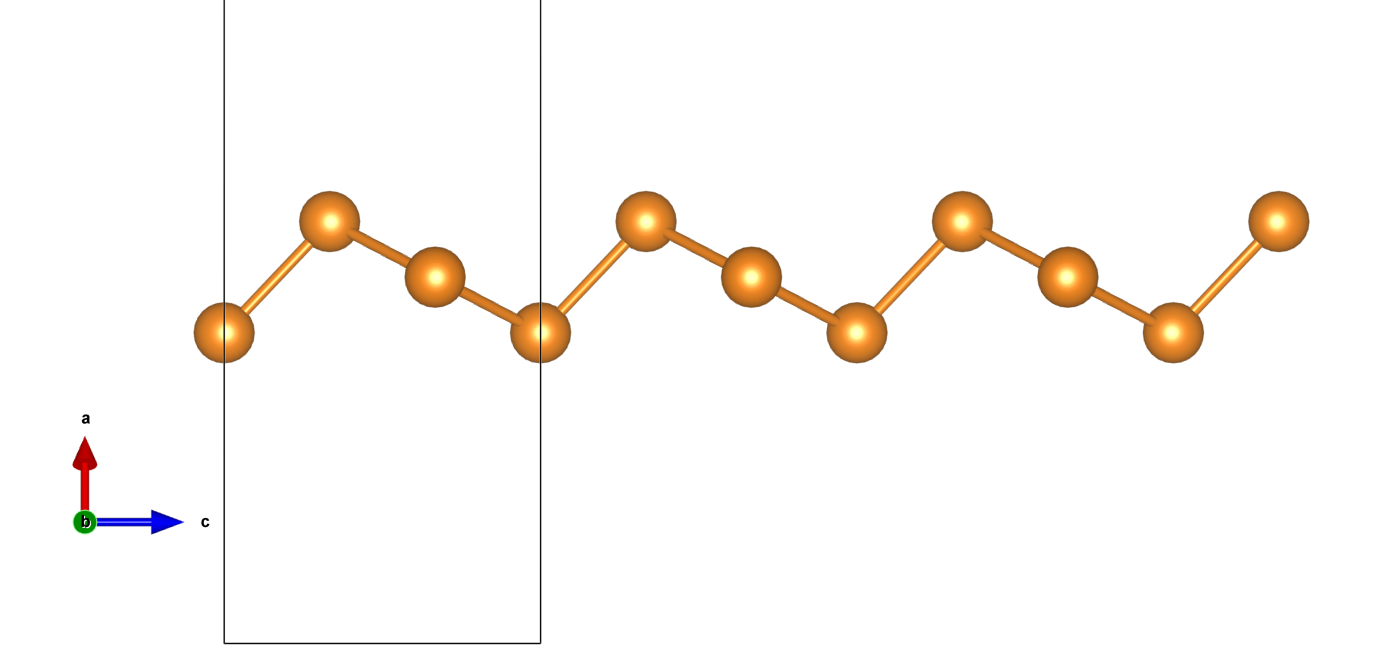}
    \end{subfigure}\\
    \begin{subfigure}[b]{0.32\columnwidth}
        \subcaption{}
        \includegraphics[width=\columnwidth]{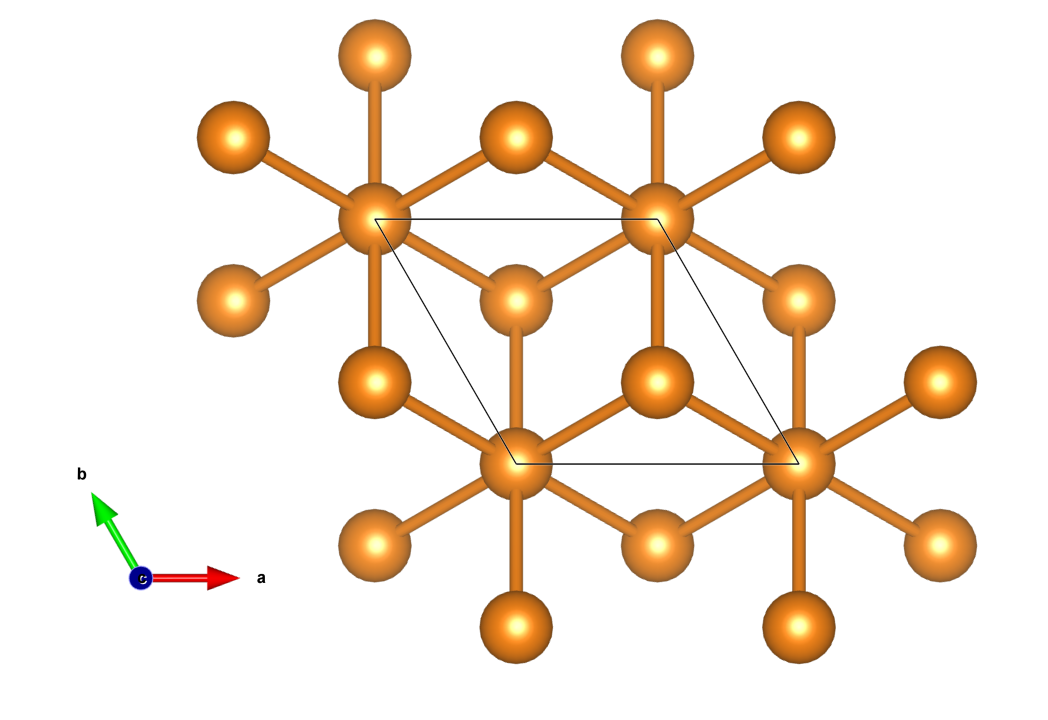}
    \end{subfigure}
    \begin{subfigure}[b]{0.32\textwidth}
        \subcaption{}
        \includegraphics[width=\textwidth]{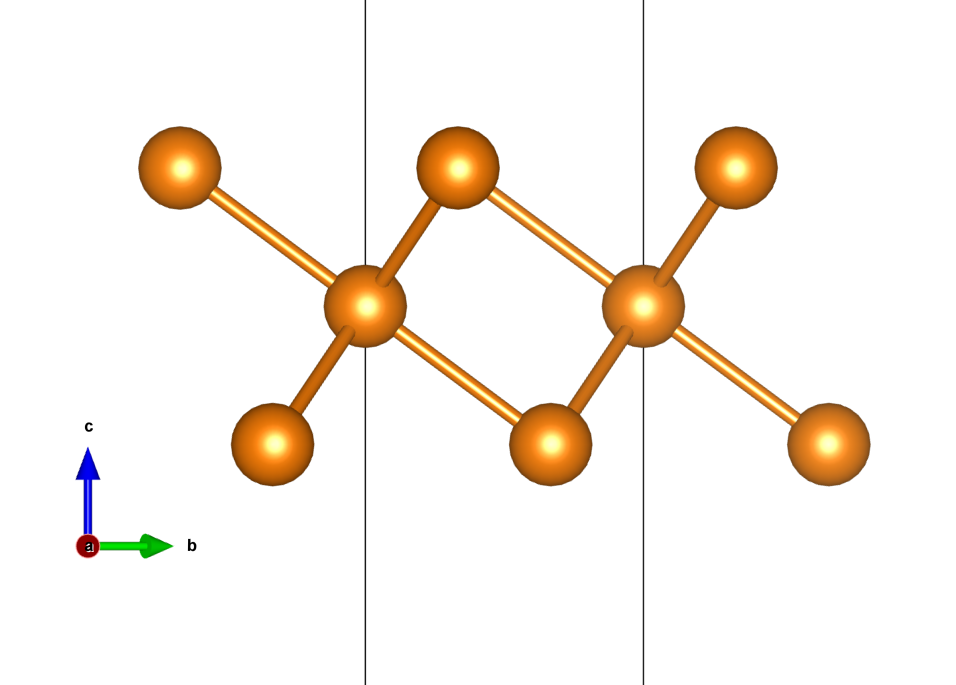}
    \end{subfigure}\\
    \begin{subfigure}[b]{0.32\columnwidth}
        \subcaption[]{}
        \includegraphics[width=\columnwidth]{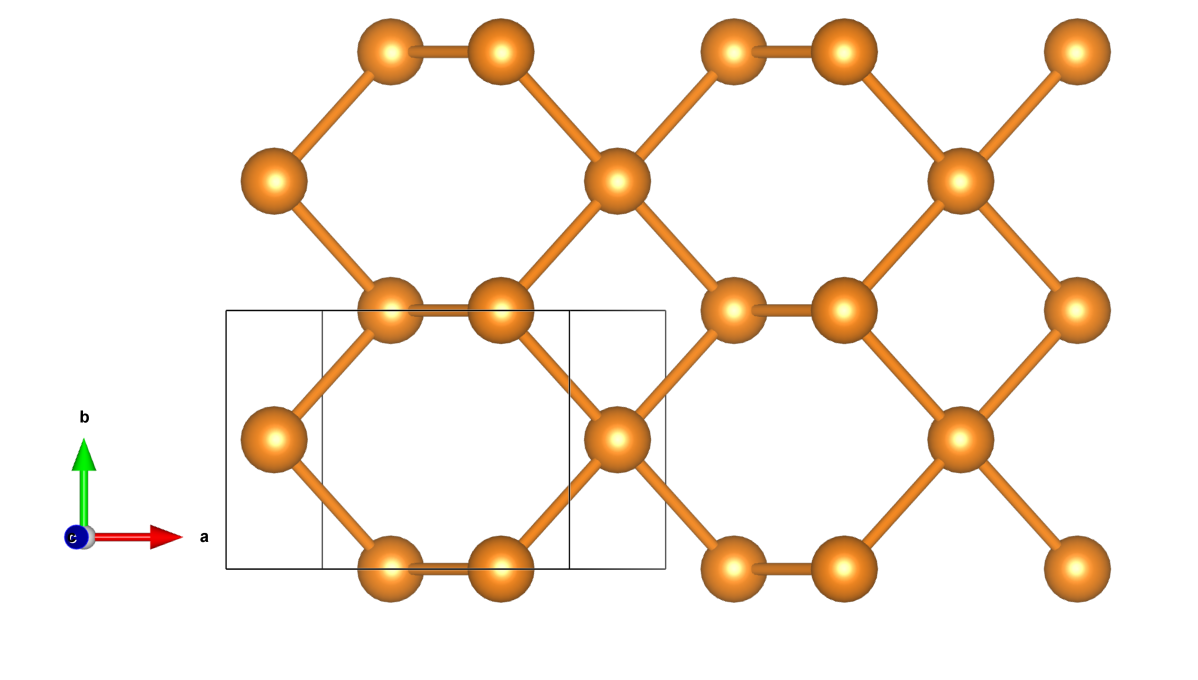}
    \end{subfigure}
    \hspace{0.1cm}
    \begin{subfigure}[b]{0.32\columnwidth}
        \subcaption[]{}
        \includegraphics[width=\columnwidth]{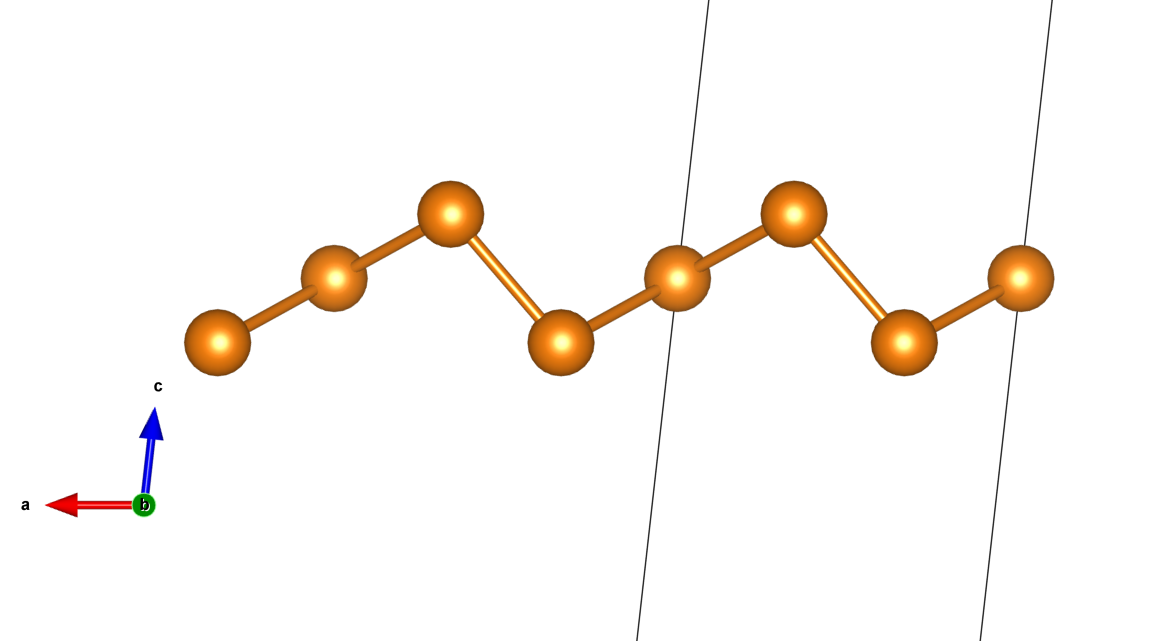}
    \end{subfigure}\\
    \begin{subfigure}[b]{0.32\textwidth}
        \subcaption{}
        \includegraphics[width=\textwidth]{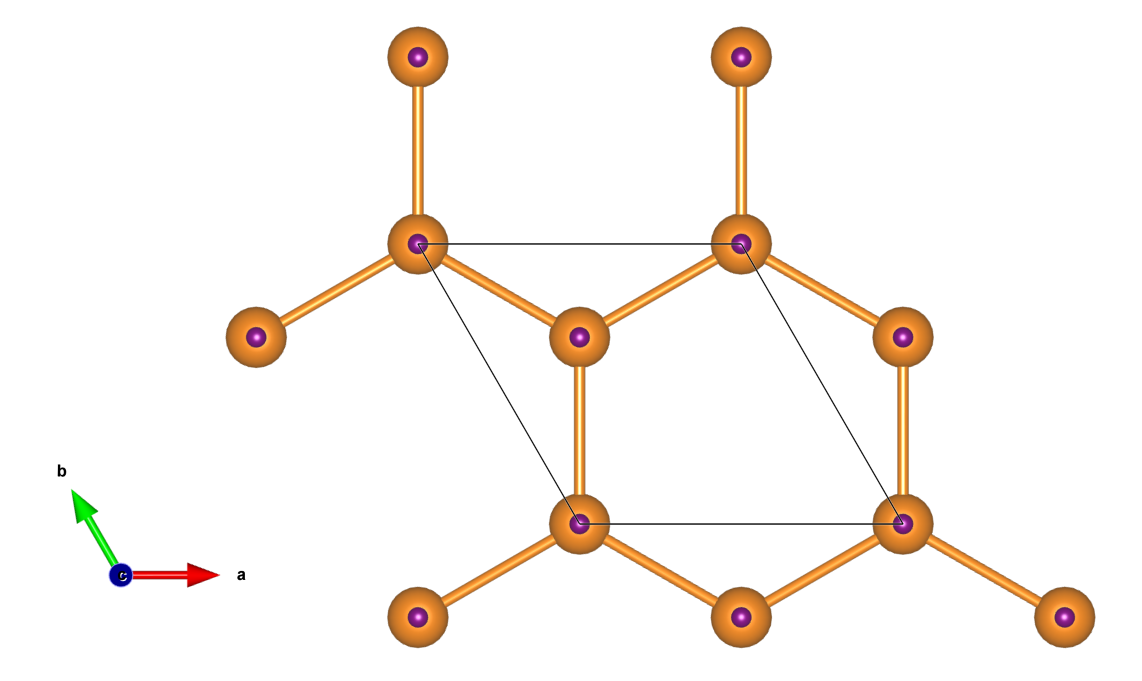}
    \end{subfigure}
    \begin{subfigure}[b]{0.32\textwidth}
        \subcaption{}
        \includegraphics[width=\textwidth]{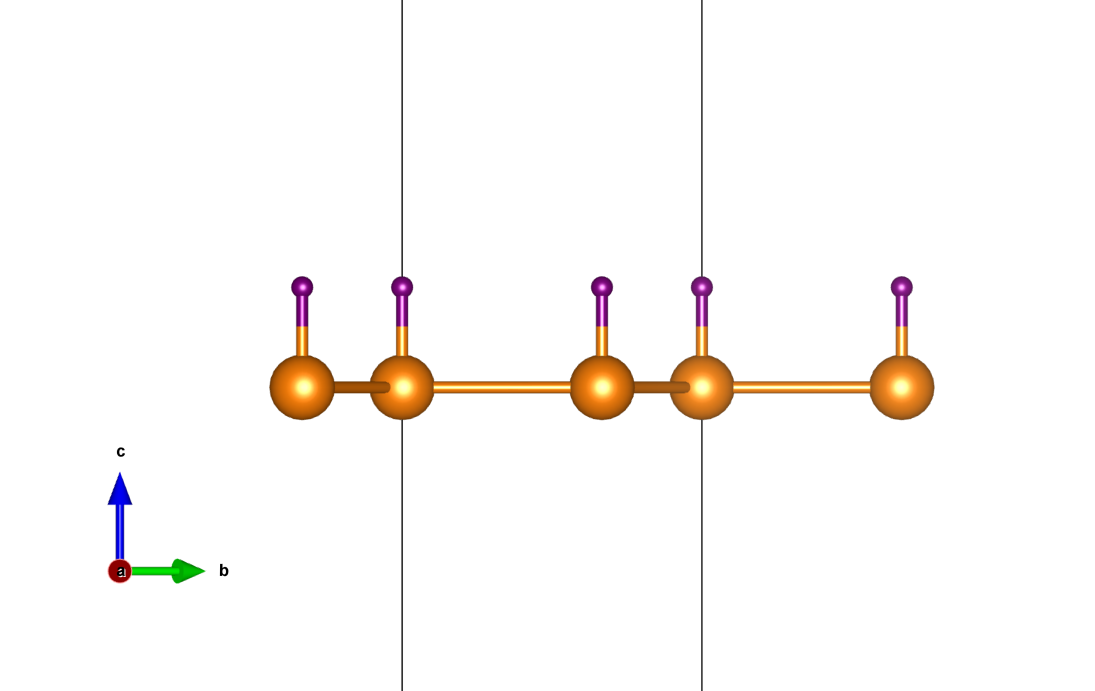}
    \end{subfigure}    \\
    \begin{subfigure}[b]{0.32\textwidth}
        \subcaption{}
        \includegraphics[width=\textwidth]{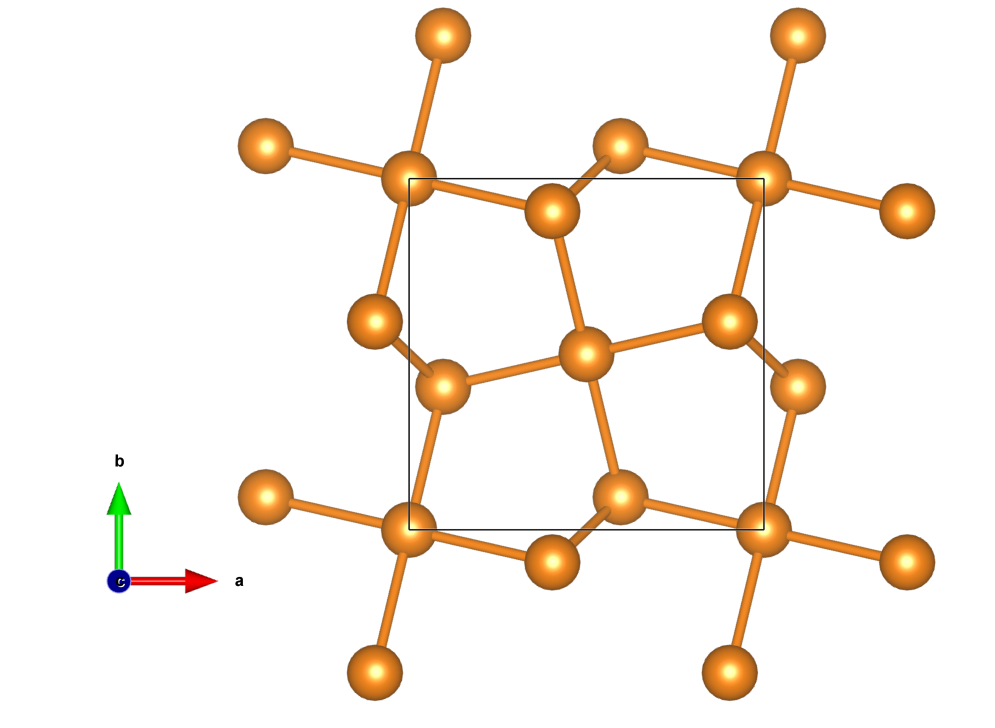}
    \end{subfigure}
    \begin{subfigure}[b]{0.32\textwidth}
        \subcaption{}
        \includegraphics[width=\textwidth]{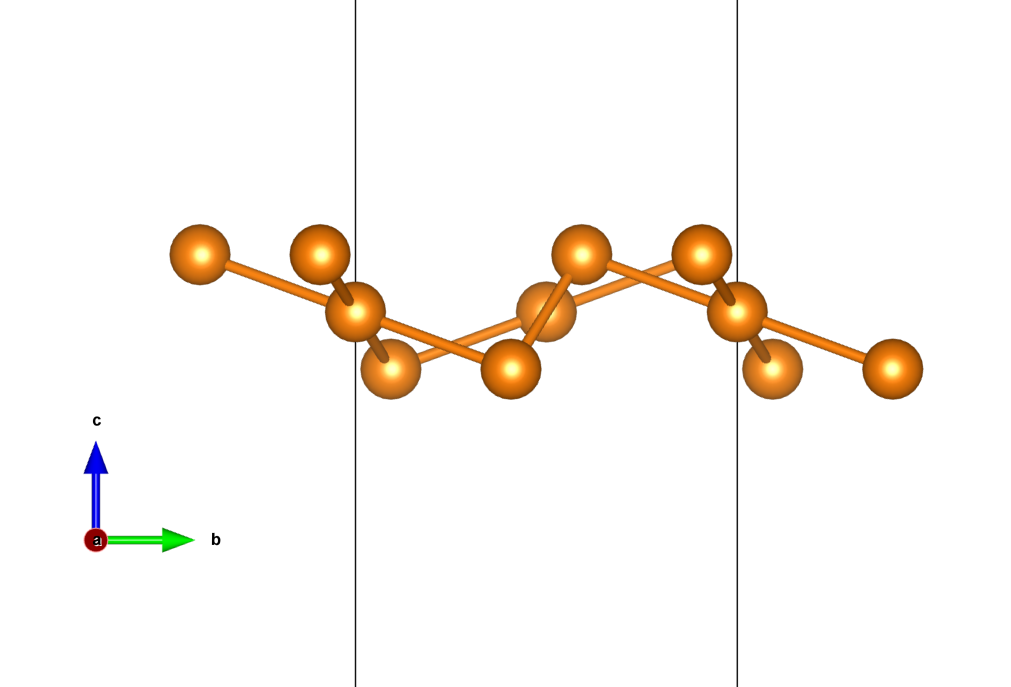}
    \end{subfigure}
\caption{Relaxed crystal structures of tellurium phases: (a,b) tellurium nanowire (Te-h), (c,d) $\alpha$-Te, (e,f) $\beta$-Te, (g,h) planar hexagonal, (i,j) buckled pentagonal.}
\label{fig:geometries}
\end{figure}

The tellurium polymorphs investigated in this work are illustrated in Fig.~\ref{fig:geometries} the $\alpha$- and $\beta$-tellurene monolayers, hydrogen-passivated hexagonal tellurene, buckled pentagonal tellurene, and the one-dimensional helical Te nanowire. Their optimized lattice parameters and characteristic Te--Te bond lengths are summarized in Table~\ref{tab:Te_structures}.

The structural, energetic, and electronic properties of these phases have been thoroughly investigated in our previous work~\cite{araujo-2026}. Therefore, we restrict the present discussion to a brief structural overview, emphasizing only the features that are relevant for understanding the quasiparticle, optical, and excitonic properties presented below.

These polymorphs cover atomically thin two-dimensional sheets and a one-dimensional helical nanowire. Such structural diversity is accompanied by distinct crystal symmetries and bonding environments that strongly influence the dielectric screening, quantum confinement, and the localization of electronic states. These factors ultimately govern the quasiparticle corrections, optical transitions, and excitonic characteristics discussed in the following sections.

\begin{table}[!htb]
\centering
\caption{Optimized lattice parameters and Te--Te bond lengths of tellurene phases and Te nanowire obtained using the PBE functional. Experimental and previous theoretical values from other works are included for comparison.}
\begin{tabular}{lccccc}
\toprule
Structure & Source & $a$ (\AA) & $b$ (\AA) & $c$ (\AA) & $d_{\rm Te-Te}$ (\AA) \\
\midrule
$\alpha$-Te 
& This work & 4.22 & 4.22 & -- & 3.03 \\
& Exp.~\cite{alpha_exp_peil,alpha_exp} & 4.23 & 4.23 & -- & 3.02 \\
\midrule
$\beta$-Te 
& This work & 5.61 & 4.22 & -- & 3.03, 2.76 \\
& Exp.~\cite{beta_exp} & 5.64 & 4.23 & -- & 3.00, 2.75 \\
\midrule
Hexagonal Te (passivated)
& This work & 5.79 & 5.79 & -- & 3.34 \\
& Exp.~\cite{hex_exp} & $6.7$ & $6.7$ & -- & -- \\
\midrule
Pentagonal Te (buckled)
& This work & 6.92 & 6.85 & -- & 2.76 \\
\midrule
Te nanowire 
& This work & -- & -- & 5.67 & 2.74 \\
& Other works \cite{teh_stab,Kramer} & -- & -- & 5.68 & 2.76 \\
\bottomrule
\end{tabular}
\label{tab:Te_structures}
\end{table}

\subsection{Quasiparticle electronic structure}

The electronic and topological properties of the tellurium polymorphs considered in this work have been comprehensively investigated in our previous study~\cite{araujo-2026}, including their electronic band structures, topological invariants, and the influence of spin--orbit coupling (SOC). Therefore, the present discussion focuses exclusively on the quasiparticle electronic structure required for the subsequent analysis of the optical and excitonic properties.

The fundamental band gaps obtained at different levels of theory are summarized in Table~\ref{tab:bandgaps}. The inclusion of many-body effects within the $G_{4}W_{0}$ approximation systematically increases the quasiparticle gaps with respect to the HSE06 results, although the magnitude of the correction depends on the structural phase. The smallest quasiparticle gaps are found for the two-dimensional $\alpha$- and hydrogen-passivated hexagonal tellurene phases, whereas the one-dimensional helical nanowire exhibits the largest gap owing to strong quantum confinement.

Beyond the gap values themselves, the dispersion of the electronic states near the valence-band maximum (VBM) and conduction-band minimum (CBM) plays a decisive role in determining the optical response. As demonstrated in our previous work, SOC induces pronounced quasi-flat bands near the band edges of $\beta$-tellurene, whereas the remaining phases exhibit considerably more dispersive frontier states. The reduced dispersion of the $\beta$ phase implies a larger effective mass and lower carrier kinetic energy\cite{araujo-2026}, conditions that favor stronger electron--hole interactions.

These quasiparticle characteristics provide the microscopic framework for understanding the optical spectra. In particular, the combination of quasiparticle gap renormalization and band-edge localization governs both the spectral position of the optical transitions and the strength of the excitonic effects obtained from the Bethe--Salpeter equation.

\begin{table}[!htb]
\caption{\label{tab:bandgaps}Quasiparticle fundamental band gaps $G_{0}W_{0}$, and $G_4W_0$ for the investigated tellurium polymorphs. Previous experimental and theoretical values are included for comparison. The labels (d) and (i) indicate the character of the band gap, corresponding to direct and indirect band gaps, respectively.}
\begin{tabular}{lccccc}
\toprule
& \multicolumn{5}{c}{QP-energy gap (eV)} \\
\cline{2-6}
Phase 
& HSE06 
& $G_{0}W_{0}$@PBE 
& $G_{4}W_{0}$@PBE 
& $G_{0}W_{0}$@HSE06 
& $G_{4}W_{0}$@HSE06 \\
\midrule

$\alpha$-Te 
& 0.75(i) 
& 0.76(i) 
& 0.77(i) 
& 0.84(i) 
& 0.85(i) \\

$\beta$-Te  
& 1.44(d) 
& 1.63(d) 
& 1.70(d) 
& 1.86(d) 
& 1.91(d) \\

Hexagonal Te (passivated) 
& 0.44(i) 
& 0.44(i) 
& 0.50(i) 
& 0.40(i) 
& 0.50(i) \\

Pentagonal Te (buckled) 
& 0.97(d) 
& 1.17(d) 
& 1.25(d) 
& 1.38(d) 
& 1.44(d) \\

Te nanowire (Te-h) 
& 2.23(i) 
& 3.10(i) 
& 3.77(i) 
& 3.75(i) 
& 4.17(i) \\
\bottomrule
\end{tabular}
\end{table}

\subsection{Optical response}

\begin{figure}[!htb]
    \centering
    \begin{subfigure}[b]{0.42\textwidth}
        \subcaption{$\alpha$-Te}
        \includegraphics[width=\textwidth]{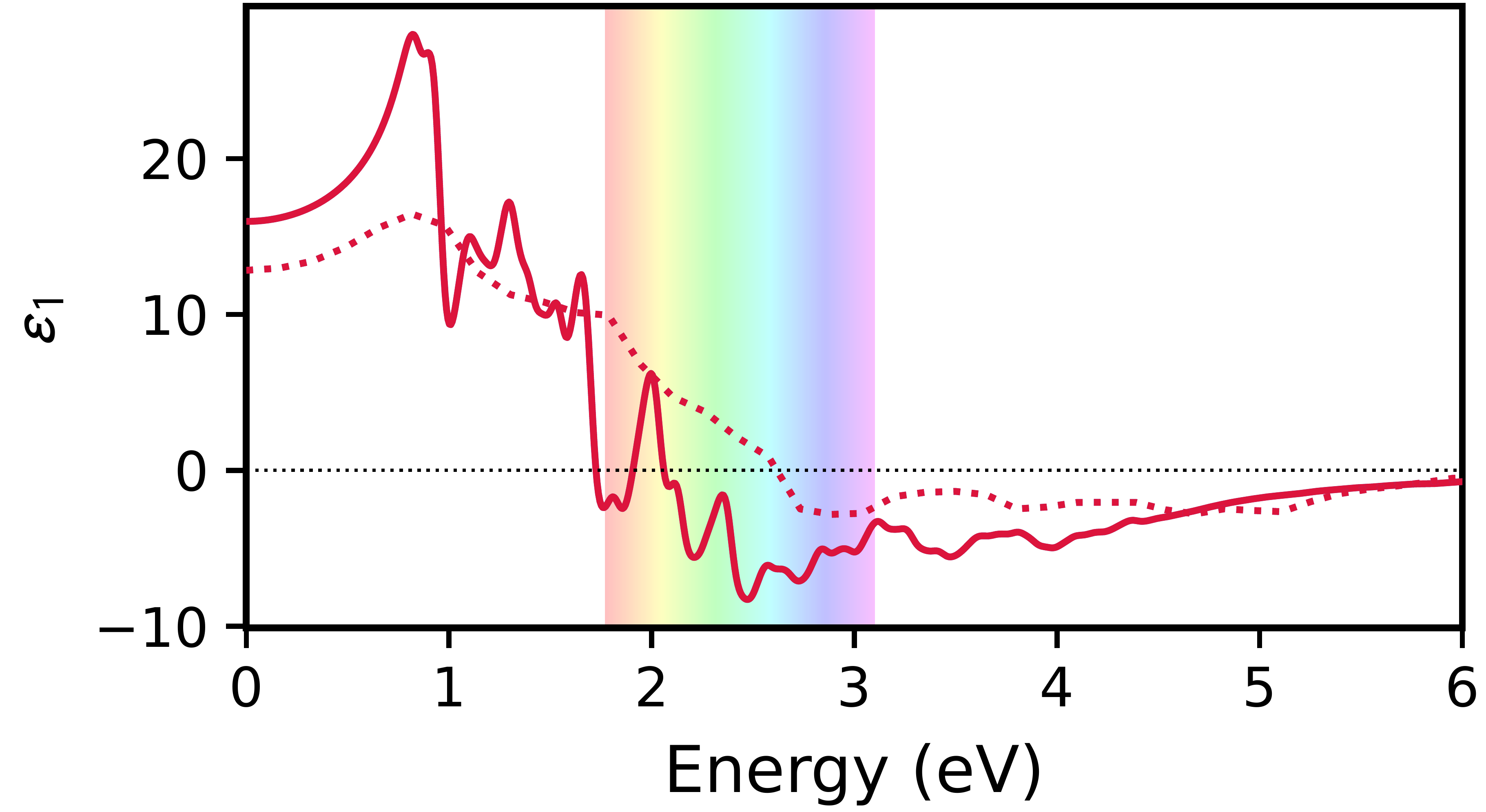}
    \end{subfigure}
    \begin{subfigure}[b]{0.42\textwidth}
        \subcaption{$\alpha$-Te}
        \includegraphics[width=\textwidth]{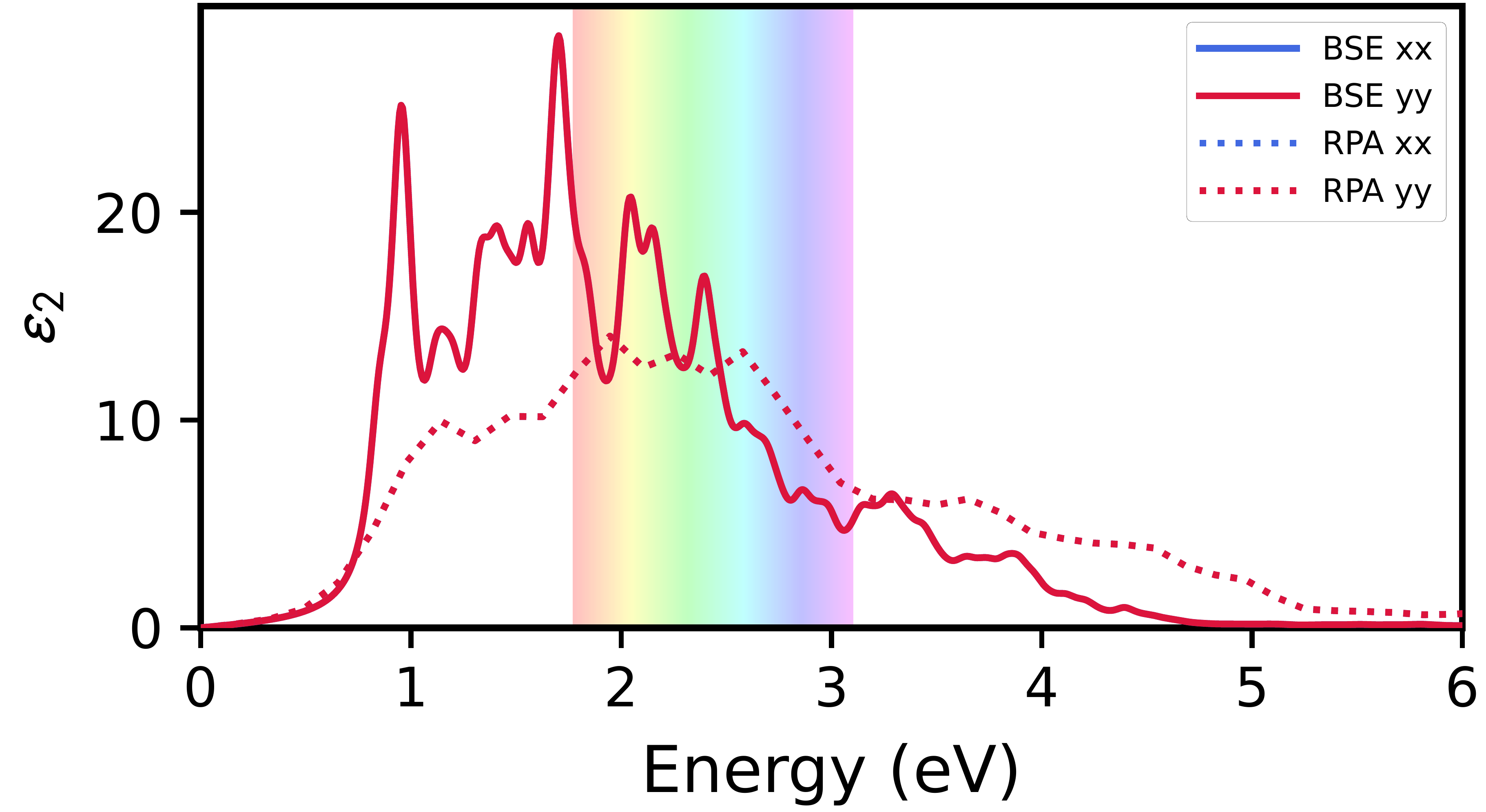}
    \end{subfigure}
    
    \begin{subfigure}[b]{0.42\textwidth}
        \subcaption{$\beta$-Te}
        \includegraphics[width=\textwidth]{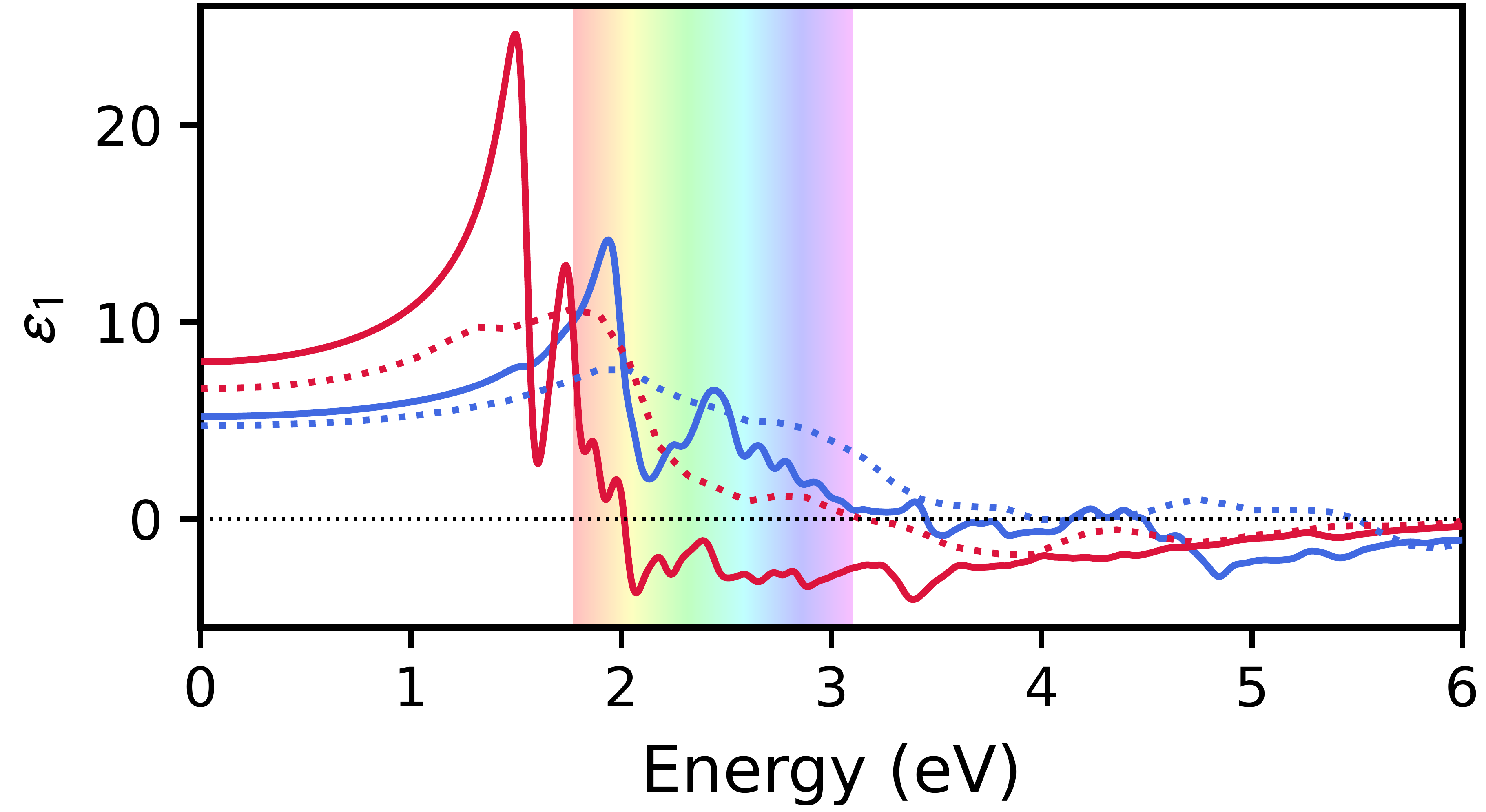}
    \end{subfigure}
    \begin{subfigure}[b]{0.42\textwidth}
        \subcaption{$\beta$-Te}
        \includegraphics[width=\textwidth]{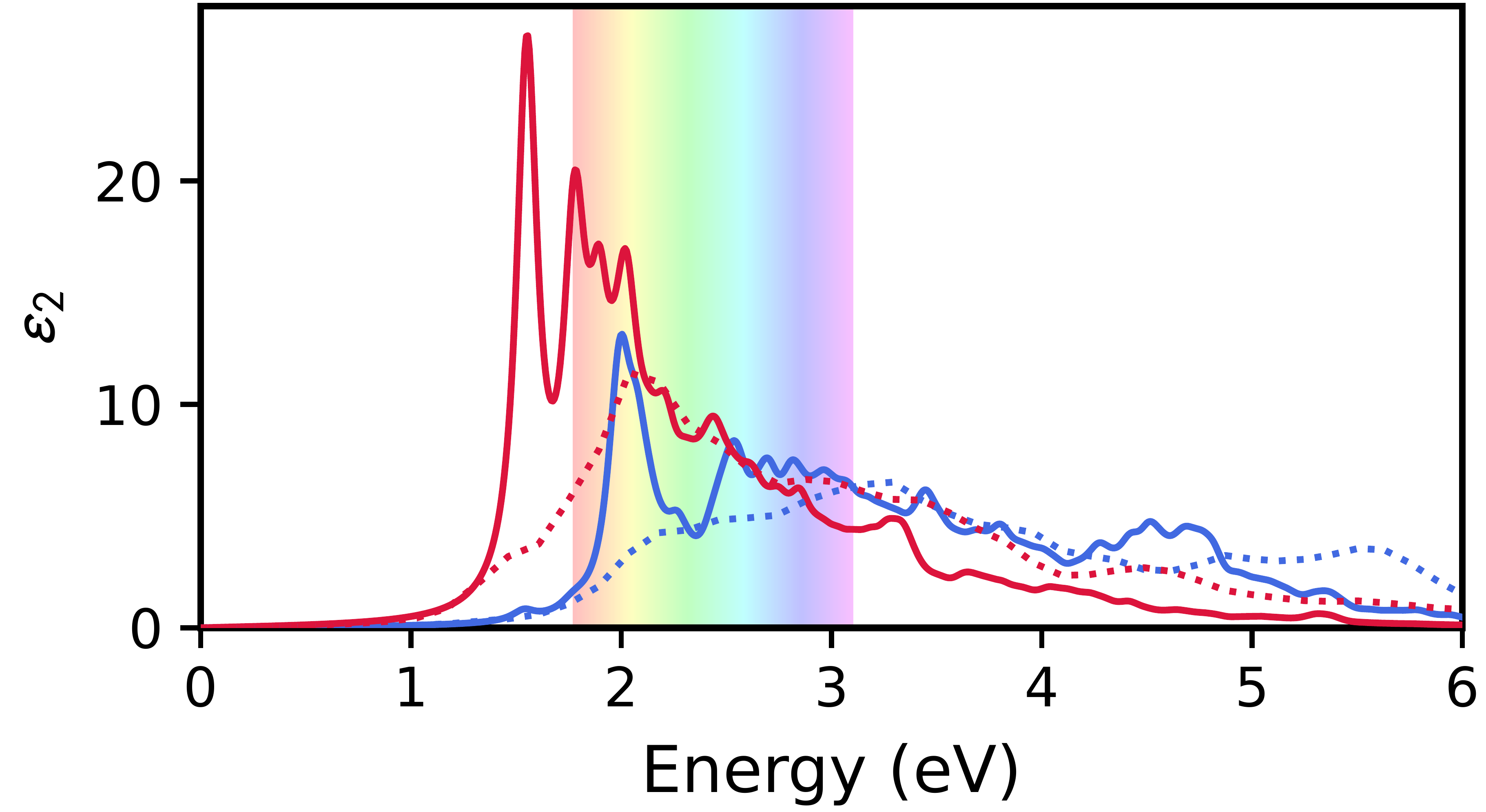}
    \end{subfigure}

    \begin{subfigure}[b]{0.42\textwidth}
        \subcaption{Hexagonal (pass.)}
        \includegraphics[width=\textwidth]{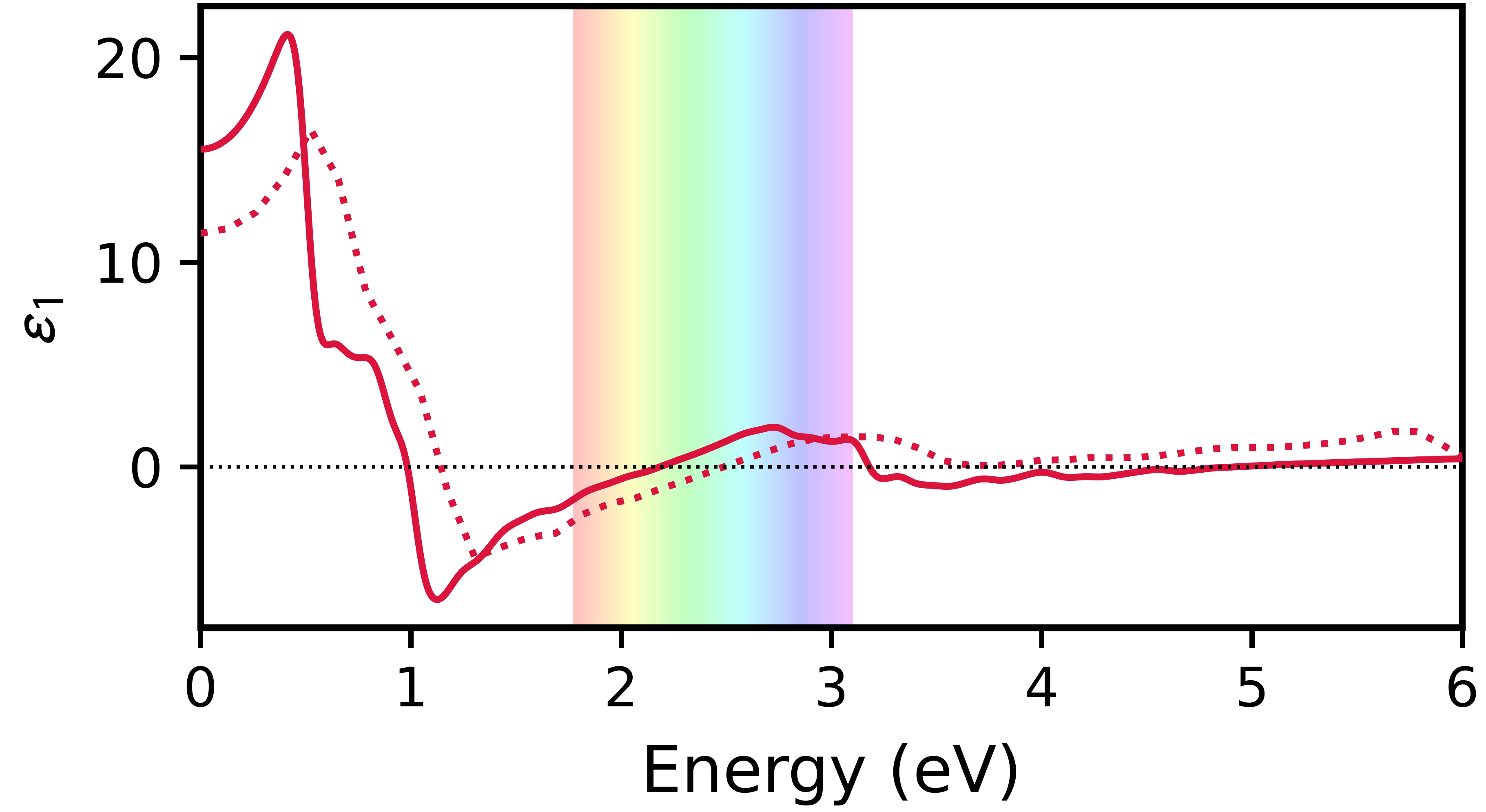}
    \end{subfigure}
    \begin{subfigure}[b]{0.42\textwidth}
        \subcaption{Hexagonal (pass.)}
        \includegraphics[width=\textwidth]{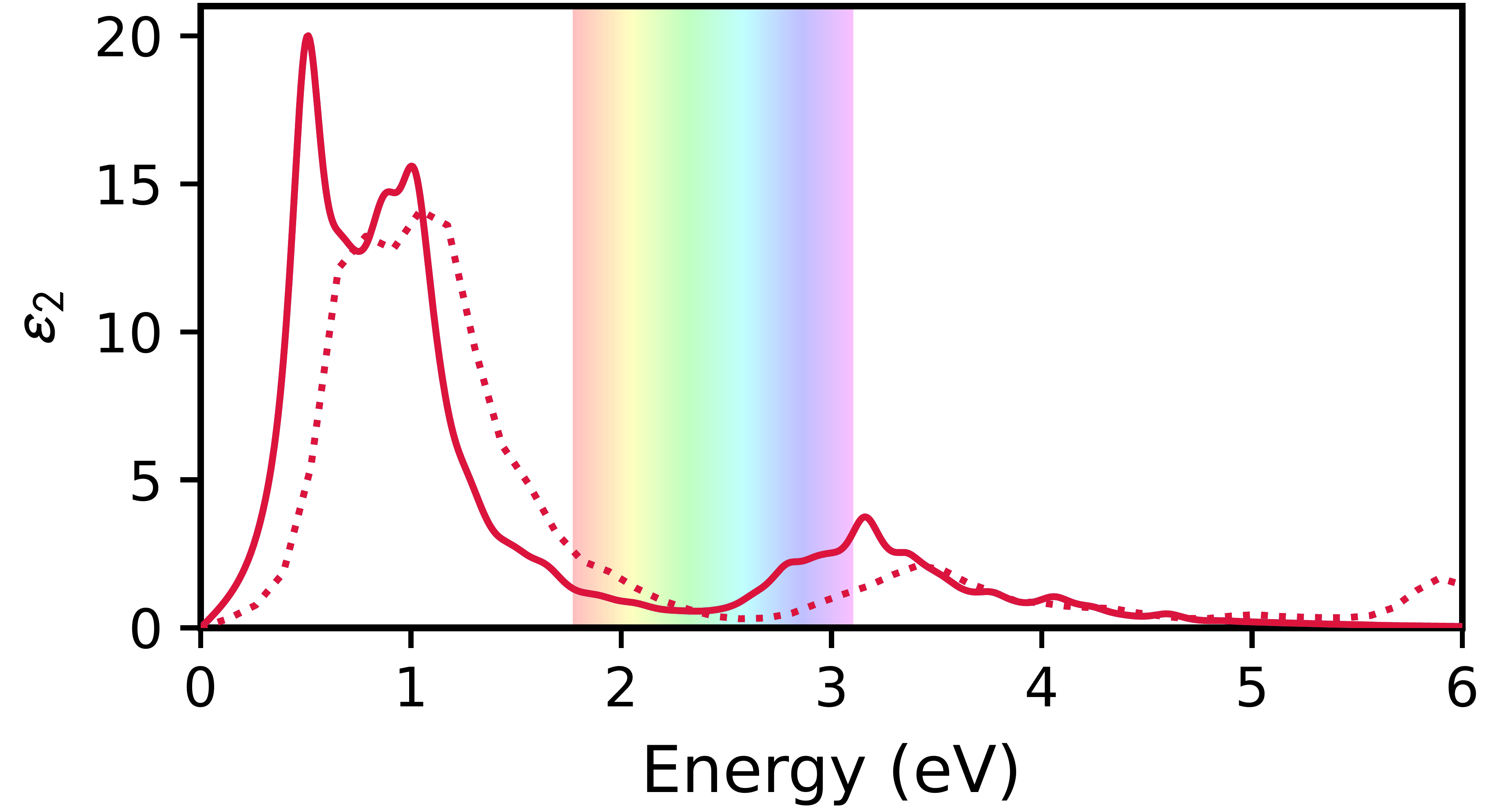}
    \end{subfigure}

    \begin{subfigure}[b]{0.42\textwidth}
        \subcaption{Pentagonal}
        \includegraphics[width=\textwidth]{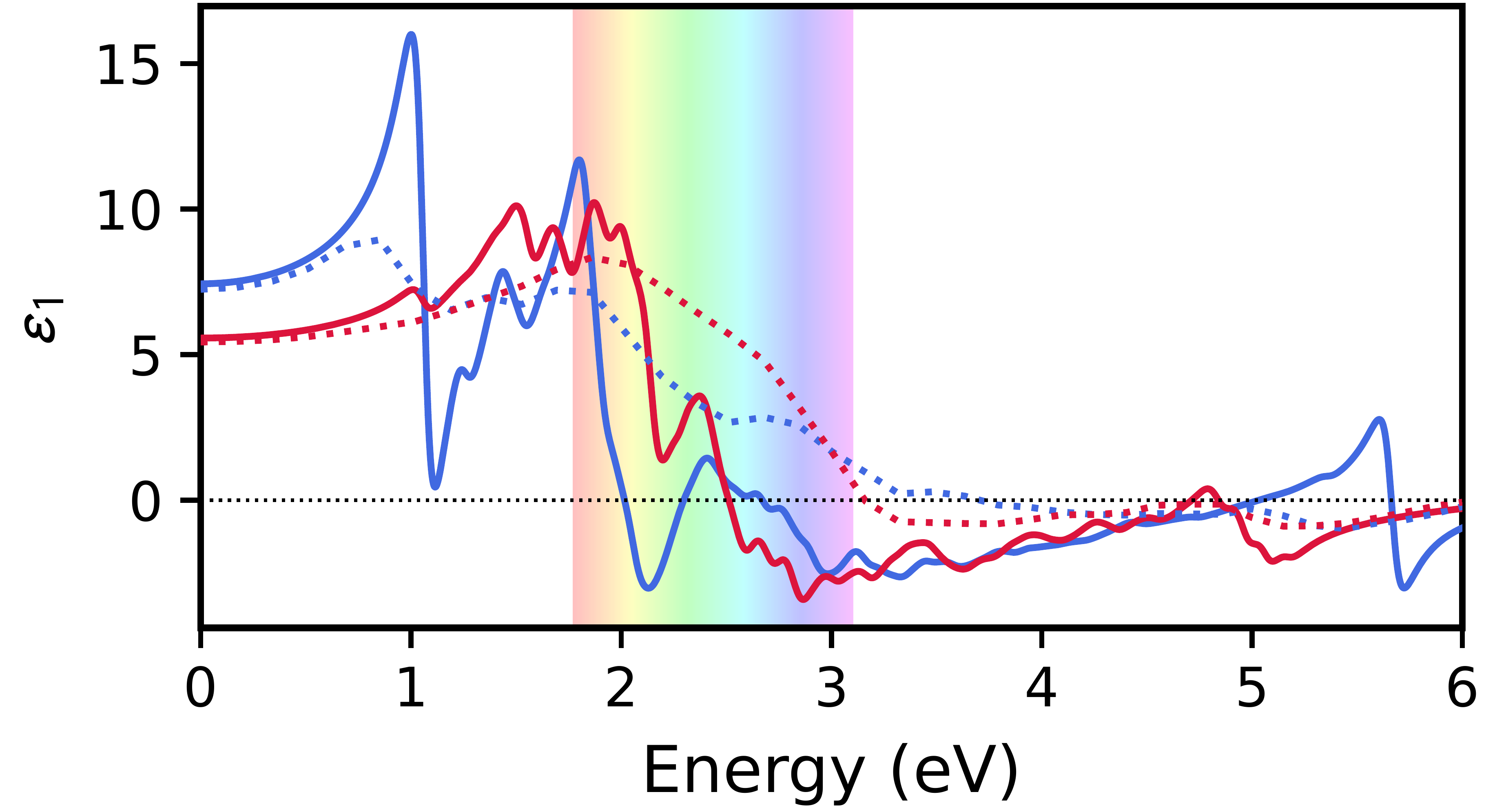}
    \end{subfigure}
    \begin{subfigure}[b]{0.42\textwidth}
        \subcaption{Pentagonal}
        \includegraphics[width=\textwidth]{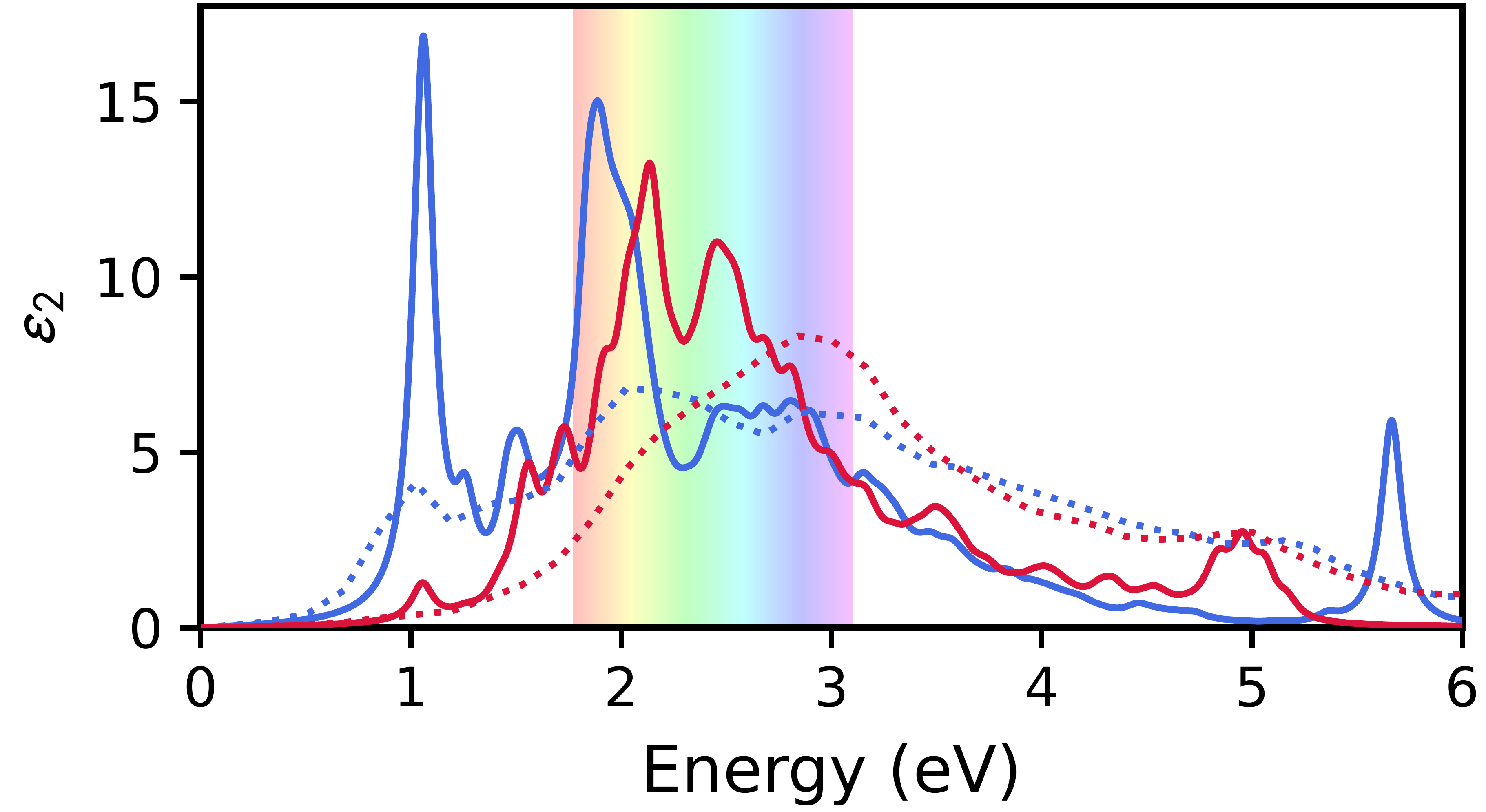}
    \end{subfigure}           
    \begin{subfigure}[b]{0.435\textwidth}
        \subcaption{Te nanowire (Te-h)}
        \includegraphics[width=\textwidth]{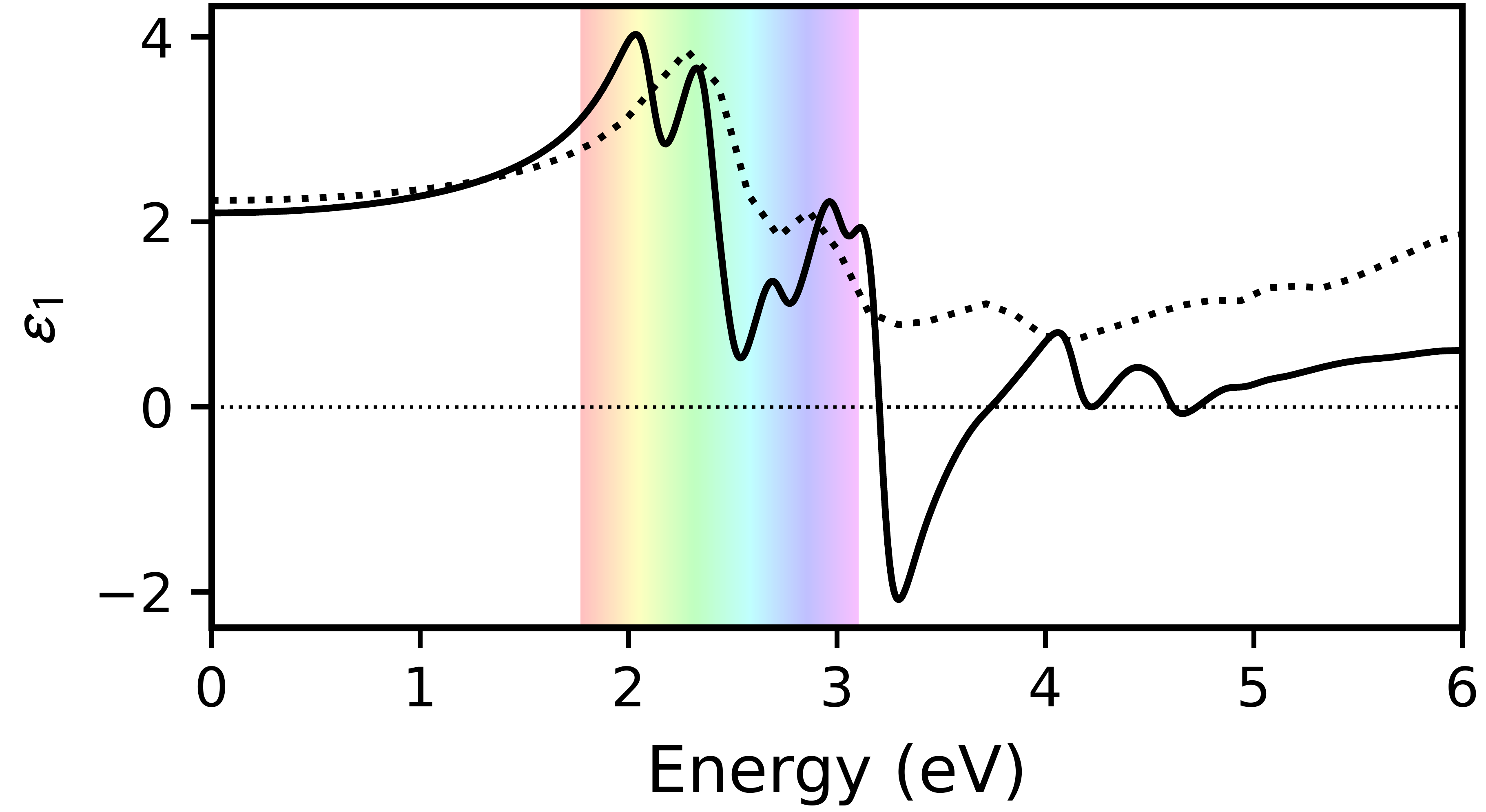}
    \end{subfigure}
    \begin{subfigure}[b]{0.42\textwidth}
        \subcaption{Te nanowire (Te-h)}
        \includegraphics[width=\textwidth]{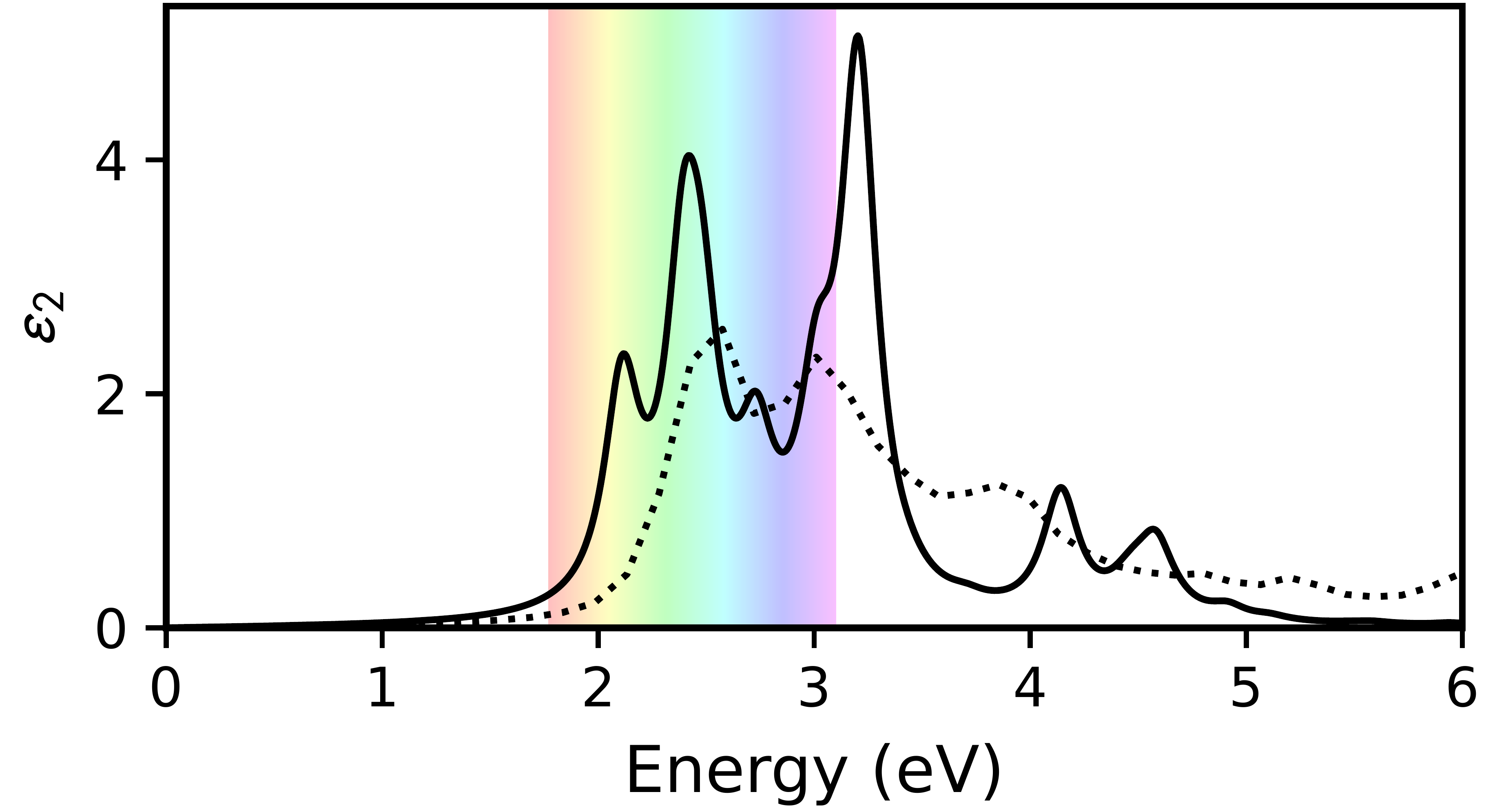}
    \end{subfigure}
    \caption{Real and imaginary parts of the dielectric function ($\varepsilon_1, \varepsilon_2$) calculated within BSE and RPA for the different tellurium phases: (a,b) $\alpha$-Te, (c,d) $\beta$-Te, (e,f) passivated hexagonal, (g,h) pentagonal, and (i,j) Te nanowire (Te-h).}
    \label{fig:dielectric_functions}
\end{figure}

The optical response of the investigated tellurium polymorphs was obtained by solving the Bethe--Salpeter equation (BSE), which explicitly accounts for electron--hole interactions. The resulting lowest bright optical transitions ($E_{\rm exc}$), quasiparticle band gaps ($E_{\rm GW}$), and exciton binding energies ($E_b=E_{\rm GW}-E_{\rm exc}$) are summarized in Table~\ref{tab:optical_properties_full}. The calculated dielectric function is presented in Fig.~\ref{fig:dielectric_functions}, while the absorption coefficient, reflectivity, electron energy-loss spectra (EELS), and refractive index are shown in Figs.~S1-S4 of the Supporting Information.

The complex dielectric function provides the fundamental description of the optical properties of the investigated tellurium polymorphs. The imaginary part, $\varepsilon_2(\omega)$, reflects the allowed interband optical transitions, whereas the real part, $\varepsilon_1(\omega)$, governs the dispersive optical response and determines the energy regions where collective electronic excitations may emerge. Comparison between the independent-particle (RPA) and BSE calculations demonstrates that electron--hole interactions significantly redistribute the optical spectral weight, modifying both the intensity and the energy position of the lowest optical transitions. These many-body effects are particularly evident in the low-energy region, where excitonic resonances reshape the dielectric response of all investigated structures.

The dielectric function also reveals that crystal symmetry strongly influences the polarization dependence of the optical response. Whereas $\alpha$- and $\beta$-tellurene exhibit pronounced in-plane optical anisotropy, the hydrogen-passivated hexagonal phase displays an almost isotropic dielectric response owing to its higher structural symmetry. The buckled pentagonal structure presents an intermediate behavior, whereas the one-dimensional helical nanowire naturally responds only to light polarized along its axial direction. These results demonstrate that structural polymorphism alone provides an efficient route for controlling the polarization dependence of light--matter interaction.

The absorption spectra shown in Fig.~S1 reveal distinct spectral regimes in the tellurium family. $\alpha$-tellurene exhibit relatively broad absorption bands extending from the near-infrared to the visible region, indicating that the optical oscillator strength is distributed over several interband transitions. In contrast, $\beta$-tellurene displays a narrow and intense near-infrared absorption peak, concentrating a large fraction of the oscillator strength within a limited energy interval. The hydrogen-passivated hexagonal and buckled pentagonal phases exhibit broader absorption profiles covering a substantial portion of the visible spectrum, whereas the one-dimensional nanowire undergoes a pronounced blueshift, with the absorption edge shifting almost entirely into the ultraviolet because of its considerably larger quasiparticle band gap.

The reflectivity spectra shown in Fig.~S2 follow the evolution of the dielectric function and further emphasize the influence of crystal symmetry. The strong optical anisotropy of $\beta$-tellurene produces markedly different reflectivities for the two in-plane polarization directions, whereas the hydrogen-passivated hexagonal phase exhibits nearly identical reflectance spectra along both crystallographic axes. The nanowire presents well-defined ultraviolet reflectivity features associated with its enlarged optical gap, while the remaining polymorphs display broader reflectance bands extending over wider energy intervals.

The calculated refractive-index spectra shown in Fig.~S3 also reflect the distinct dielectric responses of the investigated phases. Significant differences in the low-energy refractive index are observed, indicating substantial variations in electronic polarizability among the different crystal structures. In agreement with the dielectric function, the optical anisotropy is strongest in $\beta$-tellurene, reduced in the pentagonal phase, and becomes nearly negligible in hydrogen-passivated hexagonal tellurene. The nanowire exhibits the smallest low-energy refractive index, consistent with its large optical gap and reduced electronic polarizability.

Additional information is provided by the electron energy-loss spectra shown in Fig.~S4. Rather than exhibiting similar plasmonic responses, the investigated polymorphs display remarkably distinct distributions of collective electronic excitations.  $\alpha$-tellurene presents relatively broad loss features extending over wide energy intervals, whereas $\beta$-tellurene exhibits well-defined resonances associated with its highly anisotropic dielectric response. The hydrogen-passivated hexagonal and buckled pentagonal phases display multiple loss channels distributed over broad energy windows, while the one-dimensional nanowire develops a series of sharp and well-resolved loss peaks, reflecting the highly structured dielectric response characteristic of quasi-one-dimensional systems.

In summary, the calculated dielectric function and the derived optical constants consistently demonstrate that crystal symmetry and dimensionality profoundly modify the interaction between light and tellurium. Rather than simply shifting the optical absorption edge, the structural polymorphism changes the entire dielectric response, allowing the optical behavior to be continuously tuned from the near-infrared to the ultraviolet spectrum while simultaneously controlling optical anisotropy, reflectivity, refractive index, and collective electronic excitations.

The corresponding dielectric response and momentum-resolved excitonic weights of bulk trigonal tellurium are shown in Figs. S5 and S6 of the Supporting Information, providing a reference for the evolution of the excitonic response with dimensional confinement.

\section{Excitonic properties}

\begin{table*}[!htb]
\centering
\caption{Excitonic properties of tellurium polymorphs: fundamental quasiparticle gap ($E_g$), minimum direct quasiparticle gap ($E_g^{d}$), lowest BSE excitonic eigenvalue ($E_{\rm exc}$), direct exciton binding energy ($E_b^{d}=E_g^{d}-E_{\rm exc}$).}
\label{tab:optical_properties_full}
\begin{tabular}{lcccccccc}
\toprule
Phase & Pol. & \multicolumn{2}{c}{$E_{\mathrm{GW}}$ (eV)} & \multicolumn{2}{c}{$E_{\mathrm{exc}}$ (eV)} & \multicolumn{2}{c}{$E_b^d$ (eV)} \\
\cmidrule(lr){3-4} \cmidrule(lr){5-6} \cmidrule(lr){7-8}
& & PBE & HSE06 & PBE & HSE06 & PBE & HSE06 & \\
& & $E_g$ ($E_g^d$) & $E_g$ ($E_g^d$) & & & & & \\
\midrule

$\alpha$-Te & $yy$ & 0.77 (0.96) & 0.85 (1.05) & 0.81 & 0.85 & 0.15 & 0.20  \\
            & $xx$ & -- & -- & -- & -- & -- & --   \\
\midrule
$\beta$-Te & $yy$ & 1.70 & 1.91 & 1.42 & 1.51 & 0.28 & 0.40  \\
           & $xx$ & -- & -- & 1.43 & 1.54 & 0.27 & 0.37 & \\
\midrule
Hexagonal Te & $yy$ & 0.50 / 0.80 & 0.50 / 0.82 & 0.33 & 0.31 & 0.47 & 0.51  \\
             & $xx$ & -- & -- & -- & -- & -- & -- &  \\
\midrule
Pentagonal Te & $yy$ & -- & -- & 0.95 & 1.05 & 0.30 & 0.39   \\
              & $xx$ & 1.25 & 1.44 & 0.94 & 1.04 & 0.31 & 0.40 & \\
\midrule
Te-h & $zz$ & 3.75 / 3.76 & 4.17 / 4.18 & 1.73 & 1.86 & 2.03 & 2.32  \\
\bottomrule
\end{tabular}
\end{table*}

\begin{figure}[!htb]
    \centering
    
    \begin{subfigure}[b]{0.42\textwidth}
        \subcaption{$\alpha$-Te, $GW$@PBE}
        \includegraphics[width=\textwidth]{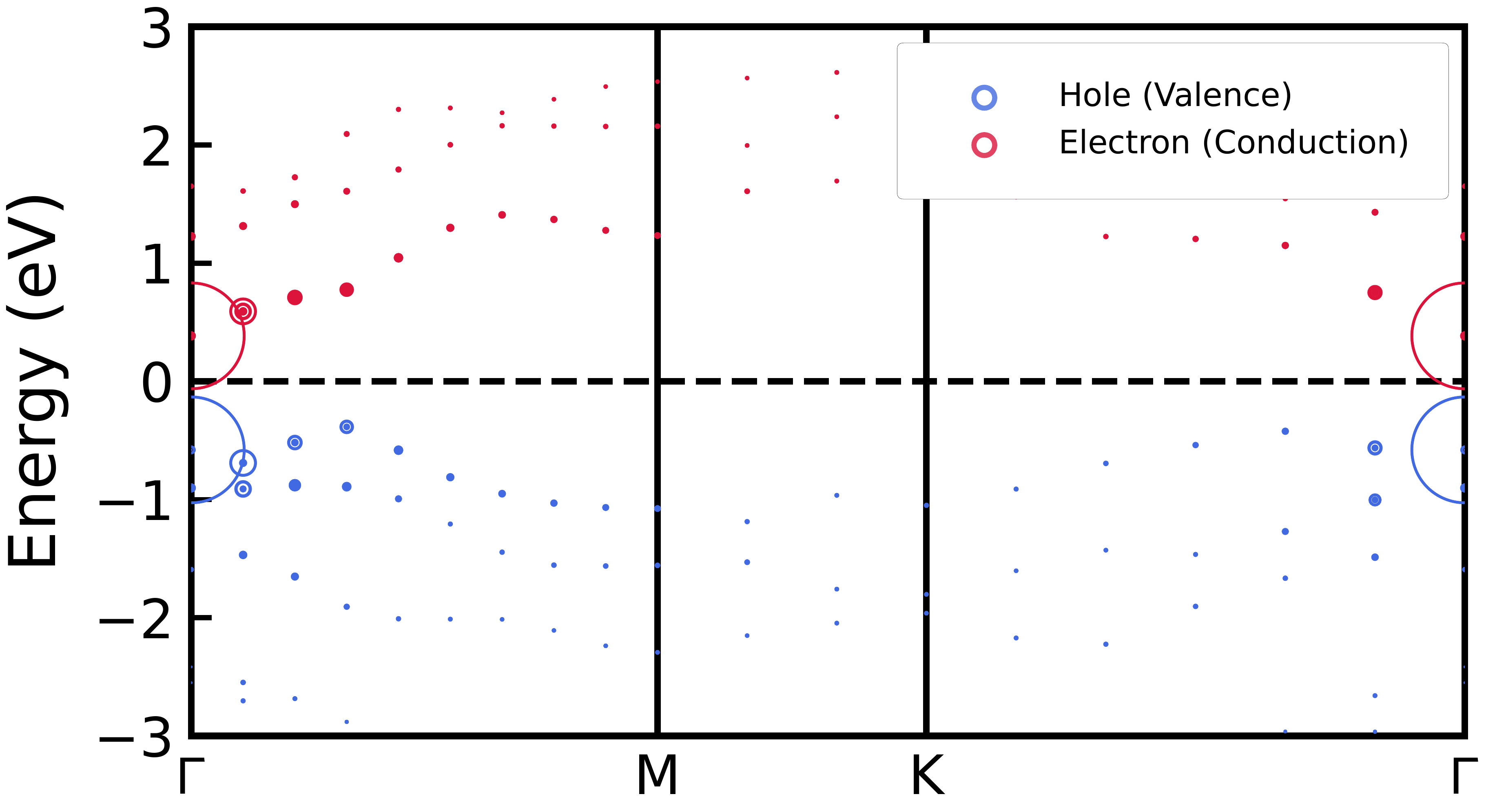}
    \end{subfigure}
    \begin{subfigure}[b]{0.42\textwidth}
        \subcaption{$\alpha$-Te, $GW$@HSE06}
        \includegraphics[width=\textwidth]{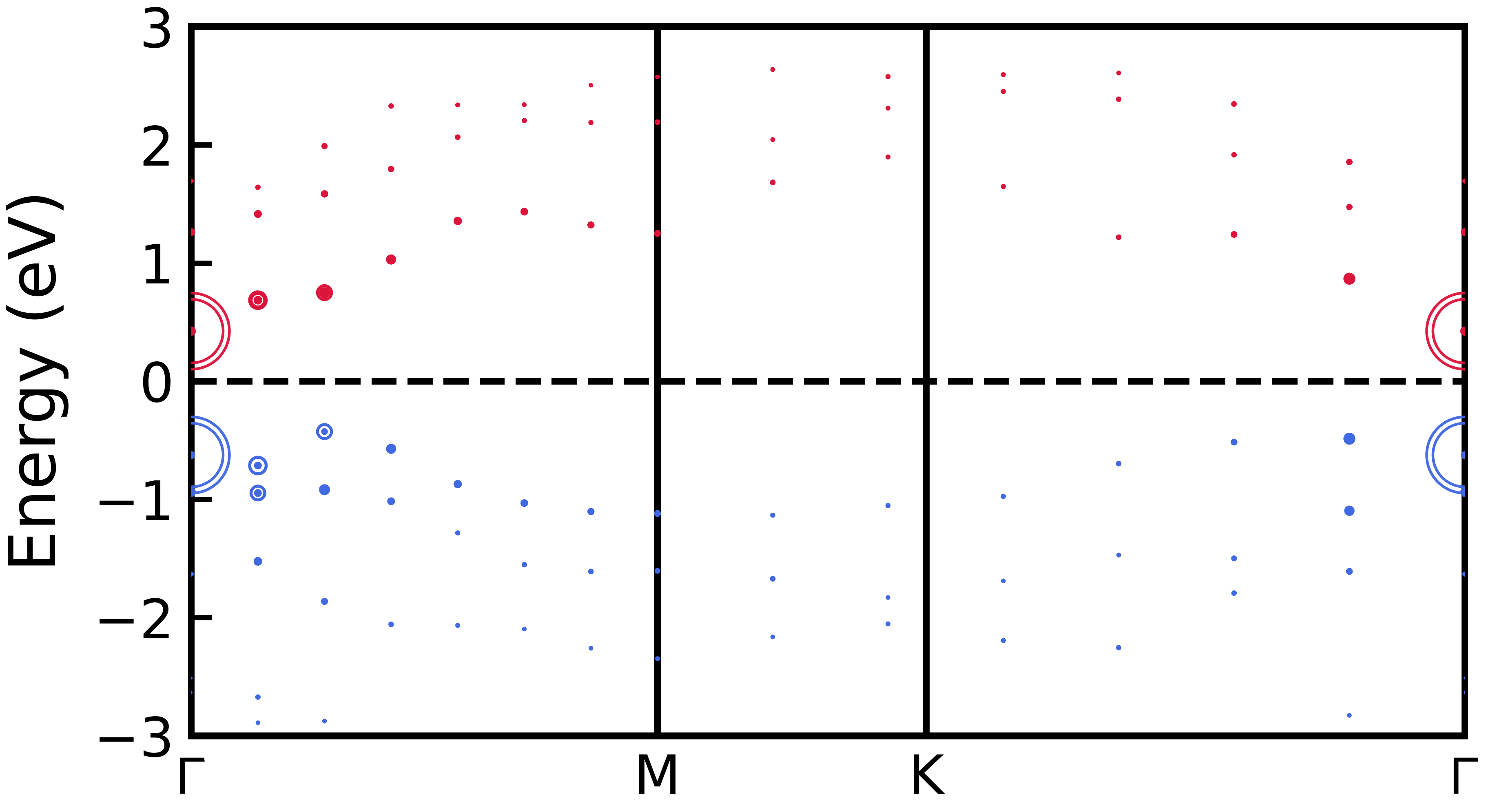}
    \end{subfigure}
    \\
    \begin{subfigure}[b]{0.42\textwidth}
        \subcaption{$\beta$-Te, $GW$@PBE}
        \includegraphics[width=\textwidth]{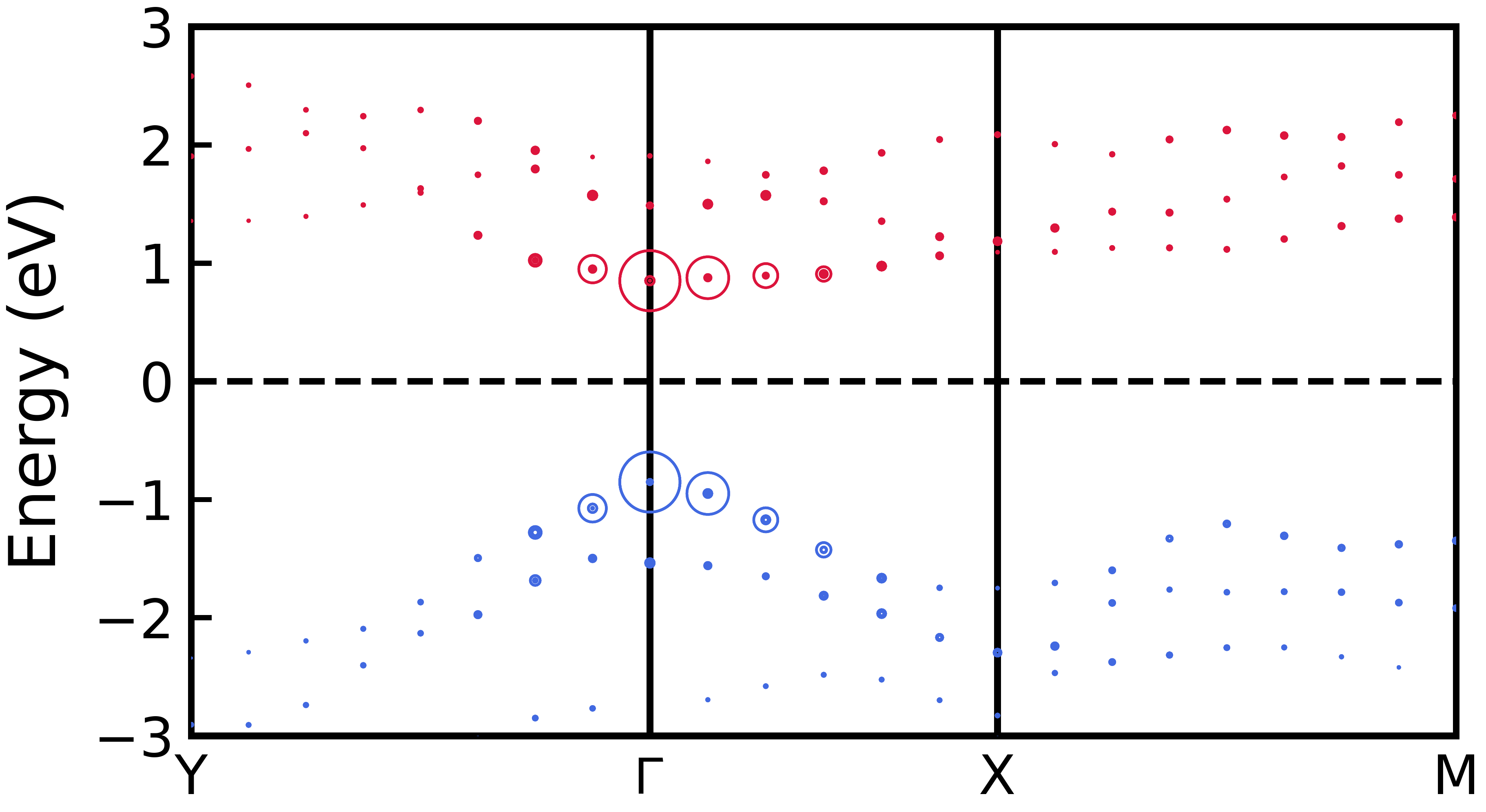}
    \end{subfigure}
    \begin{subfigure}[b]{0.42\textwidth}
        \subcaption{$\beta$-Te, $GW$@HSE06}
        \includegraphics[width=\textwidth]{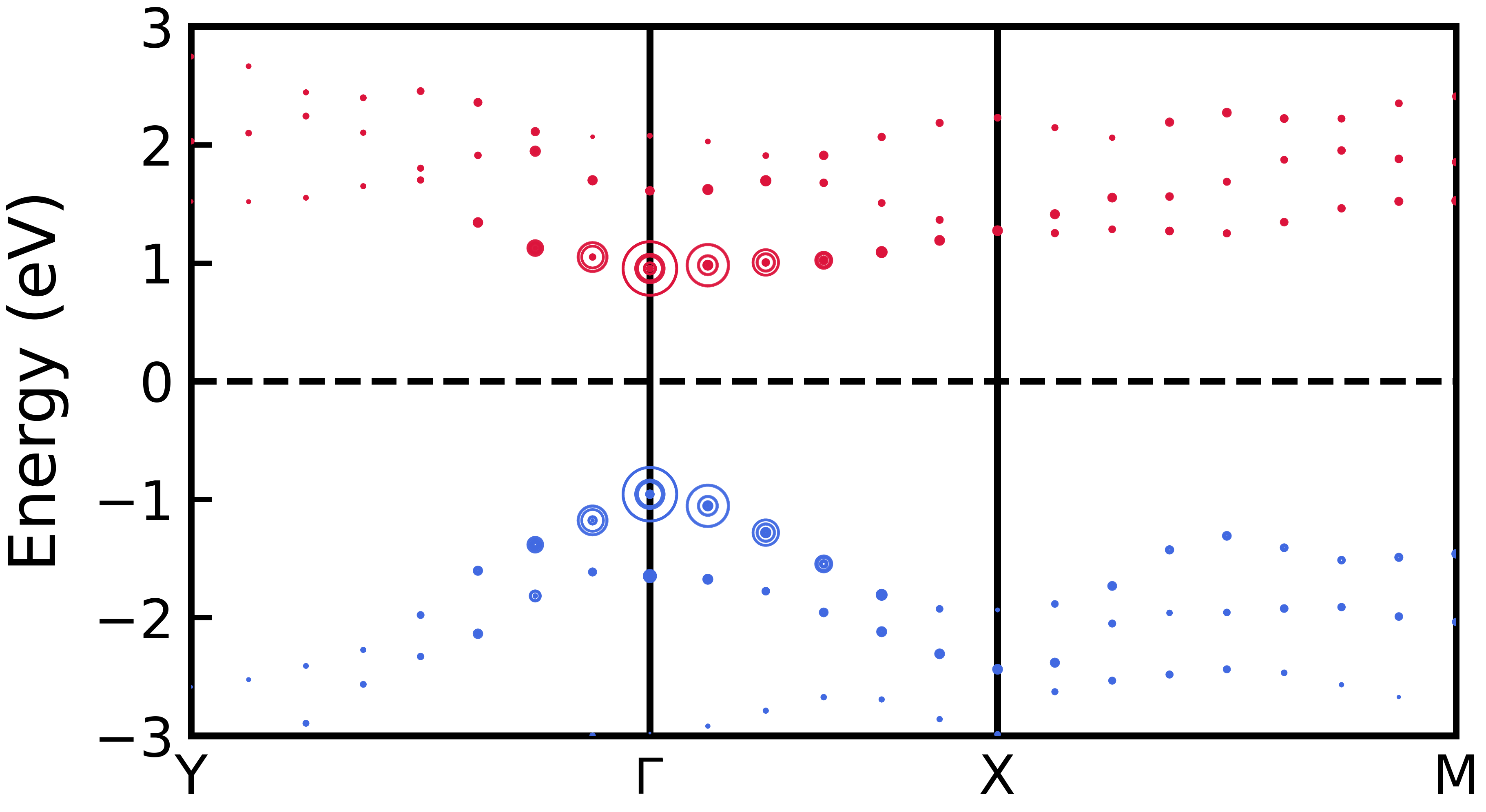}
    \end{subfigure}
    \\
    \begin{subfigure}[b]{0.42\textwidth}
        \subcaption{Hexagonal-pass., $GW$@PBE}
        \includegraphics[width=\textwidth]{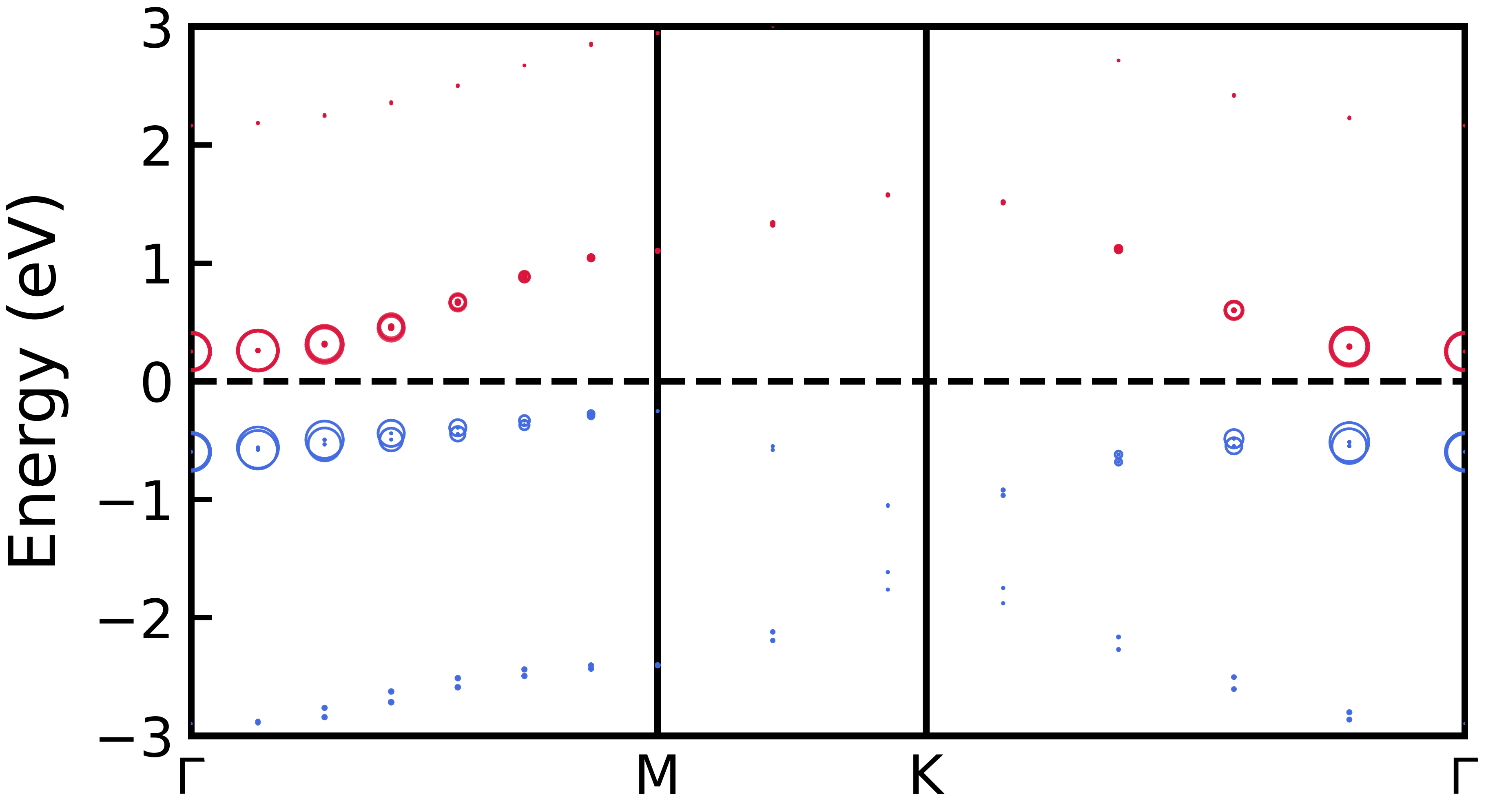}
    \end{subfigure}
    \begin{subfigure}[b]{0.42\textwidth}
        \subcaption{Hexagonal-pass., $GW$@HSE06}
        \includegraphics[width=\textwidth]{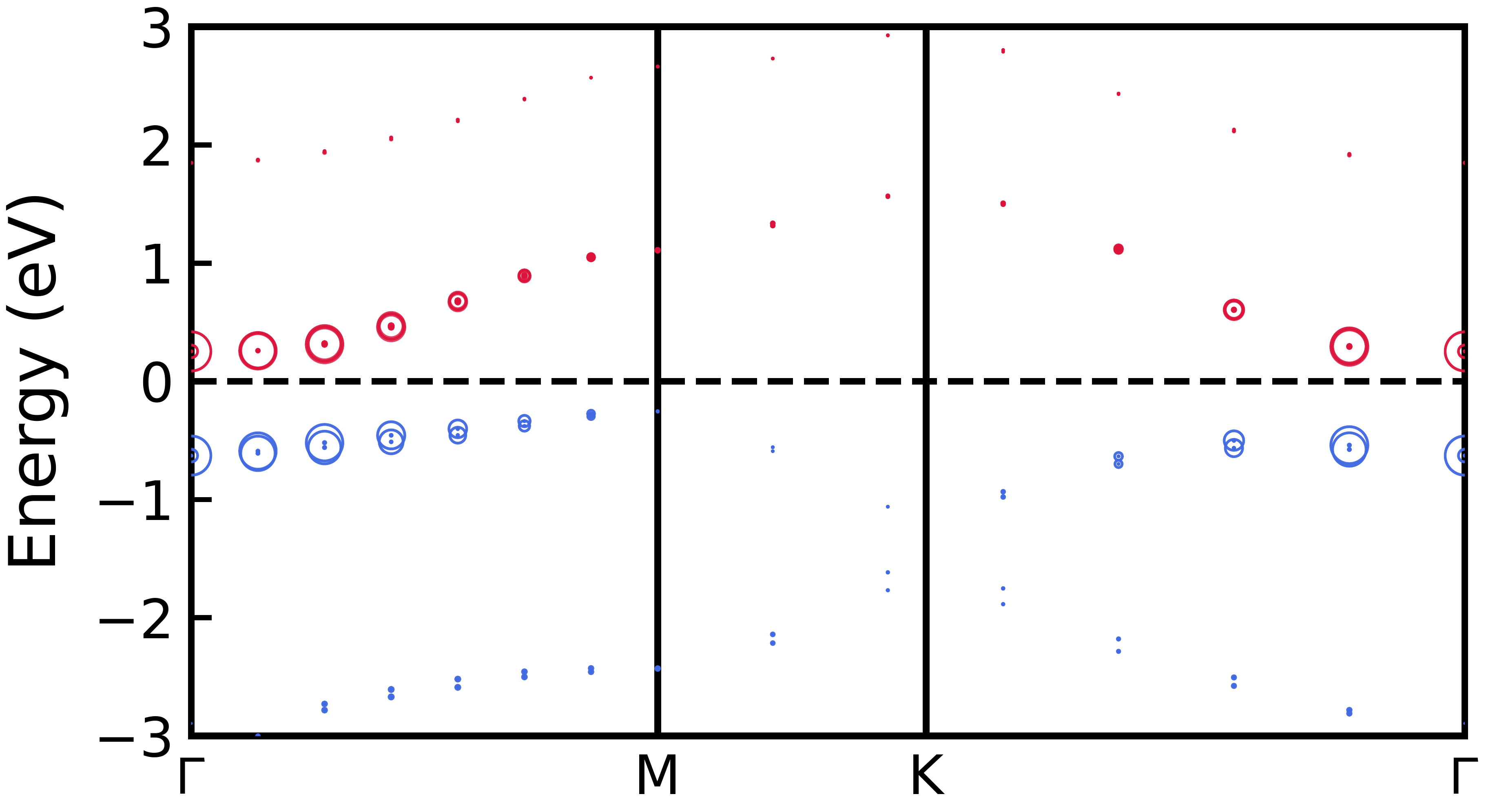}
    \end{subfigure}
    \\
    \begin{subfigure}[b]{0.42\textwidth}
        \subcaption{Pentagonal, $GW$@PBE}
        \includegraphics[width=\textwidth]{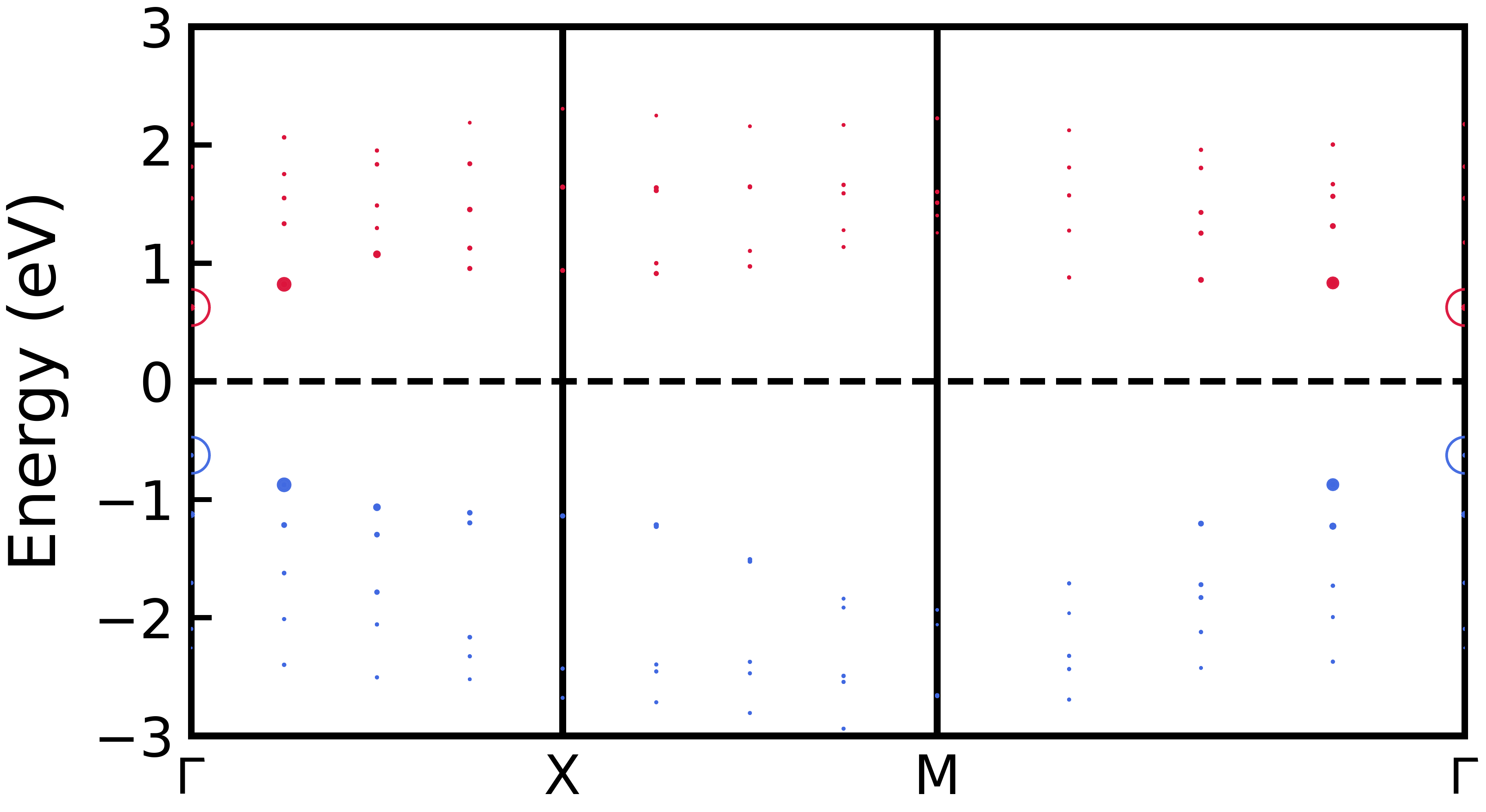}
    \end{subfigure}
    \begin{subfigure}[b]{0.42\textwidth}
        \subcaption{Pentagonal, $GW$@HSE06}
        \includegraphics[width=\textwidth]{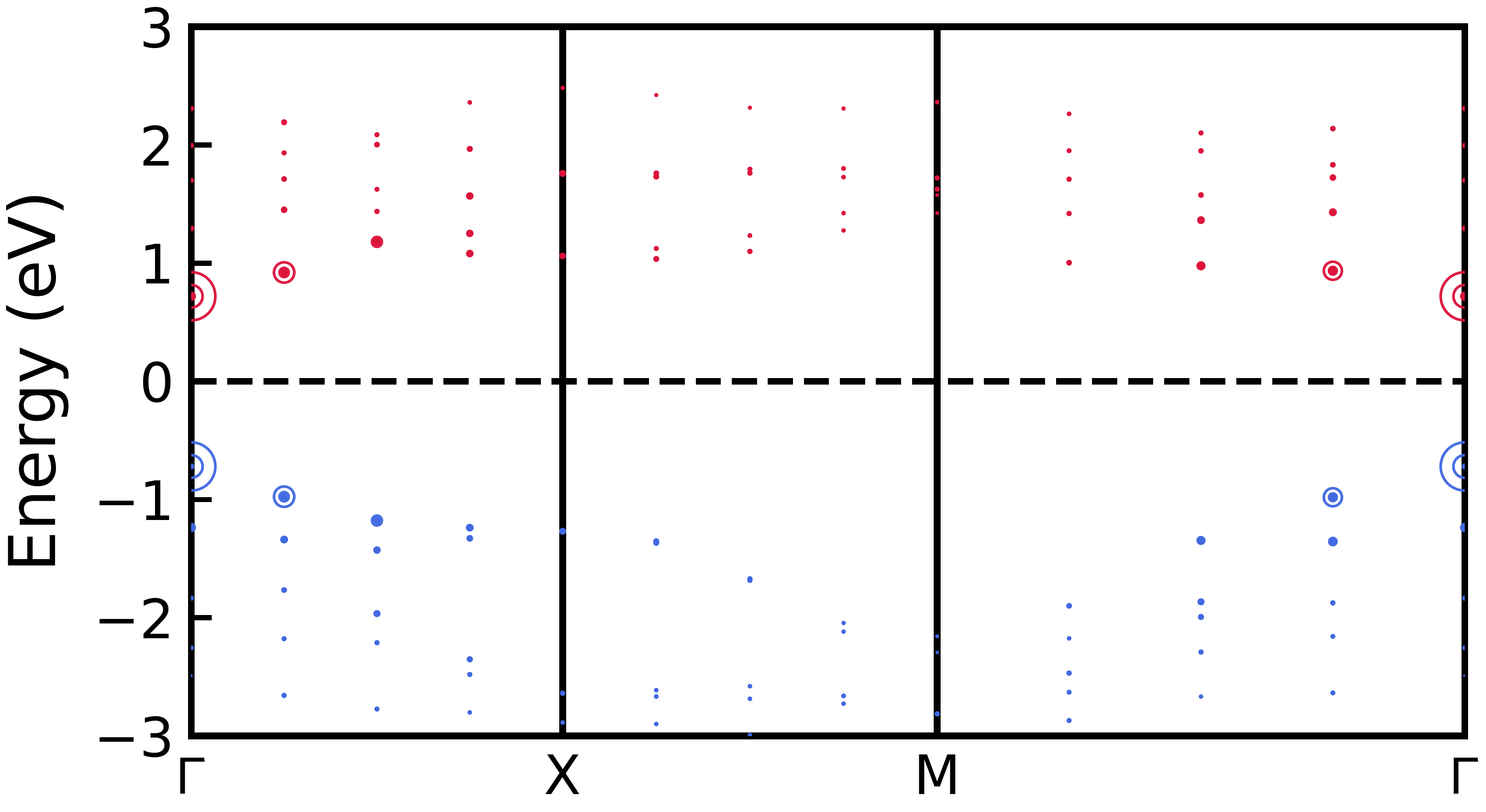}
    \end{subfigure}
    \\
    \begin{subfigure}[b]{0.42\textwidth}
        \subcaption{Te-h, $GW$@PBE}
        \includegraphics[width=\textwidth]{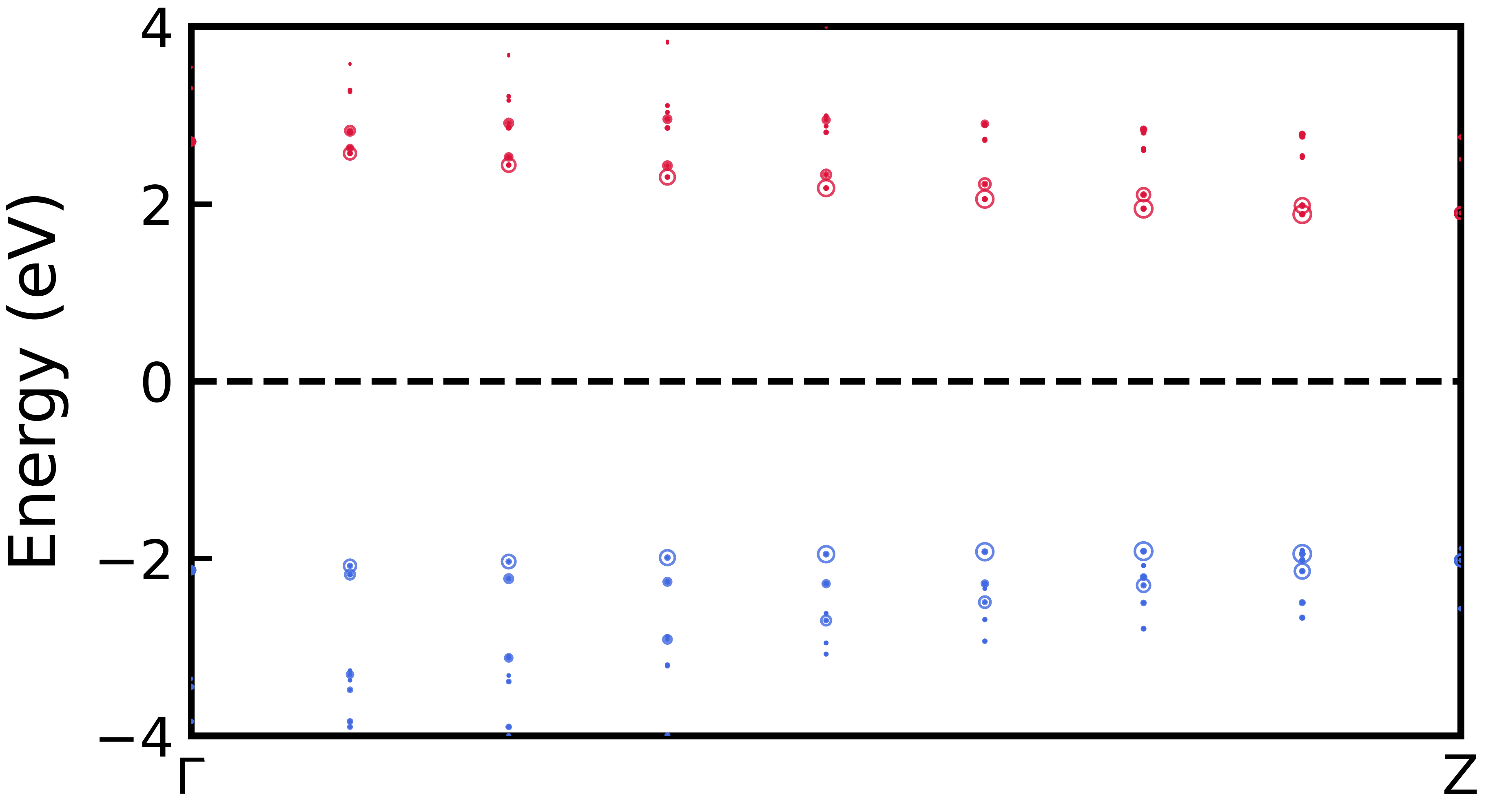}
    \end{subfigure}
    \begin{subfigure}[b]{0.42\textwidth}
        \subcaption{Te-h, $GW$@HSE06}
        \includegraphics[width=\textwidth]{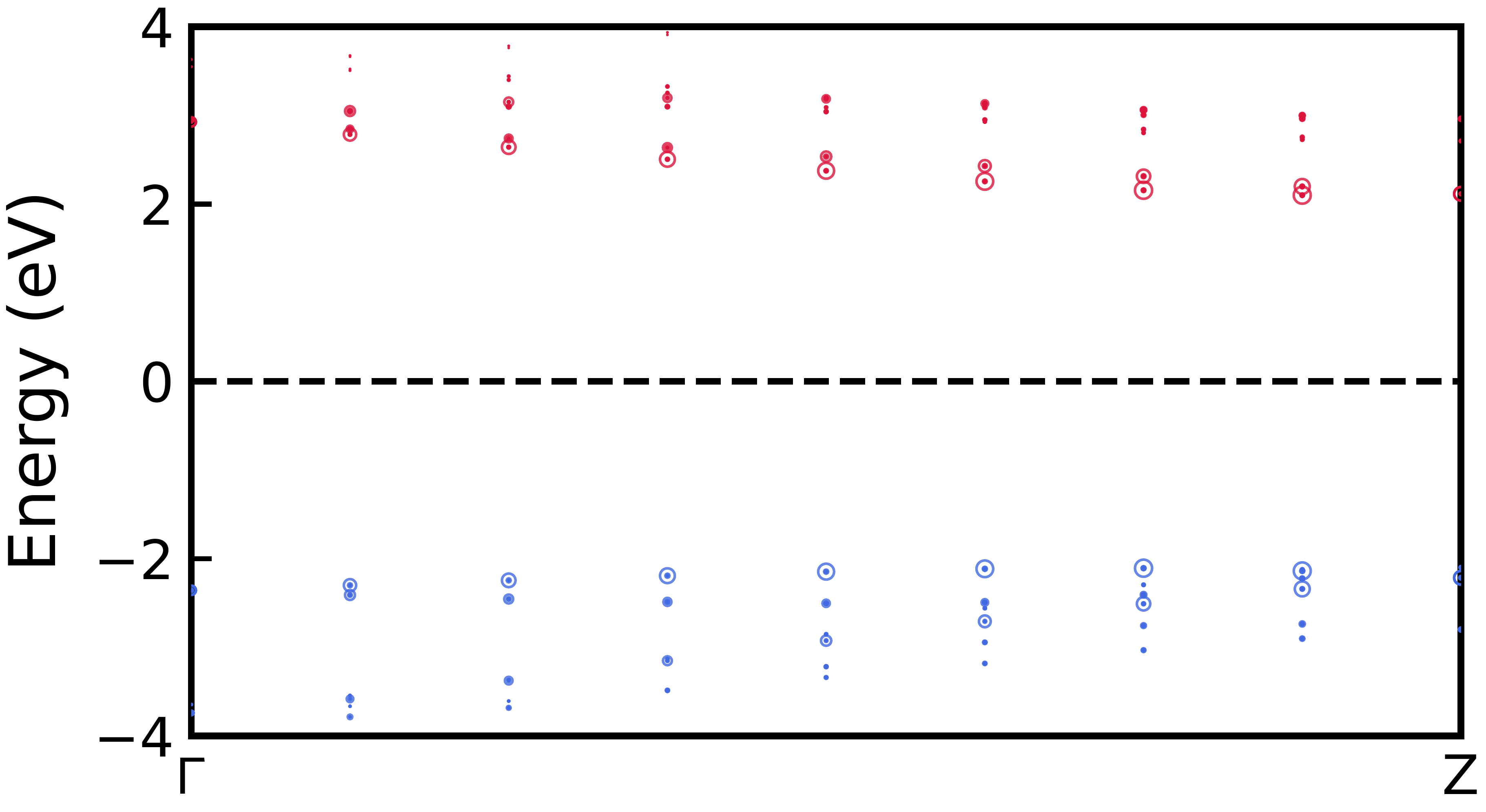}
    \end{subfigure}
    \caption{Momentum-resolved excitonic distributions obtained from the BSE eigenvectors for the lowest BSE excitonic eigenvalues. The size of each marker is proportional to the excitonic weight, allowing direct visualization of the degree of localization in reciprocal space.}
    \label{fig:exciton_fatbands}
\end{figure}

\begin{figure}[!htb]
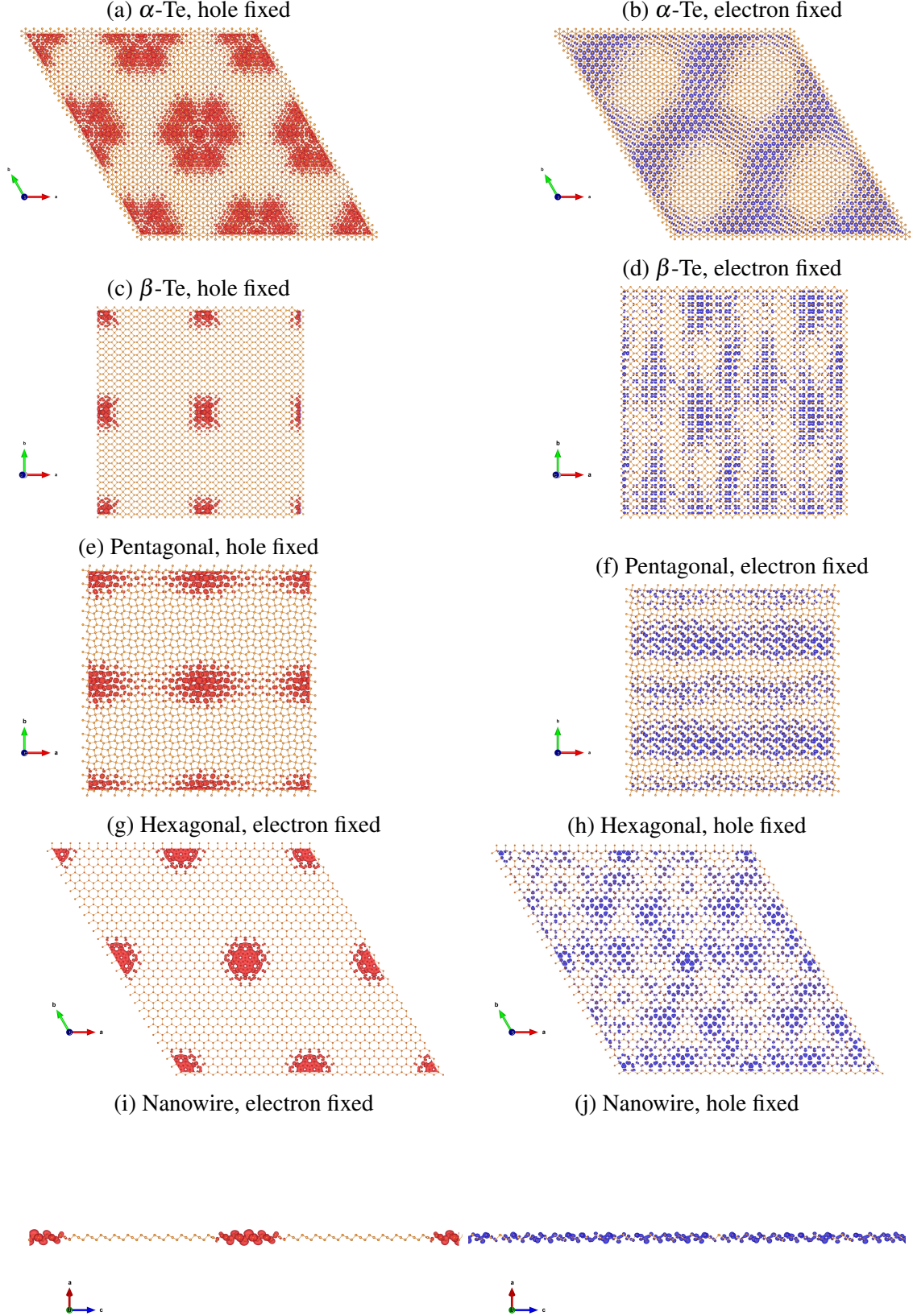

    \centering
    \begin{subfigure}[b]{0.45\textwidth}
        \subcaption{$\alpha$-Te, hole fixed}
        \includegraphics[width=\linewidth]{WF_ELECTRON_alpha.pdf}
    \end{subfigure}
    \hfill
    \begin{subfigure}[b]{0.45\textwidth}
        \centering
        \subcaption{$\alpha$-Te, electron fixed}
        \includegraphics[width=\textwidth]{WF_HOLE_alpha.pdf}    
    \end{subfigure}
    \begin{subfigure}[b]{0.45\textwidth}
        \centering
        \subcaption{$\beta$-Te, hole fixed}
        \includegraphics[width=\textwidth]{WF_ELECTRON_beta.pdf}
    \end{subfigure}
    \hfill    
    \begin{subfigure}[b]{0.45\textwidth}
        \centering
        \subcaption{$\beta$-Te, electron fixed}
        \includegraphics[width=\textwidth]{WF_HOLE_beta.pdf}
    \end{subfigure}
    \begin{subfigure}[b]{0.45\textwidth}
        \centering
        \subcaption{Pentagonal, hole fixed}
        \includegraphics[width=\textwidth]{WF_ELECTRON_penta.pdf}    
    \end{subfigure}
    \hfill
    \begin{subfigure}[b]{0.45\textwidth}
        \centering
        \subcaption{Pentagonal, electron fixed}
        \includegraphics[width=\textwidth]{WF_HOLE_penta.pdf}    
    \end{subfigure}
    \begin{subfigure}[b]{0.45\textwidth}
        \centering
        \subcaption{Hexagonal, electron fixed}
        \includegraphics[width=\textwidth]{WF_ELECTRON_hexa.pdf}    
    \end{subfigure}
    \begin{subfigure}[b]{0.45\textwidth}
        \centering
        \subcaption{Hexagonal, hole fixed}
        \includegraphics[width=\textwidth]{WF_HOLE_hexa.pdf}    
    \end{subfigure}
    \\
    \begin{subfigure}[b]{0.45\textwidth}
        \centering
        \subcaption{Nanowire, electron fixed}
        \includegraphics[width=\textwidth]{WF_ELECTRON_fio.pdf}    
    \end{subfigure}
    \begin{subfigure}[b]{0.45\textwidth}
        \centering
        \subcaption{Nanowire, hole fixed}
        \includegraphics[width=\textwidth]{WF_HOLE_fio.pdf}    
    \end{subfigure}
   \caption{Real-space exciton wavefunctions $\Psi_S(\mathbf{r}_e, \mathbf{r}_h)$ for the lowest bright optical transitions in the 2D tellurene polymorphs. The isosurfaces represent the conditional probability distribution of one charge carrier given the fixed position of the other. In panels where the hole is fixed (blue point), the red isosurface maps the spatial distribution of the excited electron; conversely, when the electron is fixed (red point), the blue isosurface maps the distribution of the hole.}
\label{fig:exciton_wavefunction}
\end{figure}

\begin{table}[!htb]
\centering
\caption{Characteristic excitonic radii obtained from the momentum-resolved BSE eigenvectors. The values along the principal crystallographic directions (R$_x$, R$_y$,  and R$_z$) quantify the spatial extension and anisotropy of the lowest BSE excitons for tellurium polymorph.}
\begin{tabular}{lcccc} \toprule
Phase & R (\AA)  & R$_x$ (\AA) & R$_y$ (\AA) & R$_z$ (\AA)\\ \midrule
$\alpha$-Te &  $28.2$ & 42.1 & 40.6 & -- \\ 
$\beta$-Te  &  $9.0$  & 11.5 & 14.4 & -- \\
Hexagonal Te  &  $8.1$ & 11.4 & 11.4 & --\\ 
Pentagonal Te &  $14.5$ & 31.0 & 16.4& --\\ 
Te-h & 4.12 & -- & -- & 4.12 \\
\bottomrule
\end{tabular}
\label{tab:exciton_radius}
\end{table}

The calculated binding energies reveal substantial variations in the
strength of the electron--hole interaction among the investigated
tellurium phases. At the $G_{4}W_{0}$@HSE06 level, $\alpha$-tellurene
exhibits the smallest direct exciton binding energy,
$E_b^{d}=0.20$~eV, whereas $\beta$- and pentagonal tellurene reach
approximately 0.40~eV. The hydrogen-passivated hexagonal phase
displays the largest value among the two-dimensional structures,
$E_b^{d}=0.51$~eV. The one-dimensional nanowire lies in a markedly
stronger confinement regime, with a substantially larger binding
energy of 2.32~eV.

The reciprocal-space distributions of the lowest BSE excitons provide further insight into the origin of these differences. As shown in Fig.~\ref{fig:exciton_fatbands}, the excitonic amplitudes exhibit distinct degrees of localization within the Brillouin zone depending on the crystal structure. In $\alpha$-tellurene, the excitonic weight extends over a relatively broad region around the band extrema, indicating that several electronic transitions contribute to the lowest optical excitation. In contrast, $\beta$-tellurene exhibits a much more concentrated distribution, demonstrating that the lowest exciton is dominated by a limited number of electronic transitions. The hydrogen-passivated hexagonal and pentagonal structures exhibit
distinct spatial characteristics: the hexagonal phase displays a
compact and nearly isotropic in-plane excitonic distribution, whereas
the pentagonal phase retains a pronounced directional anisotropy.

The real-space excitonic wave functions shown in Fig.~\ref{fig:exciton_wavefunction} corroborate the trends inferred from the reciprocal-space analysis. When the hole is fixed at a selected atomic site, the electron probability density remains broadly distributed throughout the lattice in $\alpha$-tellurene, consistent with its comparatively smaller binding energy. Conversely, $\beta$-tellurene exhibits a compact electron distribution surrounding the hole, reflecting the strong electron--hole correlation responsible for its large binding energy. The hydrogen-passivated hexagonal and pentagonal structures exhibit
distinct spatial characteristics. The hexagonal phase displays a compact
and nearly isotropic in-plane excitonic distribution, whereas the
pentagonal phase exhibits a more extended and strongly anisotropic
exciton.

This qualitative picture is quantified in Table~\ref{tab:exciton_radius}. The effective exciton radius is defined from the directional extents as
\begin{equation}
R =
\left(
\frac{1}{R_x^2}
+
\frac{1}{R_y^2}
\right)^{-1/2}.
\end{equation}
The characteristic exciton radii further distinguish the spatial character of the excitonic states. 
$\alpha$-tellurene exhibits the largest effective radius, $R=28.2$~\AA, consistent with a spatially extended exciton. 
In contrast, $\beta$- and hydrogen-passivated hexagonal tellurene exhibit considerably smaller effective radii of 9.0 and 8.1~\AA, respectively. 
Despite their comparable spatial confinement, their directional behavior is markedly different: $\beta$-tellurene is anisotropic, with $R_x=11.5$~\AA\ and $R_y=14.4$~\AA, whereas the hexagonal phase exhibits an isotropic in-plane extension, $R_x=R_y=11.4$~\AA. 
The pentagonal phase occupies an intermediate regime, with $R=14.5$~\AA\ and a pronounced directional anisotropy.

The comparison between $\beta$-tellurene and hydrogen-passivated
hexagonal tellurene is particularly relevant in view of their distinct
electronic topology. In our previous work~\cite{araujo-2026}, we showed
that hydrogen-passivated hexagonal tellurene realizes a quantum spin
Hall phase characterized by a nontrivial $\mathbb{Z}_2$ invariant,
whereas $\beta$-tellurene is topologically trivial and develops
SOC-induced quasi-flat states near the band edges. The present results
show that these markedly different electronic structures support
excitons with comparable spatial confinement, but with distinct
binding strengths and directional character.

At the $G_{4}W_{0}$@HSE06 level, the topological hexagonal phase
exhibits a direct exciton binding energy of $E_b^{d}=0.51$~eV and an
effective exciton radius of $R=8.1$~\AA, whereas $\beta$-tellurene
exhibits $E_b^{d}=0.40$~eV and $R=9.0$~\AA. Thus, the nontrivial phase
does not exhibit weaker electron--hole correlations or a more
spatially extended exciton. Instead, the two phases display excitonic
states of comparable spatial extent, with the hexagonal phase showing
a slightly larger binding energy.

A more pronounced difference appears in the directional extension of
the excitonic states. In hydrogen-passivated hexagonal tellurene,
$R_x=R_y=11.4$~\AA, indicating an essentially isotropic in-plane
exciton. In contrast, $\beta$-tellurene exhibits $R_x=11.5$~\AA\ and
$R_y=14.4$~\AA, revealing a clear anisotropy. This contrast is
consistent with the different symmetry and band-edge electronic
structure of the two phases: the SOC-induced quasi-flat states of
$\beta$-tellurene favor a strongly direction-dependent excitonic
distribution, whereas the higher in-plane symmetry of the hexagonal
phase leads to a nearly isotropic spatial extension.

Importantly, these results show that the $\mathbb{Z}_2$ invariant
alone does not determine the strength or spatial extent of the
electron--hole correlation. Exciton binding and localization also
depend on the dispersion and orbital character of the band-edge
states, dielectric screening, spin--orbit coupling, and crystal
symmetry. The coexistence of a nontrivial quantum spin Hall electronic
structure with a strongly bound and spatially compact exciton in
hydrogen-passivated hexagonal tellurene therefore demonstrates that
strong electron--hole correlations are fully compatible with the
topological phase. Within the present calculations, the clearest
distinction between the topological hexagonal phase and the trivial
$\beta$ phase is not a suppression of exciton binding or localization,
but the markedly different spatial symmetry of their excitonic states.

The contrasting excitonic behavior of the investigated tellurium polymorphs can be interpreted in light of their previously established electronic and topological properties~\cite{araujo-2026}. In our earlier work, we demonstrated that $\beta$-tellurene develops SOC-induced quasi-flat bands near the band edges, whereas hydrogen-passivated hexagonal tellurene realizes a quantum spin Hall phase. The present results show that these distinct electronic structures lead to markedly different excitonic characteristics.

\section{Conclusions}

In this work, we investigated the quasiparticle electronic structure, optical response, and excitonic properties of two-dimensional tellurene polymorphs and the one-dimensional helical tellurium nanowire using many-body $GW$ and Bethe--Salpeter equation calculations. The combination of quasiparticle band structures, dielectric functions, momentum-resolved excitonic eigenvectors, and real-space exciton wave functions enabled a comprehensive microscopic description of how the crystal structure governs the optical properties across the tellurium family.

Our results demonstrate that the excitonic response is controlled not only by the dimensionality but also by the dispersion and orbital character of the electronic states forming the valence- and conduction-band edges. In particular, the quasi-flat band-edge states of $\beta$-tellurene induced by SOC concentrate the excitonic weight within a restricted region of reciprocal space, giving rise to a strongly localized near-infrared exciton with one of the highest binding energies among the investigated polymorphs ($0.40$~eV at HSE06 level). In contrast, the more dispersive band edges of $\alpha$-tellurene produce weakly bound excitations. Our results demonstrate that the excitonic response is controlled not
only by dimensionality but also by the dispersion and orbital character
of the electronic states forming the valence- and conduction-band
edges. In particular, the SOC-induced quasi-flat band-edge states of
$\beta$-tellurene concentrate the excitonic weight within a restricted
region of reciprocal space, giving rise to a strongly bound and
anisotropic near-infrared exciton with a direct binding energy of
0.40~eV at the $G_{4}W_{0}$@HSE06 level. In contrast, the more
dispersive band edges of $\alpha$-tellurene produce considerably weaker
electron--hole correlations. Remarkably, hydrogen-passivated hexagonal
tellurene combines a nontrivial quantum spin Hall electronic structure
with a strongly bound exciton ($E_b^d=0.51$~eV) and a compact,
nearly isotropic spatial distribution. This comparison demonstrates
that strong electron--hole correlations are fully compatible with
nontrivial topology, while exciton binding, localization, and
anisotropy remain governed by the detailed band-edge electronic
structure, dielectric screening, and crystal symmetry.

The calculated dielectric functions and derived optical constants further reveal that structural polymorphism provides an efficient mechanism for engineering optical anisotropy, spectral selectivity, and the operating spectral window from the near-infrared to the ultraviolet without changing chemical composition. The one-dimensional tellurium nanowire represents the strong-confinement limit of this behavior, exhibiting a large quasiparticle gap together with highly structured optical and energy-loss spectra.

Beyond the specific case of tellurium, this work establishes a direct microscopic connection between the band-edge electronic structure, the formation of excitons, and the macroscopic optical response in low-dimensional materials. These findings offer a microscopic framework for understanding and engineering excitonic and optical properties in low-dimensional semiconductors through crystal polymorphism and band-structure engineering.

\section{Acknowledgements}

Authors acknowledge financial support from the Brazilian funding agency CNPq under grant numbers 305174/2023-1, 408144/2022-0, and 305952/2023-4, 444069/2024-0 and 444431/2024-1.
FAPDF grant numbers 00193-00001817/2023-43 and 00193-00002073/2023-84, FAPDF/CAPES Centro-Oeste grant number 00193-00000867/2024-94 and CAPES-COFECUB grant number 88881.188740/2025-01.

We thank computational resources from Supercomputers LaMCAD/UFG, Santos Dumont/LNCC, and CENAPAD/Unicamp.

\section{Conflict of Interests}

There are no conflicts to declare.

\clearpage
\newpage

{\bf Supporting Information\\
Engineering Excitons through Polymorphism and Dimensional Confinement in Low-Dimensional Tellurium}

\section{Optical properties}

Additional optical response functions derived from the BSE dielectric
function are presented in Figs.~S1--S4. These results complement the
dielectric-function analysis discussed in the main text and illustrate
the phase- and polarization-dependent absorption, reflectivity,
electron energy-loss function, and refractive index of the investigated
tellurium structures.

\begin{figure}[!htb]
\centering
\begin{subfigure}[b]{0.19\textwidth}
    \includegraphics[width=\textwidth]{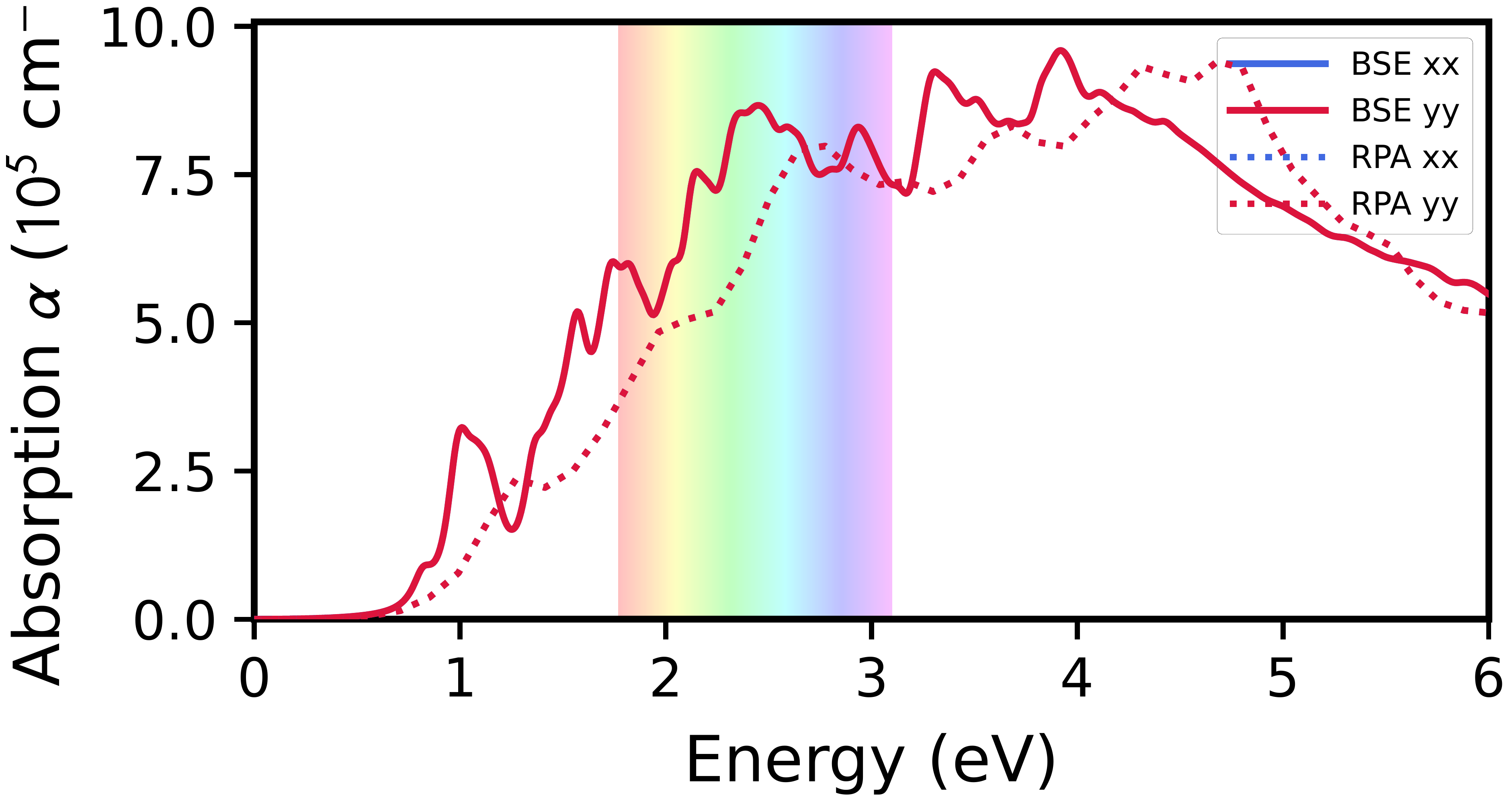}
    \subcaption{$\alpha$-Te}
\end{subfigure}
\begin{subfigure}[b]{0.19\textwidth}
    \includegraphics[width=\textwidth]{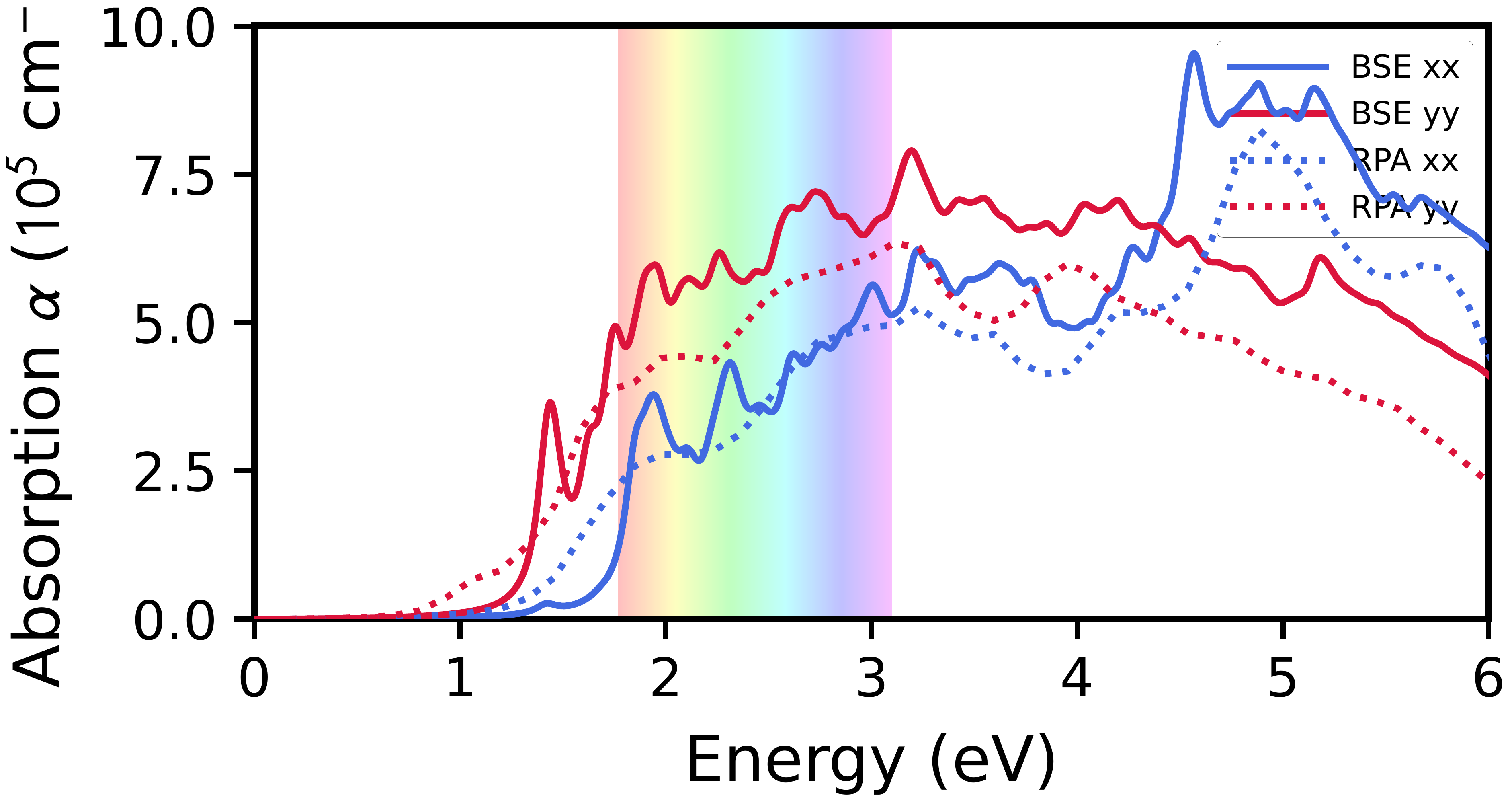}
    \subcaption{$\beta$-Te}
\end{subfigure}
\begin{subfigure}[b]{0.19\textwidth}
    \includegraphics[width=\textwidth]{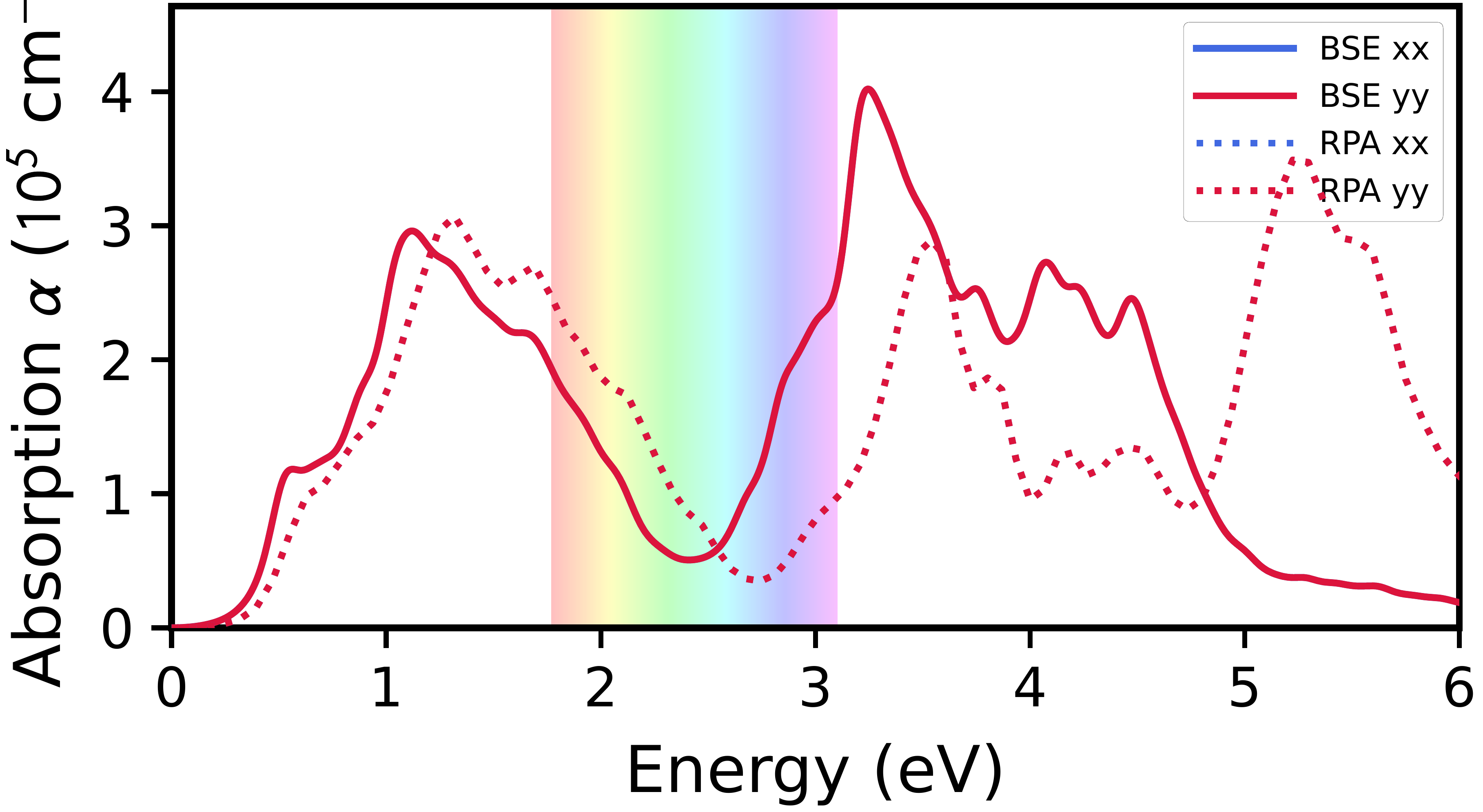}
    \subcaption{Hexagonal}
\end{subfigure}
\begin{subfigure}[b]{0.19\textwidth}
    \includegraphics[width=\textwidth]{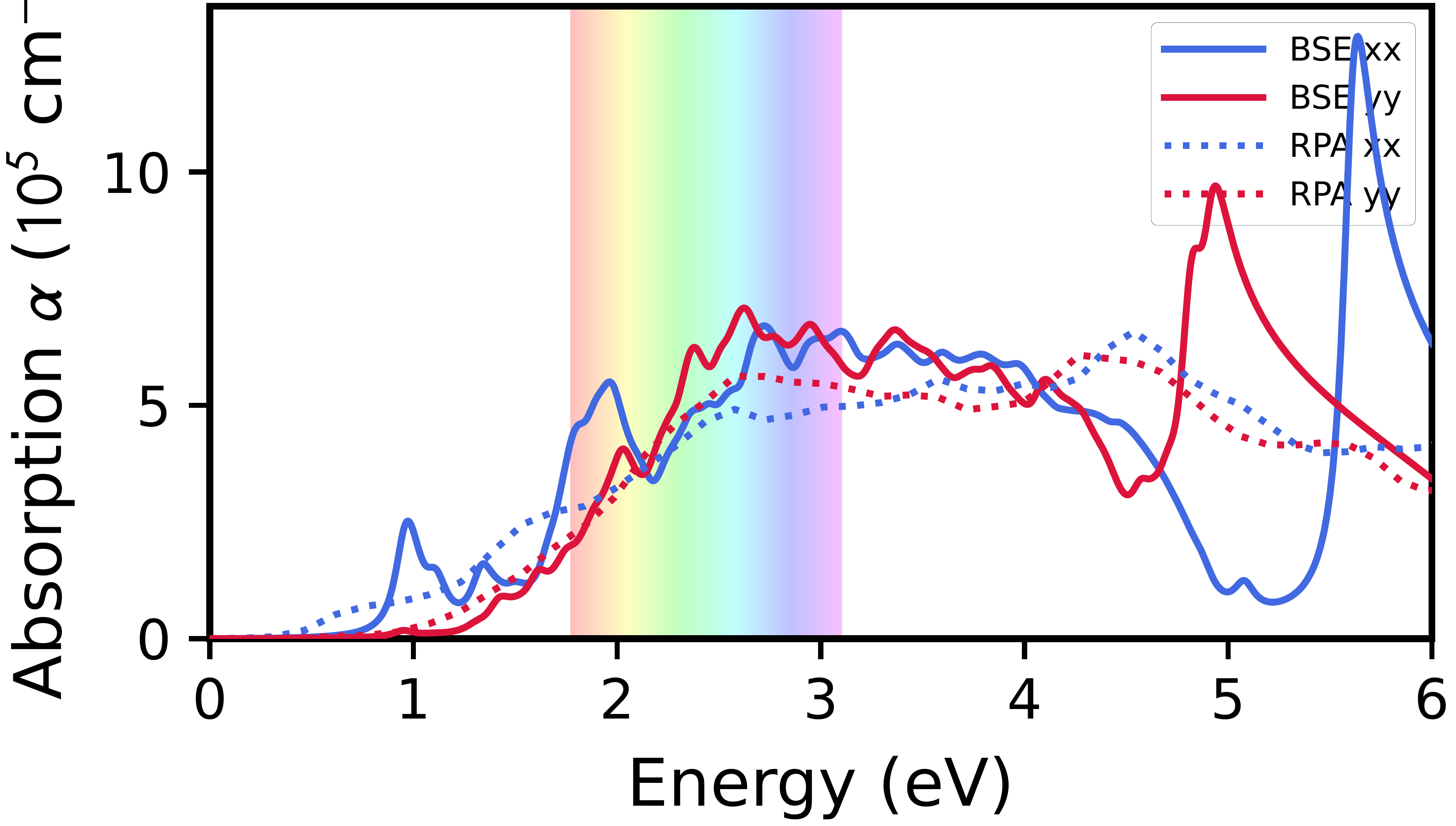}
    \subcaption{Pentagonal}
\end{subfigure}
\begin{subfigure}[b]{0.19\textwidth}
    \includegraphics[width=\textwidth]{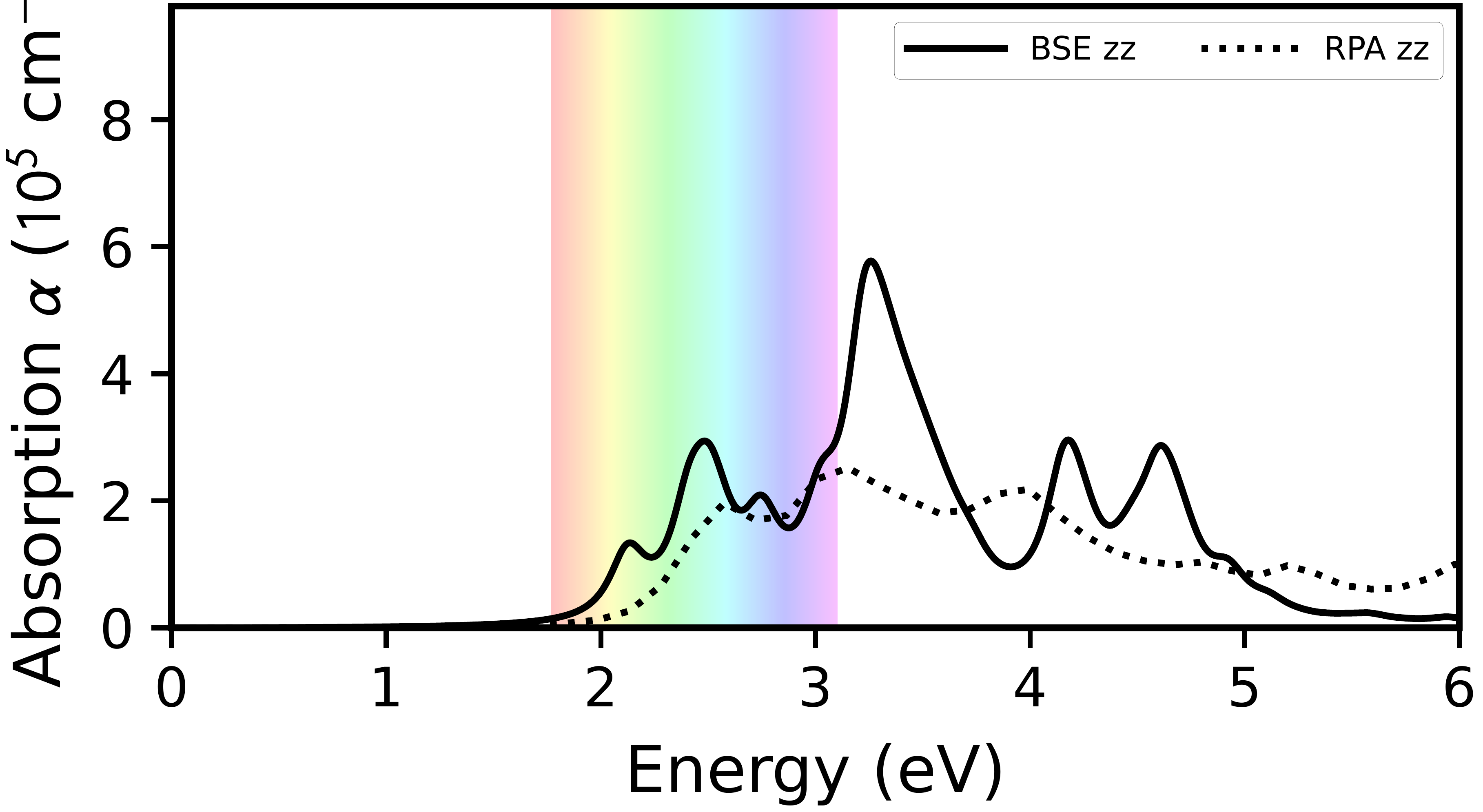}
    \subcaption{Te-h}
\end{subfigure}

\caption{Calculated optical absorption coefficient of the five tellurium phases investigated in this work: (a) $\alpha$-Te, (b) $\beta$-Te, (c) passivated hexagonal tellurene, (d) pentagonal tellurene, and (e) tellurium nanowire (Te-h).}
\label{fig:absorption}
\end{figure}

\begin{figure}[!htb]
\centering
\begin{subfigure}[b]{0.19\textwidth}
    \includegraphics[width=\textwidth]{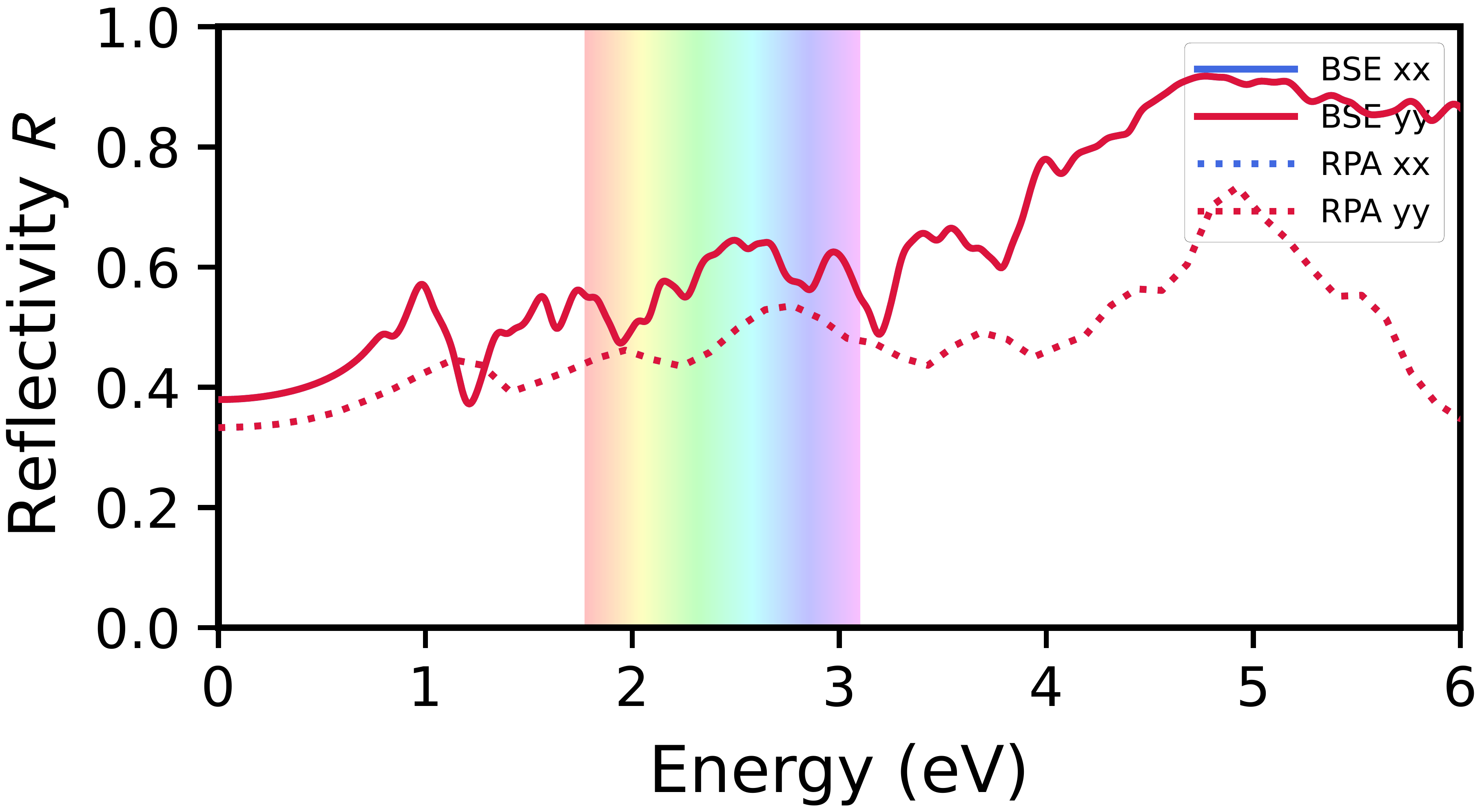}
    \subcaption{$\alpha$-Te}
\end{subfigure}
\begin{subfigure}[b]{0.19\textwidth}
    \includegraphics[width=\textwidth]{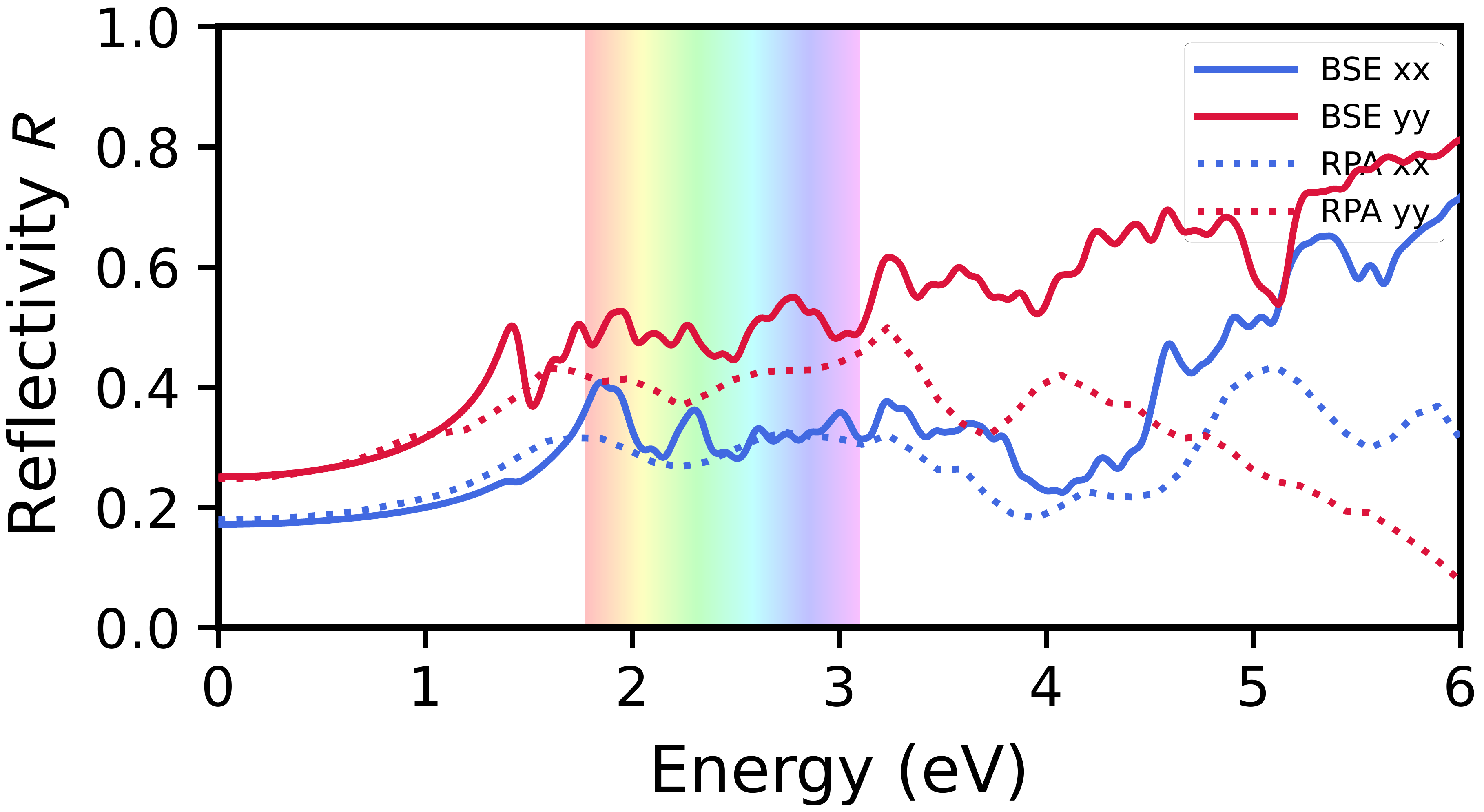}
    \subcaption{$\beta$-Te}
\end{subfigure}
\begin{subfigure}[b]{0.19\textwidth}
    \includegraphics[width=\textwidth]{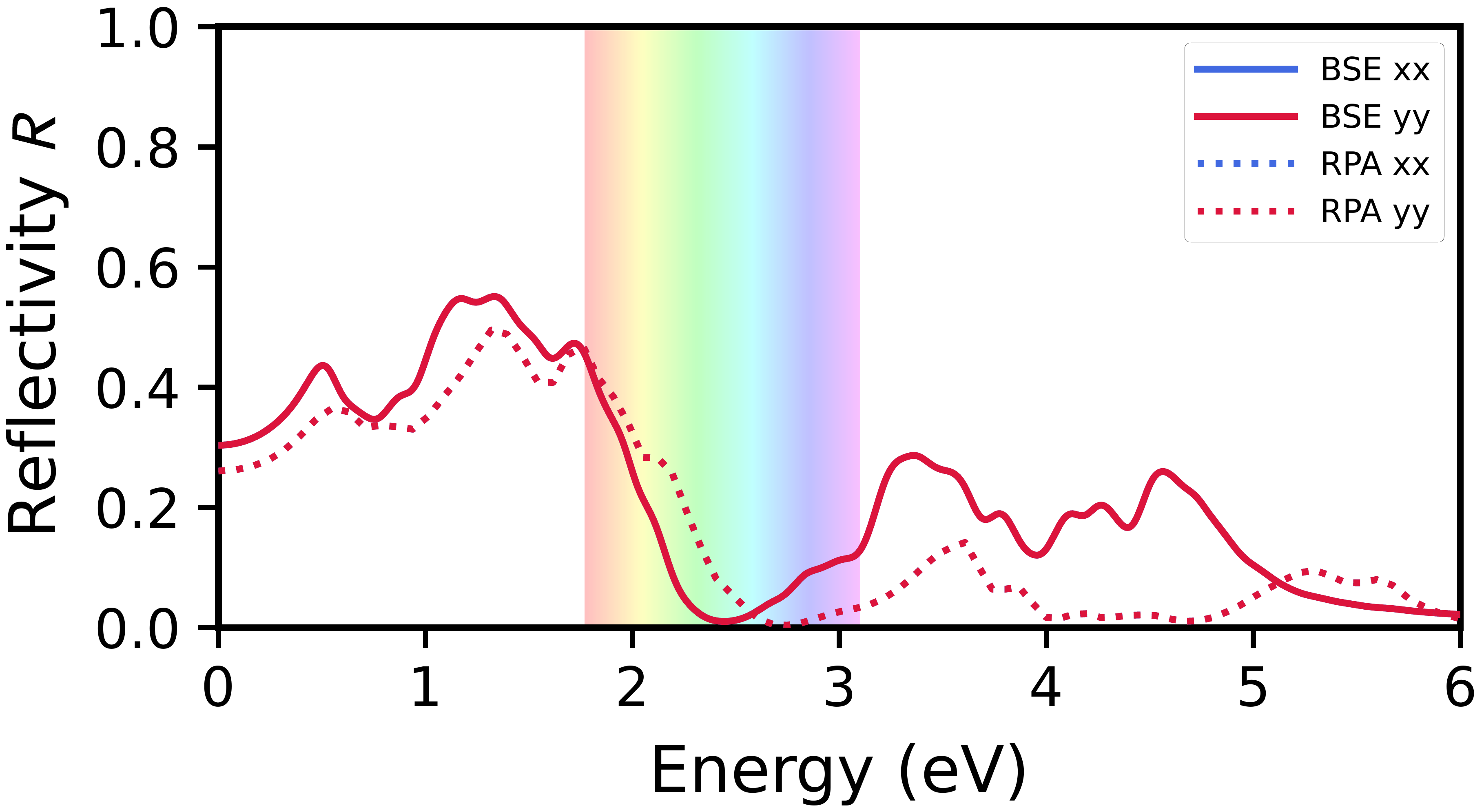}
    \subcaption{Hexagonal}
\end{subfigure}
\begin{subfigure}[b]{0.19\textwidth}
    \includegraphics[width=\textwidth]{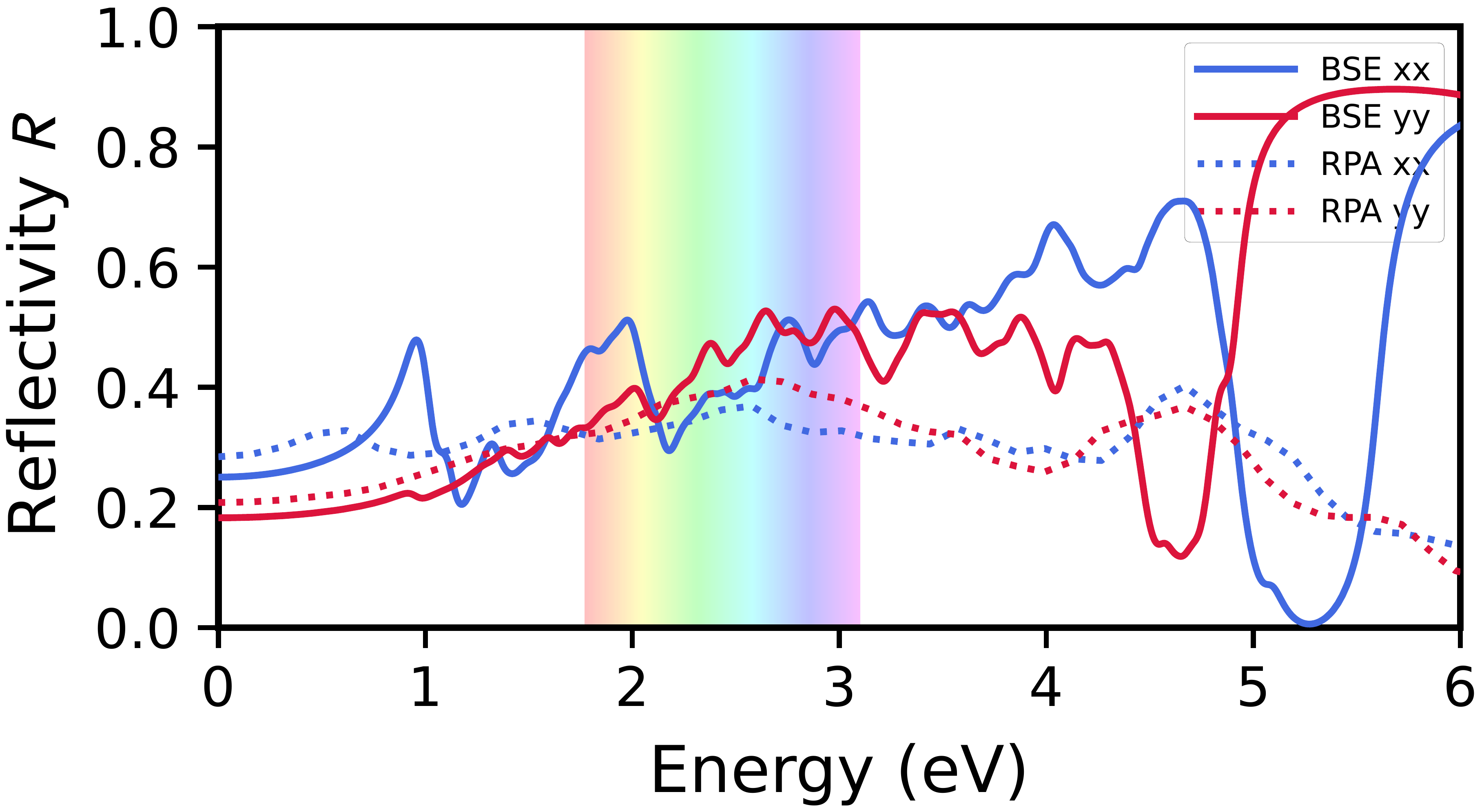}
    \subcaption{Pentagonal}
\end{subfigure}
\begin{subfigure}[b]{0.19\textwidth}
    \includegraphics[width=\textwidth]{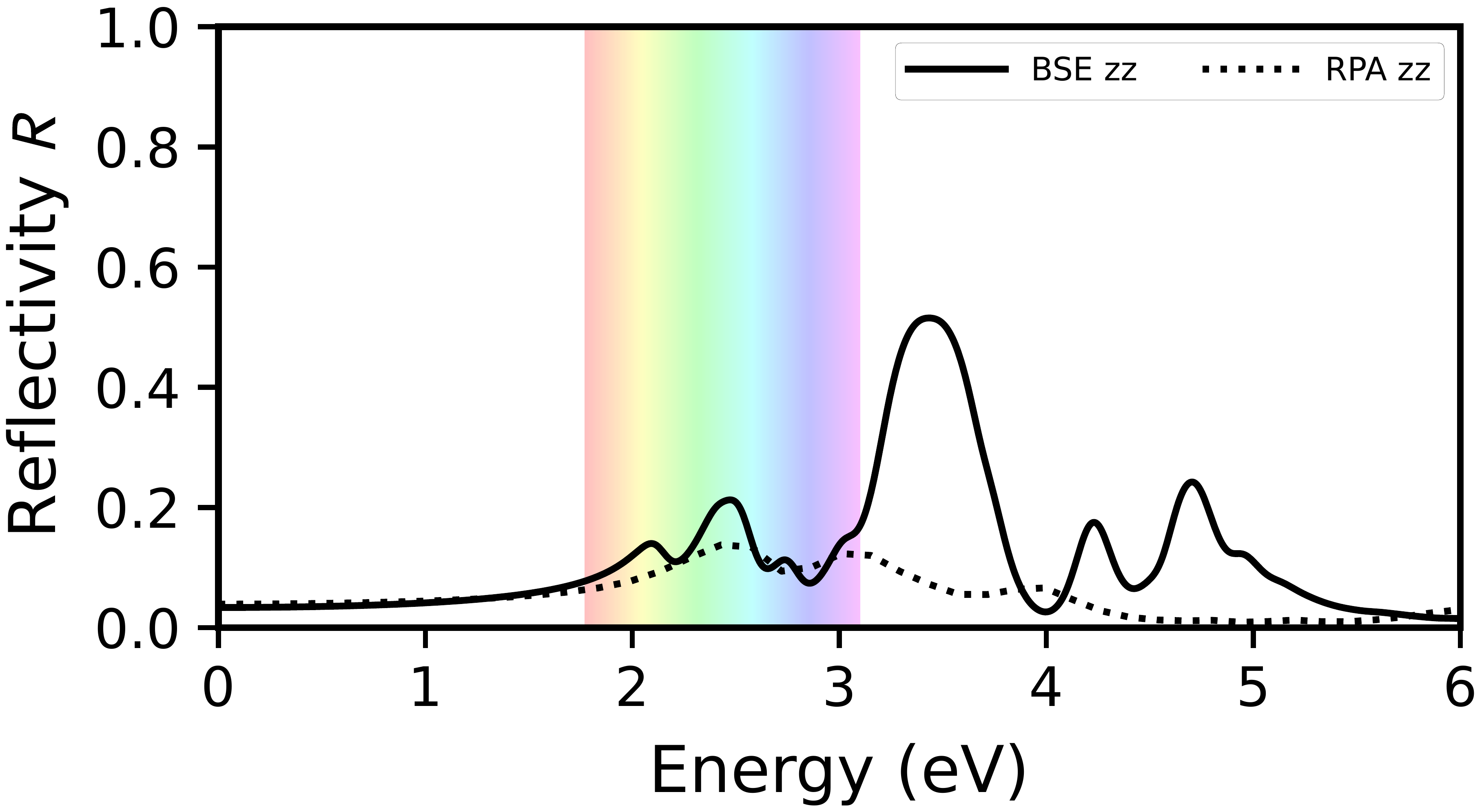}
    \subcaption{Te-h}
\end{subfigure}

\caption{Calculated reflectivity spectra of the five tellurium phases: (a) $\alpha$-Te, (b) $\beta$-Te, (c) passivated hexagonal tellurene, (d) pentagonal tellurene, and (e) tellurium nanowire (Te-h).}
\label{fig:reflectivity}
\end{figure}

\begin{figure}[!htb]
\centering
\begin{subfigure}[b]{0.19\textwidth}
    \includegraphics[width=\textwidth]{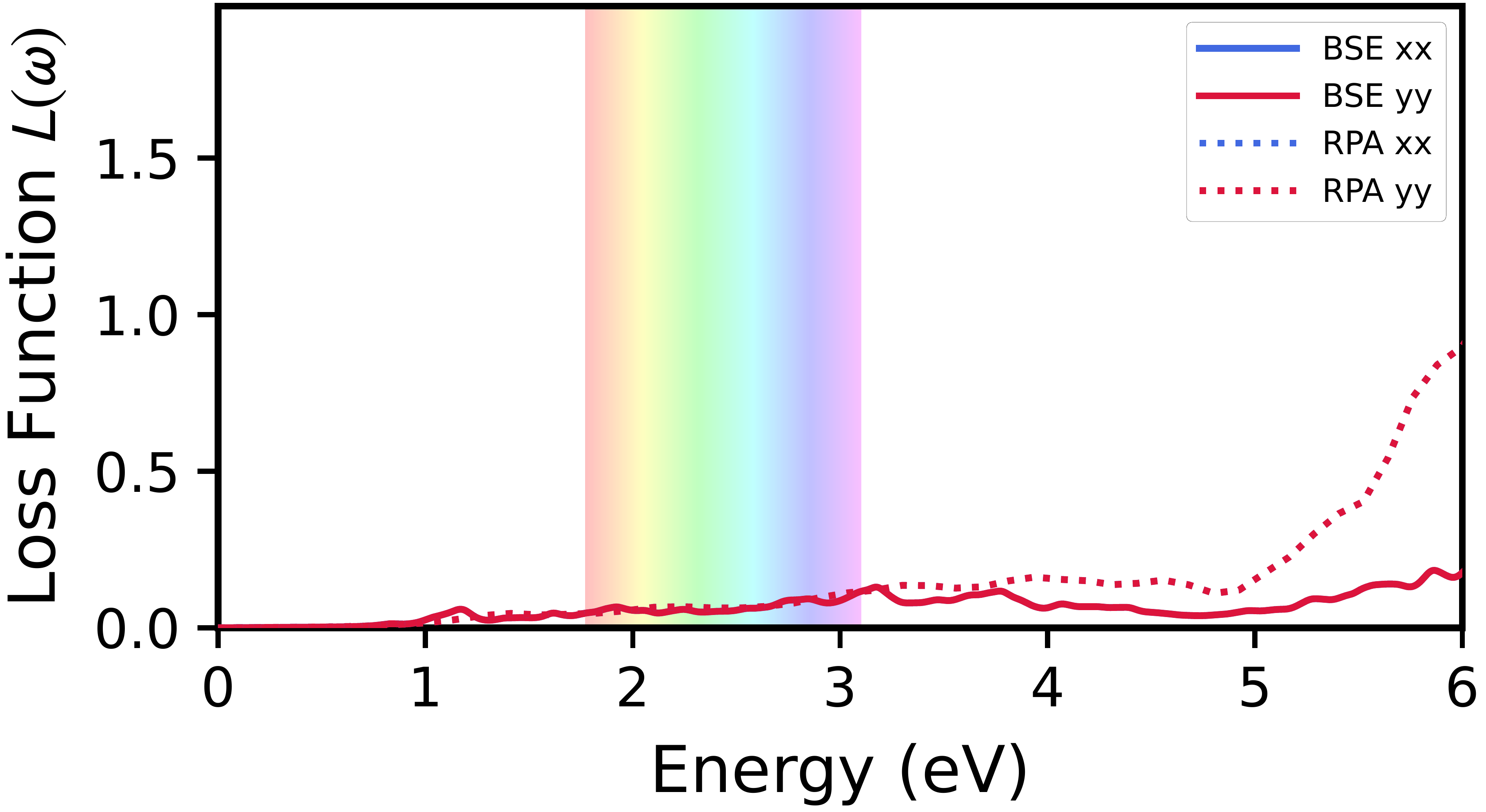}
    \subcaption{$\alpha$-Te}
\end{subfigure}
\begin{subfigure}[b]{0.19\textwidth}
    \includegraphics[width=\textwidth]{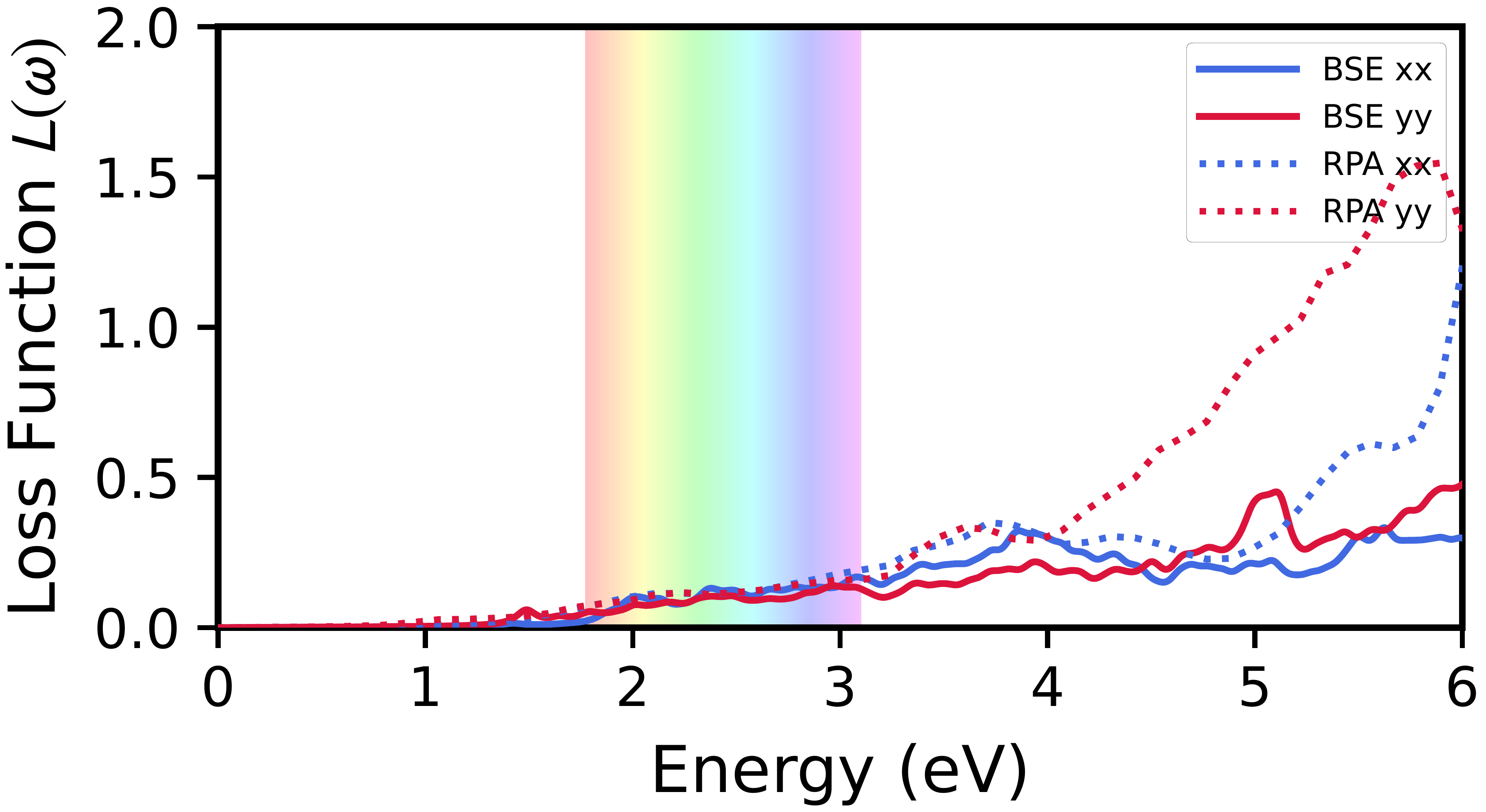}
    \subcaption{$\beta$-Te}
\end{subfigure}
\begin{subfigure}[b]{0.19\textwidth}
    \includegraphics[width=\textwidth]{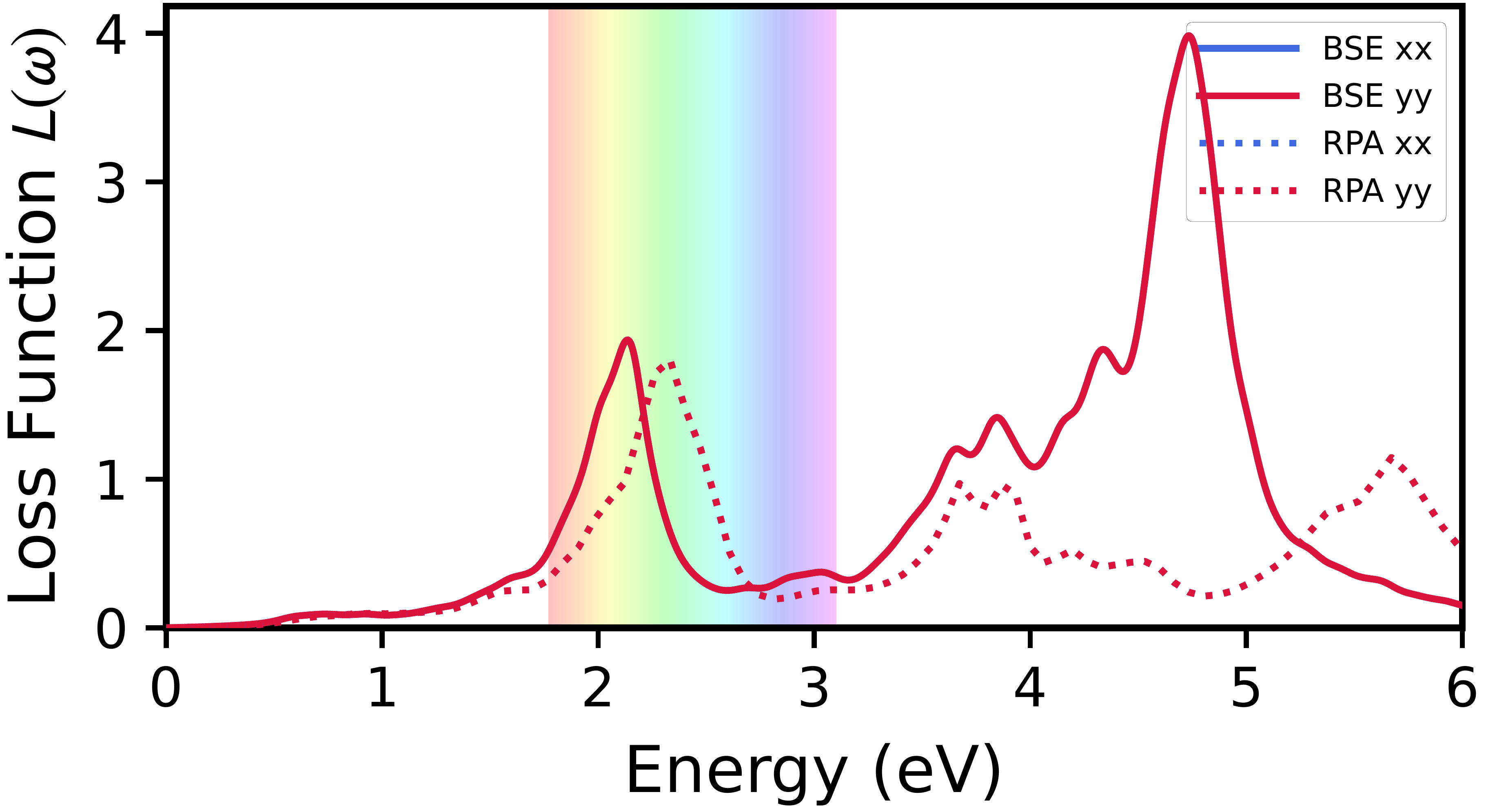}
    \subcaption{Hexagonal}
\end{subfigure}
\begin{subfigure}[b]{0.19\textwidth}
    \includegraphics[width=\textwidth]{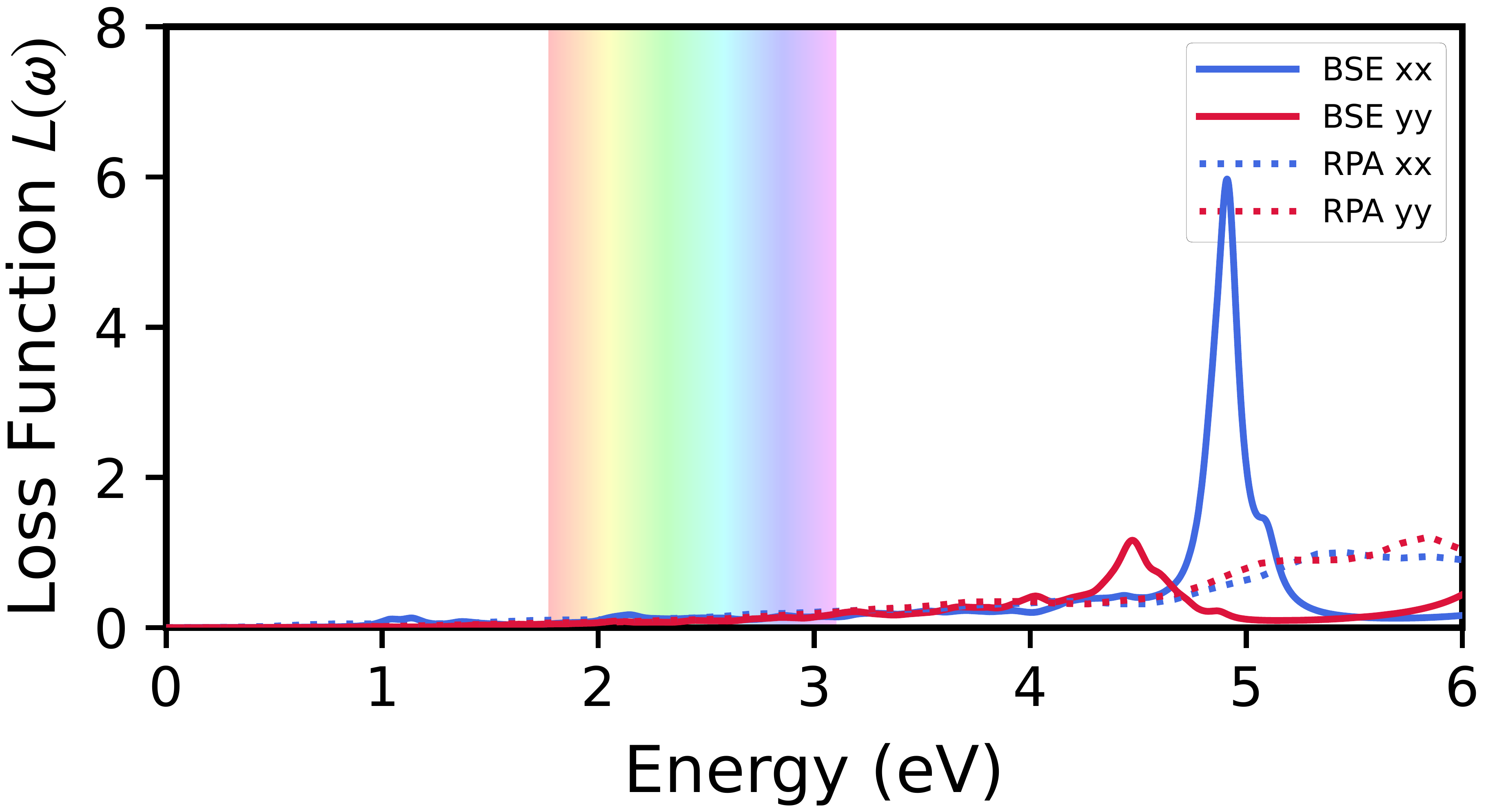}
    \subcaption{Pentagonal}
\end{subfigure}
\begin{subfigure}[b]{0.19\textwidth}
    \includegraphics[width=\textwidth]{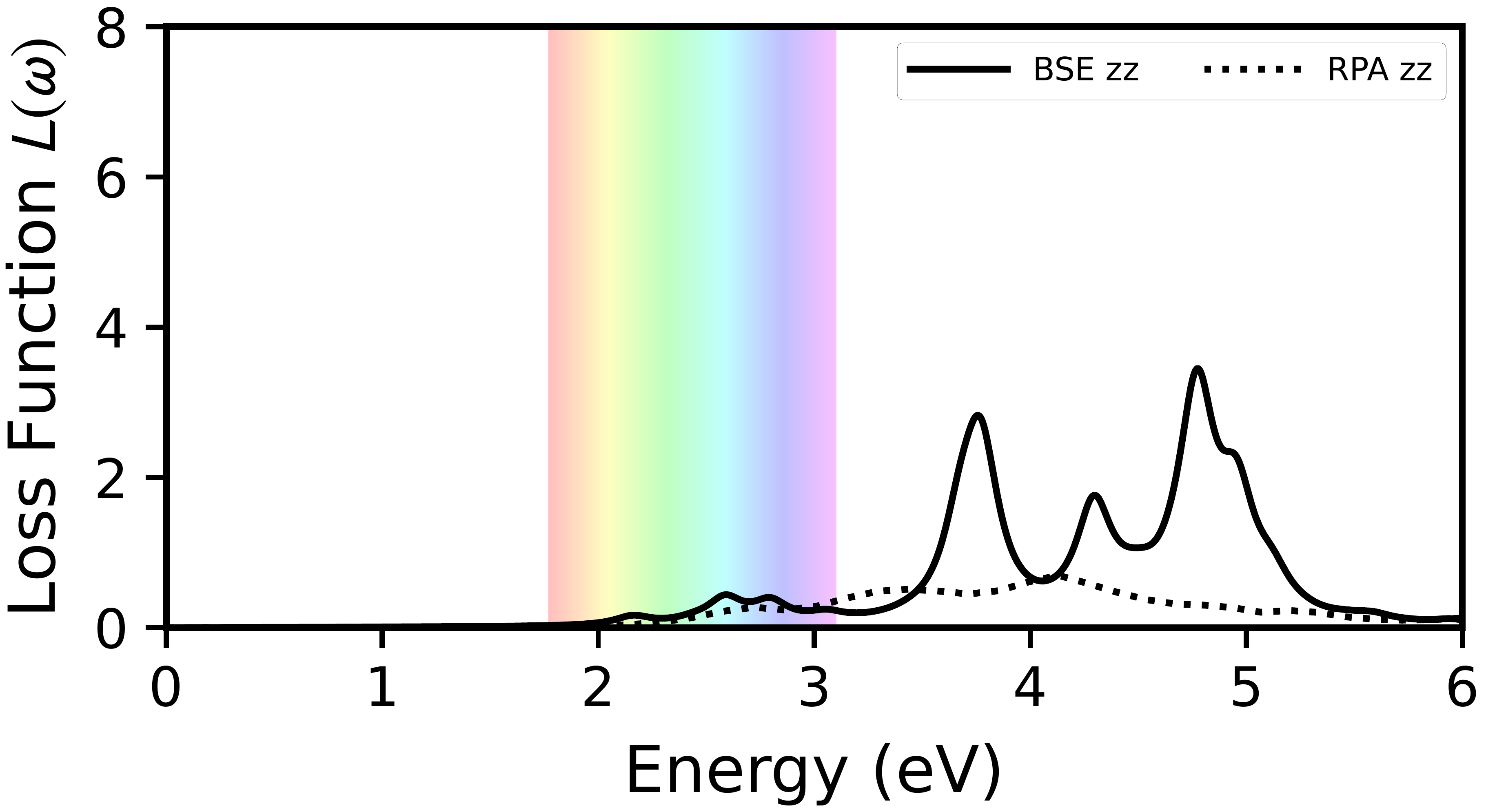}
    \subcaption{Te-h}
\end{subfigure}

\caption{Calculated electron energy-loss function of the five tellurium phases: (a) $\alpha$-Te, (b) $\beta$-Te, (c) passivated hexagonal tellurene, (d) pentagonal tellurene, and (e) tellurium nanowire (Te-h).}
\label{fig:loss}
\end{figure}
\begin{figure}[!htb]
\centering
\begin{subfigure}[b]{0.19\textwidth}
    \includegraphics[width=\textwidth]{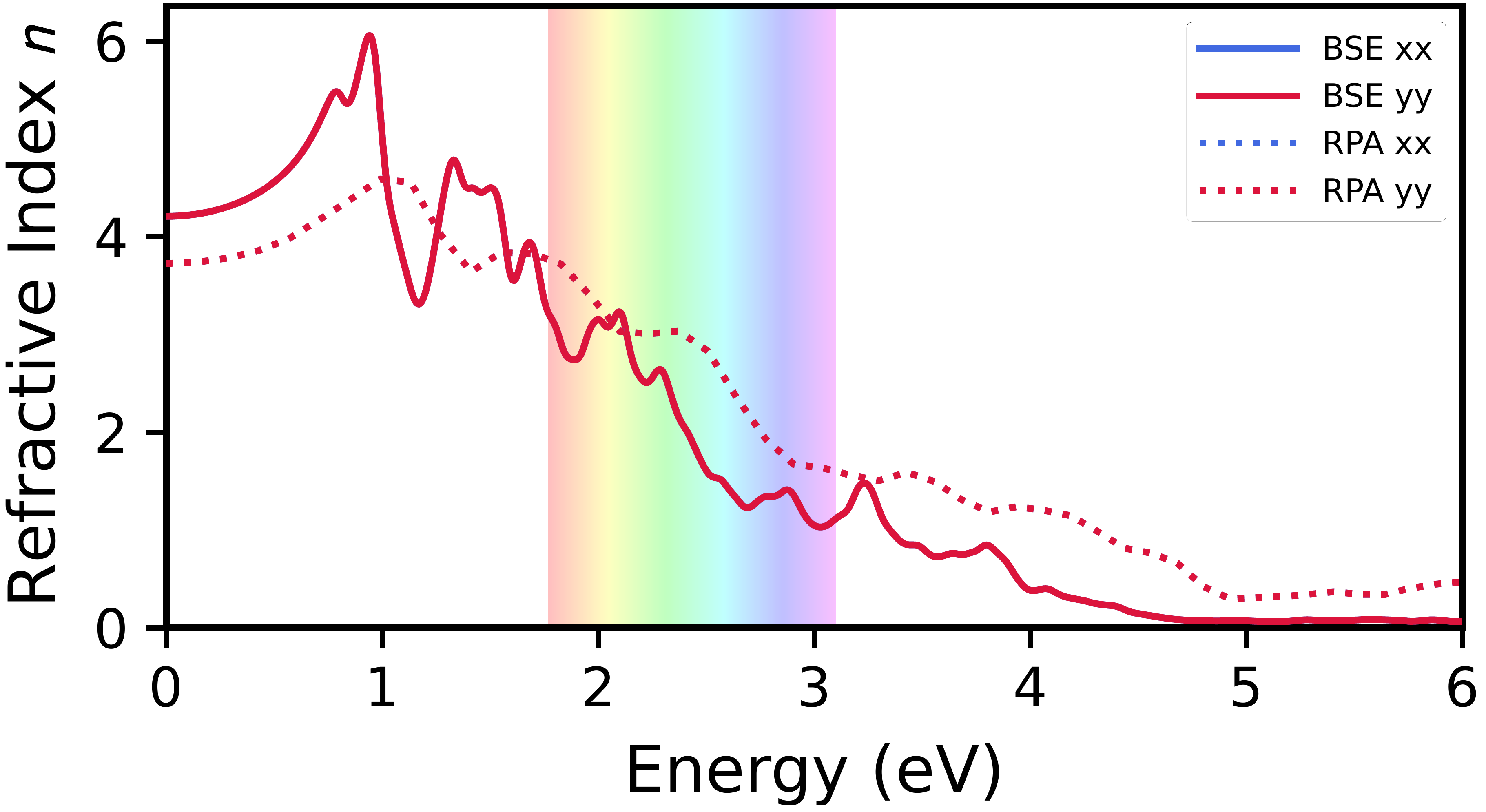}
    \subcaption{$\alpha$-Te}
\end{subfigure}
\begin{subfigure}[b]{0.19\textwidth}
    \includegraphics[width=\textwidth]{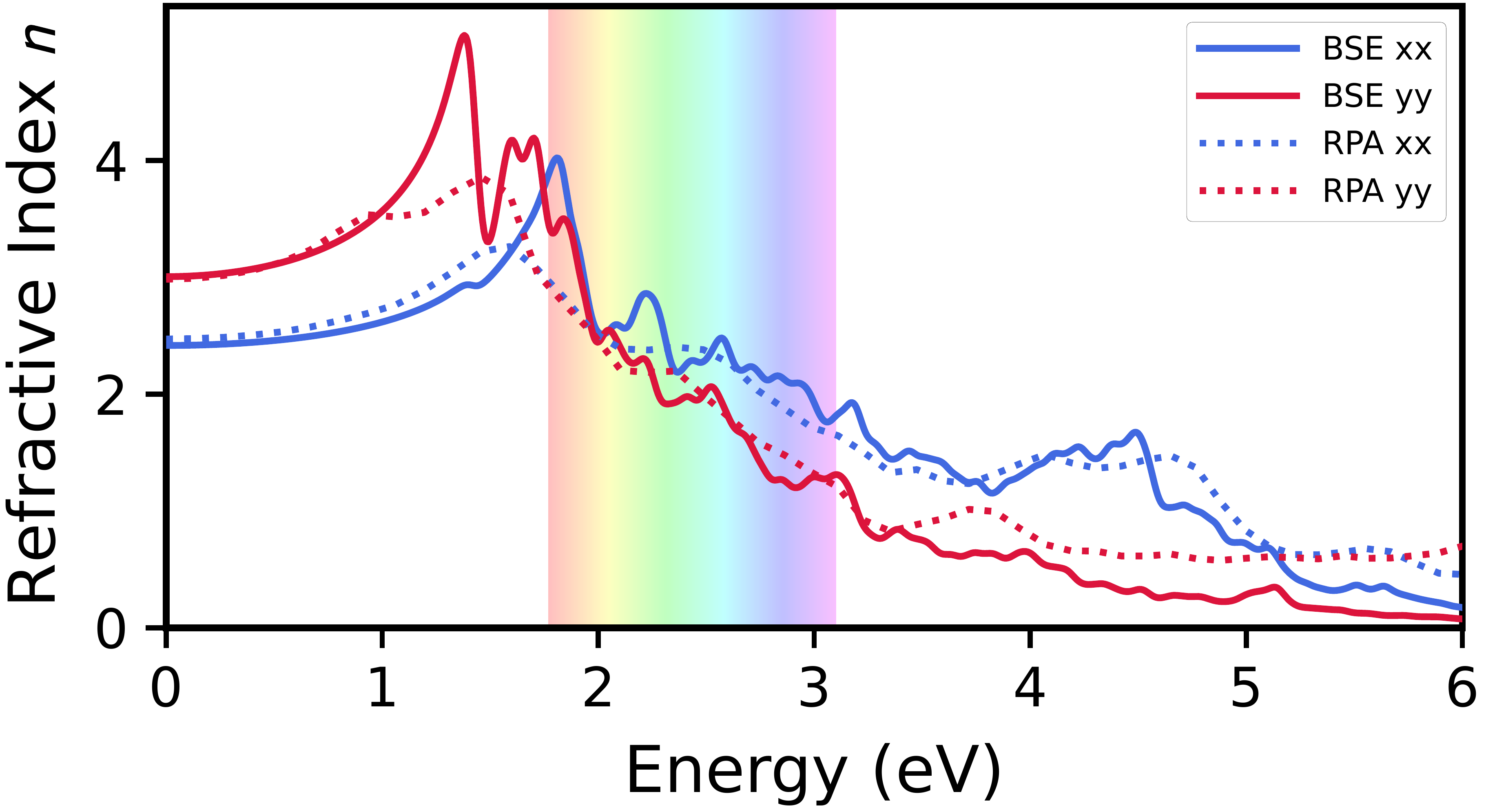}
    \subcaption{$\beta$-Te}
\end{subfigure}
\begin{subfigure}[b]{0.19\textwidth}
    \includegraphics[width=\textwidth]{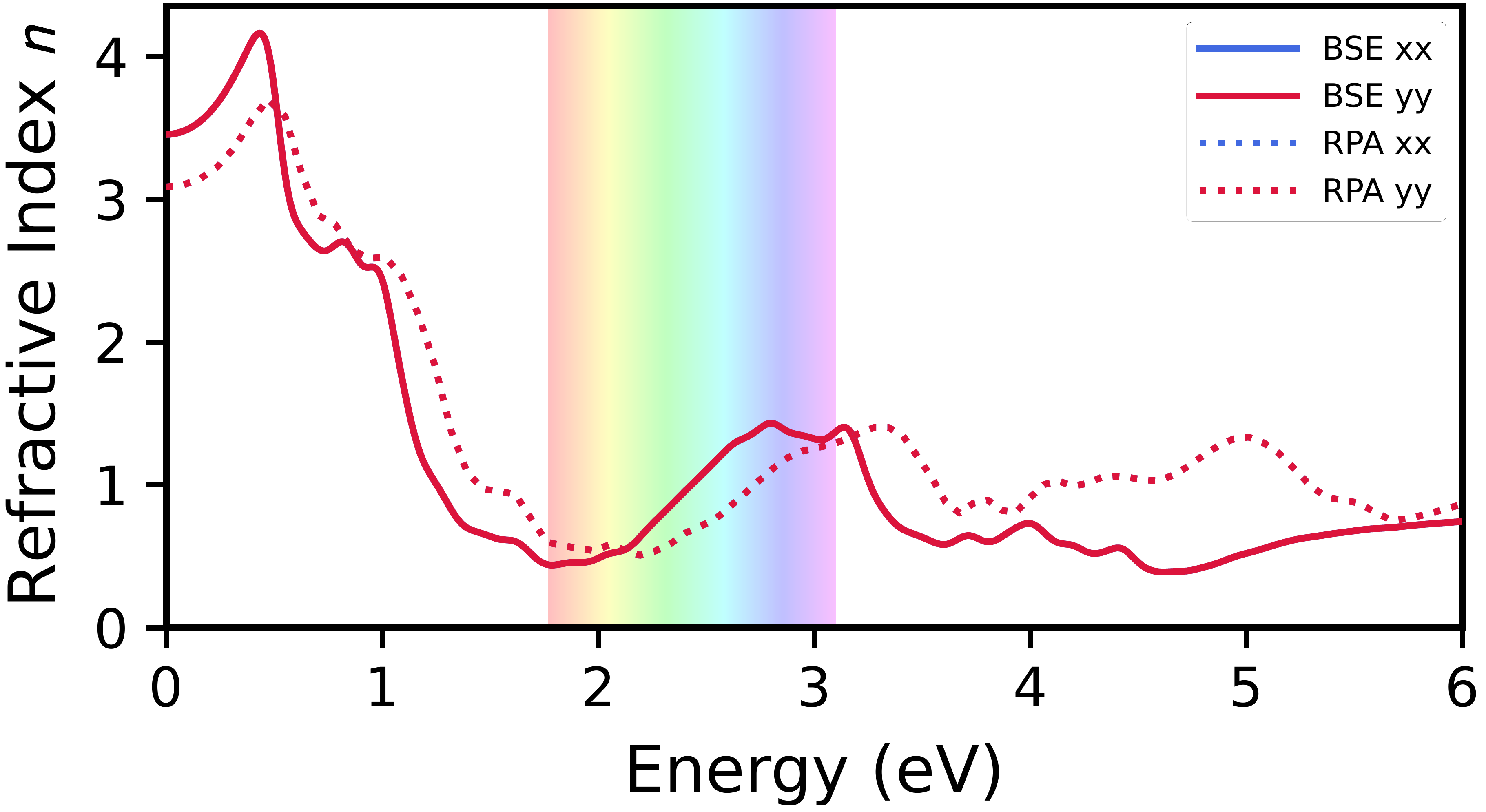}
    \subcaption{Hexagonal}
\end{subfigure}
\begin{subfigure}[b]{0.19\textwidth}
    \includegraphics[width=\textwidth]{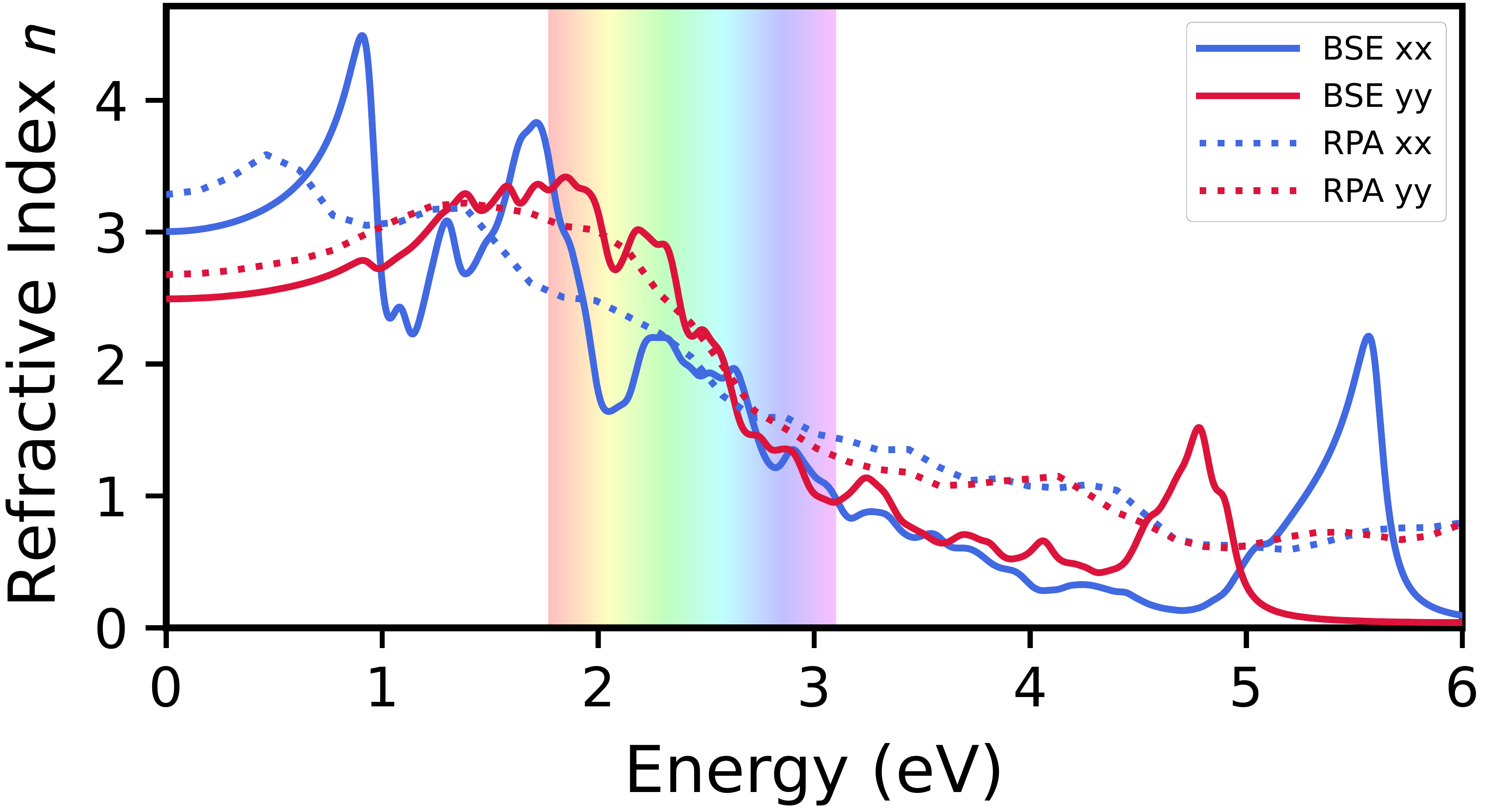}
    \subcaption{Pentagonal}
\end{subfigure}
\begin{subfigure}[b]{0.19\textwidth}
    \includegraphics[width=\textwidth]{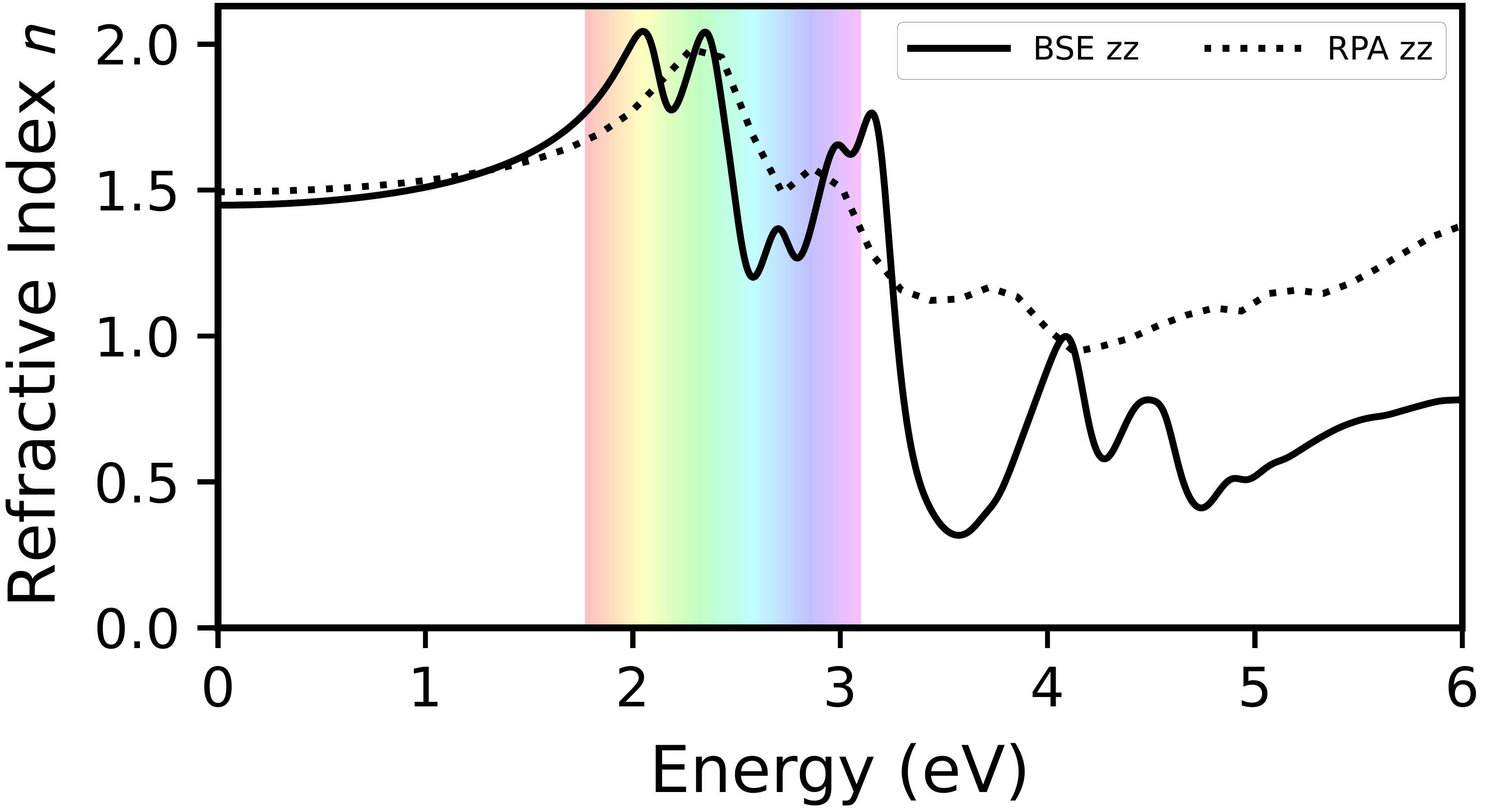}
    \subcaption{Te-h}
\end{subfigure}
\caption{Calculated refractive index of the five tellurium phases: (a) $\alpha$-Te, (b) $\beta$-Te, (c) passivated hexagonal tellurene, (d) pentagonal tellurene, and (e) tellurium nanowire (Te-h).}
\label{fig:refractive}
\end{figure}

\section{Excitons in trigonal tellurium}

\begin{figure}[!htb]
\centering
\begin{subfigure}[b]{0.6\textwidth}
\subcaption[]{}
    \includegraphics[width=\textwidth]{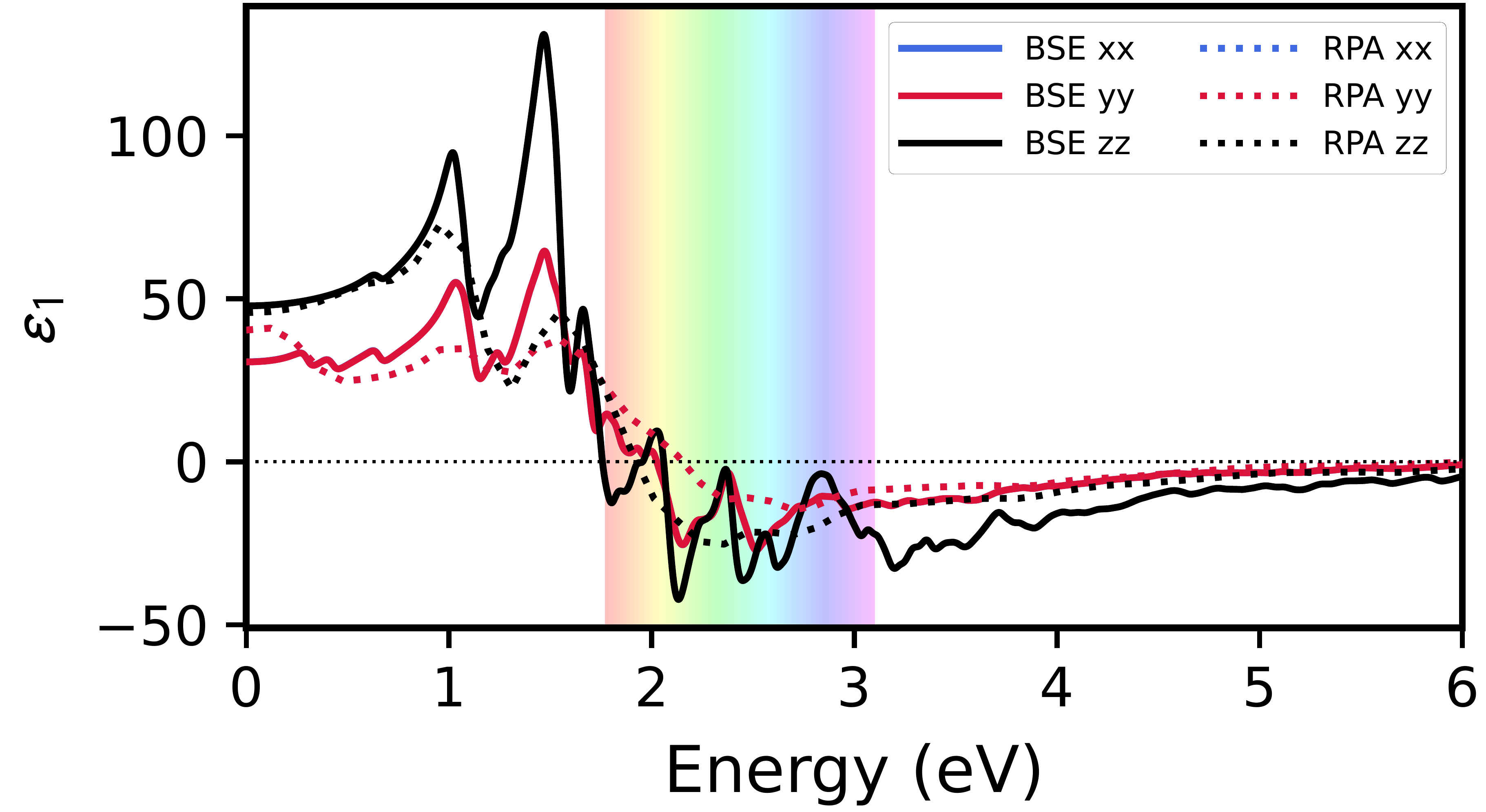} 
\end{subfigure}
\begin{subfigure}[b]{0.6\textwidth}
\subcaption[]{}
    \includegraphics[width=\textwidth]{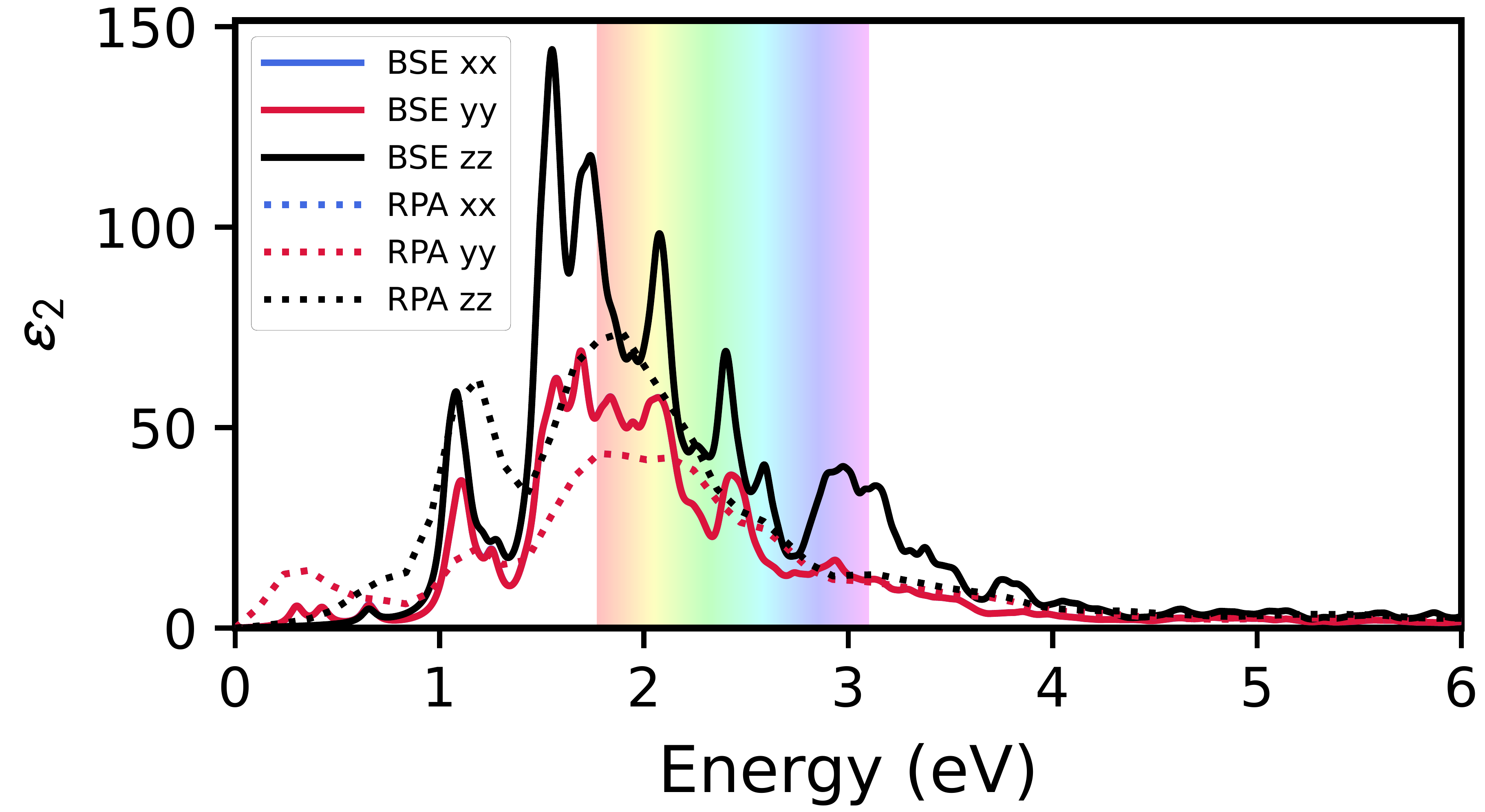}
\end{subfigure}
\caption{Real and imaginary parts of the dielectric function ($\varepsilon_1, \varepsilon_2$) calculated within BSE and RPA for trigonal tellurium.}
\end{figure}

\begin{figure}[!htb]
\centering
\begin{subfigure}[b]{0.6\textwidth}
    \includegraphics[width=\textwidth]{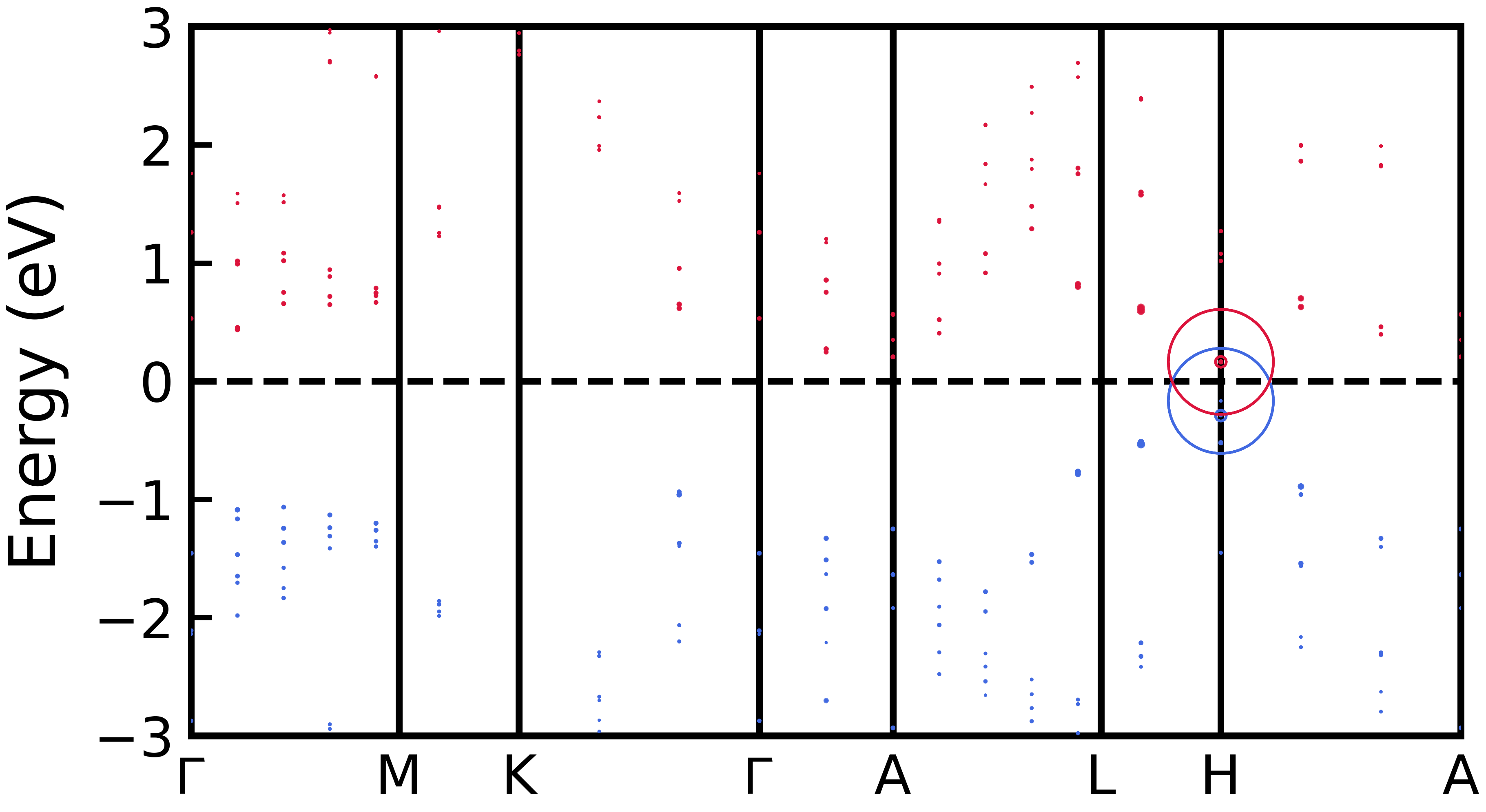}
\end{subfigure}
\caption{Momentum-resolved excitonic distributions obtained from the BSE eigenvectors for the lowest BSE excitonic eigenvalues in trigonal tellurium. The size of each marker is proportional to the excitonic weight, allowing direct visualization of the degree of localization in reciprocal space.}
\end{figure}

Bulk trigonal tellurium provides a useful reference for assessing the role of dimensional confinement. As shown in Fig. S5, its dielectric response is characterized by a broad low-energy spectral weight and pronounced polarization dependence, in contrast to the more strongly reshaped excitonic spectra found upon dimensional reduction. The momentum-resolved excitonic weights in Fig. S6 are distributed over several electronic states and \(k\)-points near the band edges, rather than displaying the strong confinement characteristic of the 1D nanowire. Overall, the trigonal phase represents the weak-confinement limit of the Te family, while the 2D polymorphs exhibit an intermediate, structure-dependent excitonic regime and the 1D nanowire displays the strongest electron–hole confinement.

\section{Momentum-resolved excitonic weights}

By exploiting the orthonormality relations of the single-particle Bloch states, defined as $\int \phi_{c\mathbf{k}}^*(\mathbf{r}_e) \phi_{c'\mathbf{k}'}(\mathbf{r}_e) \, d\mathbf{r}_e = \delta_{cc'} \delta_{\mathbf{k}\mathbf{k}'}$ and $\int \phi_{v\mathbf{k}}(\mathbf{r}_h) \phi_{v'\mathbf{k}'}^*(\mathbf{r}_h) \, d\mathbf{r}_h = \delta_{vv'} \delta_{\mathbf{k}\mathbf{k}'}$, the many-body exciton amplitudes $A_{vc\mathbf{k}}^S$ can be rigorously isolated by projecting the real-space two-particle wavefunction $\Psi_S(\mathbf{r}_e, \mathbf{r}_h)$ back onto the independent quasiparticle basis:

\begin{equation}
    A_{vc\mathbf{k}}^S = \iint \phi_{v\mathbf{k}}(\mathbf{r}_h) \phi_{c\mathbf{k}}^*(\mathbf{r}_e) \Psi_S(\mathbf{r}_e, \mathbf{r}_h) \, d\mathbf{r}_e d\mathbf{r}_h
\end{equation}

This projection filters out the individual contribution of each valence-to-conduction band transition at a specific $k$-point, establishing the direct mathematical link to the projected weights $W_{v\mathbf{k}}^S$ and $W_{c\mathbf{k}}^S$ used to construct the momentum-resolved exciton distributions. The corresponding excitonic weights on a given band are defined by tracing out the opposing degrees of freedom:

\begin{equation}
    W_{v\mathbf{k}}^S = \sum_{c} \left| A_{vc\mathbf{k}}^S \right|^2
\end{equation}
for the valence (hole) bands, and 
\begin{equation}
    W_{c\mathbf{k}}^S = \sum_{v} \left| A_{vc\mathbf{k}}^S \right|^2
\end{equation}

for the conduction (electron) bands. Here, $\left| A_{vc\mathbf{k}}^S \right|^2$ represents the probability of finding the correlated electron--hole pair at the reciprocal space point $\mathbf{k}$. The size of the scattered circles overlaid on the quasiparticle bands directly reflects these weights, $W_{v\mathbf{k}}^S$ and $W_{c\mathbf{k}}^S$, visually mapping the $k$-space localization of the many-body excitations.

For topologically nontrivial systems, the BSE excitonic states may additionally be interpreted within the framework of band quantum geometry. In this framework, the quantum geometric tensor (QGT) of the underlying Bloch states provides a connection between band topology and the spatial and dynamical properties of electron--hole pairs\cite{Kang2025}:

\begin{equation}
T_{ij}^{n}(\mathbf{k}) = g_{ij}^{n}(\mathbf{k}) - \frac{i}{2}\Omega_{ij}^{n}(\mathbf{k})
\end{equation}

where $g_{ij}^{n}(\mathbf{k}) = \text{Re}\langle \partial_{k_i} u_{n\mathbf{k}} | (1 - |u_{n\mathbf{k}}\rangle\langle u_{n\mathbf{k}}|) | \partial_{k_j} u_{n\mathbf{k}}\rangle$ represents the Fubini--Study metric, and $\Omega_{ij}^{n}(\mathbf{k}) = -2\text{Im}\langle \partial_{k_i} u_{n\mathbf{k}} | \partial_{k_j} u_{n\mathbf{k}}\rangle$ is the Berry curvature of band $n$. 

The imaginary part of the QGT introduces an effective gauge field into the relative motion of the exciton, coupling the relative orbital angular momentum $\mathbf{L}_{\text{rel}}$ to the Berry flux $\boldsymbol{\Omega}_c(\mathbf{k}) - \boldsymbol{\Omega}_v(\mathbf{k})$~\cite{Srivastava2015}:

\begin{equation}
H_{\text{rel}} = \frac{\mathbf{p}^2}{2\mu} + V_{\text{coul}}(\mathbf{r}) - \frac{e^2}{\hbar} \left[ \boldsymbol{\Omega}_c(\mathbf{k}) - \boldsymbol{\Omega}_v(\mathbf{k}) \right] \cdot \mathbf{L}_{\text{rel}}
\end{equation}

This coupling lifts the degeneracy between states with opposite orbital angular momentum ($m_{\ell} = \pm 1$), inducing a topological \textit{Lamb shift} between the $2p_+$ and $2p_-$ excitonic levels~\cite{Zhou2015}. Furthermore, the finite Berry curvature of the exciton center-of-mass wavepacket, $\boldsymbol{\Omega}_{\text{exc}}(\mathbf{Q})$, dictates an anomalous transverse velocity under an external potential gradient $\boldsymbol{\nabla} V_{\text{ext}}$:

\begin{equation}
\mathbf{v}_{\text{exc}}(\mathbf{Q}) = \frac{1}{\hbar} \boldsymbol{\nabla}_{\mathbf{Q}} E_{\text{exc}}(\mathbf{Q}) - \frac{1}{\hbar} \left( \boldsymbol{\nabla} V_{\text{ext}} \times \boldsymbol{\Omega}_{\text{exc}}(\mathbf{Q}) \right)
\end{equation}

yielding a net Exciton Hall Effect (EHE) for charge-neutral bound pairs~\cite{Yao2008}. 

Concurrently, the real part of the QGT imposes a fundamental geometric lower bound on the spatial extension of the exciton root-mean-square radius~\cite{Kang2025}:

\begin{equation}
\langle r^2 \rangle_{\text{exc}} \ge \int_{\text{BZ}} \text{Tr}\left[ g_c(\mathbf{k}) + g_v(\mathbf{k}) \right] d^2\mathbf{k}
\end{equation}

preventing spatial over-localization even in the strong-coupling regime and directly coupling the exciton size to the topology of the underlying Hilbert space.

\bibliography{referencias}
\end{document}